\documentclass[fleqn,usenatbib]{mnras}

\usepackage{newtxtext,newtxmath}
\usepackage[T1]{fontenc}

\DeclareRobustCommand{\VAN}[3]{#2}
\let\VANthebibliography\thebibliography
\def\thebibliography{\DeclareRobustCommand{\VAN}[3]{##3}\VANthebibliography}

\usepackage{graphicx}	% Including figure files
\usepackage{amsmath}	% Advanced maths commands

\usepackage{ulem} % To use \sout command

\newcommand{\kms}{\,km\,s$^{-1}$}
\newcommand{\Msun}{\,M$_{\sun}$}
\newcommand{\pc}{\,pc}
\newcommand{\kpc}{\,kpc}
\newcommand{\Myr}{\,Myr}
\newcommand{\Gyr}{\,Gyr}
\newcommand{\gcm}{\,g\,cm$^{-3}$}
\newcommand{\cmcube}{\,cm$^{-3}$}
\newcommand{\K}{\,K}
\newcommand{\erg}{\,erg}
\newcommand{\Msunkms}{\,M$_{\sun}$\,km\,s$^{-1}$}
\newcommand{\MsunYr}{\,M$_{\sun}$\,yr$^{-1}$}
\newcommand{\kpckms}{\,kpc\,km\,s$^{-1}$}

\newcommand{\MsunPc}{\,M$_{\sun}$\,pc$^{-2}$}
\newcommand{\kmskpc}{\,km\,s$^{-1}$\,kpc$^{-1}$}

\title[Efficient radial migration by GMCs]{Efficient radial migration by giant molecular clouds in the first several hundred Myr after the stellar birth}

\author[Y. Fujimoto, S.-i. Inutsuka and J. Baba]{
Yusuke Fujimoto,$^{1,2}$\thanks{E-mail: fujimoto@u-aizu.ac.jp}
Shu-ichiro Inutsuka$^{3}$
and Junichi Baba$^{4,5}$
\\
$^{1}$Department of Computer Science and Engineering, University of Aizu, Tsuruga Ikki-machi, Aizu-Wakamatsu, Fukushima 965-8580, Japan\\
$^{2}$Earth and Planets Laboratory, Carnegie Institution for Science, 5241 Broad Branch Road, NW, Washington, DC 20015, USA\\
$^{3}$Department of Physics, Nagoya University, Furo-cho, Chikusa-ku, Nagoya, Aichi 464-8602, Japan\\
$^{4}$Kagoshima University, Graduate School of Science and Engineering, Kagoshima 890-0065, Japan\\
$^{5}$National Astronomical Observatory of Japan, Mitaka, Tokyo 181-8588, Japan\\
}

\date{Accepted 2023 May 24. Received 2023 May 23; in original form 2023 May 9}

\pubyear{2022}

\begin{document}
\label{firstpage}
\pagerange{\pageref{firstpage}--\pageref{lastpage}}
\maketitle

\begin{abstract}
%This is a simple template for authors to write new MNRAS papers.
%The abstract should briefly describe the aims, methods, and main results of the paper.
%It should be a single paragraph not more than 250 words (200 words for Letters).
%No references should appear in the abstract.

Stars in the Galactic disc, including the Solar system, have deviated from their birth orbits and have experienced radial mixing and vertical heating. By performing hydrodynamical simulations of a galactic disc, we investigate how much tracer particles, which are initially located in the disc to mimic newborn stars and the thin and thick disc stars, are displaced from initial near-circular orbits by gravitational interactions with giant molecular clouds (GMCs). To exclude the influence of other perturbers that can change the stellar orbits, such as spiral arms and the bar, we use an axisymmetric form for the entire galactic potential. First, we investigate the time evolution of the radial and vertical velocity dispersion $\sigma_R$ and $\sigma_z$ by comparing them with a power law relation of $\sigma \propto t^{\beta}$. Although the exponents $\beta$ decrease with time, they keep large values of 0.3 $\sim$ 0.6 for 1\Gyr, indicating fast and efficient disc heating. Next, we find that the efficient stellar scattering by GMCs also causes a change in angular momentum for each star and, therefore, radial migration. This effect is more pronounced in newborn stars than old disc stars; nearly 30 per cent of stars initially located on the galactic mid-plane move more than 1\kpc\ in the radial direction for 1\Gyr. The dynamical heating and radial migration drastically occur in the first several hundred Myr. As the amplitude of the vertical oscillation increases, the time spent in the galactic plane, where most GMCs are distributed, decreases, and the rate of an increase in the heating and migration slows down.

\end{abstract}


%%%%%%%%%%%%%%%%%%%%%%%%%%%%%%%%%%%%%%%%%%%%%%%%%%

%%%%%%%%%%%%%%%%% BODY OF PAPER %%%%%%%%%%%%%%%%%%

\section{Introduction}

Radial mixing and vertical heating of stars in a galactic disc and physical processes that cause the stellar scattering are essential for understanding the structural and kinematic properties of the Galaxy and its formation and evolution (see the review by \citealt{Bland-Hawthorn_Gerhard_2016} and references therein). They are also important for understanding the birth environment of the early Solar system 4.6 Gyr ago and its difference from the current local environment \citep[e.g.][]{Fujimoto_Krumholz_Tachibana_2018, Fujimoto_et_al_2020, Desch_et_al_2022}.

It has long been predicted that the orbits of stars are changed by local fluctuations of the gravitational field in a galactic disc \citep[e.g.][]{Wielen_1977, Carlberg_1987}. The following two physical processes have been considered for stellar scattering; One, which is referred to as `churning', is the process that can change a star's angular momentum and, therefore, its guiding-centre radius \citep{Sellwood_Binney_2002}. The other, which is referred to as `blurring', is the process that increases a star's epicycle amplitude and, therefore, its peculiar velocity relative to the circular motion \citep{Schonrich_Binney_2009}. Therefore, the churning contributes to radial migration\footnote{Different authors use the term `radial migration' differently. This paper uses the term for a scattering event that changes a star's angular momentum regardless of a change in its epicycle motion. In contrast, some authors require the scattering event not to increase the epicycle amplitude or vertical oscillation \citep[e.g.][]{Vera-Ciro_et_al_2014, Martinez-Medina_et_al_2017}.} by changing the angular momentum, and the blurring contributes to dynamical heating of a disc by increasing the velocity dispersion.

The presence of super metal-rich stars \citep{Grenon_1972, Grenon_1989} and large dispersion in the age-metallicity relation \citep[e.g.][]{Edvardsson_et_al_1993, Haywood_2008, Casagrande_et_al_2011, Bergemann_et_al_2014, Rebassa-Mansergas_et_al_2021} in the Solar neighbourhood have suggested that many stars have mixed in the Galactic disc due to radial migration (churning) and dynamical heating (blurring). 
The Solar system also has a larger metallicity than the average metallicity of nearby stars of the same age \citep[][]{Edvardsson_et_al_1993}\footnote{We note that recent observations \citep{Haywood_2008, Gustafsson_et_al_2010, Casagrande_et_al_2011} show that the excess of Solar metallicity relative to the mean of same-age stars is not as large as shown in the previous observation by \citet{Edvardsson_et_al_1993}.}, suggesting that it has travelled from the inner region of the Galaxy, where the metallicity in the interstellar medium (ISM) was larger than in the outer region \citep[][see also \citealt{Nieva_Przybilla_2012}]{Wielen_Fuchs_Dettbarn_1996}. \citet{Wielen_Fuchs_Dettbarn_1996} estimated that the Solar birth radius $R_{\rm birth}$ was $\sim 6.6$\kpc\ using nearby F and G dwarfs from \citet{Edvardsson_et_al_1993}. \citet{Frankel_et_al_2018} estimated that $R_{\rm birth} \sim 5.2$\kpc\ using red clump giants from the APOGEE DR12 data. \citet{Minchev_et_al_2018} estimated that $R_{\rm birth} \sim 7.3$\kpc\ using main-sequence turn-off and subgiant stars from the HARPS data. \citet{Feltzing_Bowers_Agertz_2020} estimated that $5.5 < R_{\rm birth} < 7.0$\kpc\ using red giant branch stars from the APOGEE DR14 and the Gaia DR2 data. \citet{Tsujimoto_Baba_2020} estimated that $R_{\rm birth} < 5$\kpc\ by comparing the elemental abundance pattern of the Sun to those of solar twins. \citet{Lu_et_al_2022} estimated that $R_{\rm birth} \sim 4.5$\kpc\ using subgiant stars from the LAMOST DR7 spectroscopic and the Gaia eDR3 astrometric data. On the theoretical side, \citet{Minchev_Chiappini_Martig_2013} made use of a zoom-in cosmological galaxy formation simulation and estimated that $4.4 < R_{\rm birth} < 7.7$\kpc, with the highest probability of 5.6\kpc. \citet{Martinez-Medina_et_al_2017} performed a chemo-dynamical simulation with an observationally motivated galactic potential and estimated that $R_{\rm birth} \sim 7.9$\kpc. Although there is some uncertainty, almost all these estimated values are smaller than the current radius of the Solar system, which is thought to be around 8\kpc.

Non-axisymmetric galactic structures such as spiral arms and a central bar structure have been candidates that cause the radial and vertical mixing of the galactic stellar disc \citep[e.g.][]{Barbanis_Woltjer_1967, Jenkins_Binney_1990}. For example, recurrent transient spiral arms change the angular momentum of each star at multiple corotation radii without increasing their orbital eccentricities, resulting in radial migration without dynamical disc heating \citep[][see also \citealt{Daniel_Wyse_2015, Daniel_Wyse_2018}]{Sellwood_Binney_2002, Roskar_et_al_2008b, Roskar_et_al_2012, Baba_Saitoh_Wada_2013}. This has been thought to be a primary agent for radial migration. The bar, on the other hand, might contribute to increasing the velocity dispersion of the disc \citep{Friedli_Benz_Kennicutt_1994, Saha_Tseng_Taam_2010, Athanassoula_2013, Grand_et_al_2016}. Moreover, coupling of resonances between the spirals and bar might also enhance the radial migration in the inner galactic disc \citep[][see also \citealt{Tsujimoto_Baba_2020}]{Minchev_Famaey_2010, Brunetti_Chiappini_Pfenniger_2011}. In addition to the internal galactic structures, interactions or mergers with satellite galaxies can also affect the galactic disc's kinematics and cause radial and vertical mixing, especially in the outer disc. \citep[e.g.][]{Quillen_et_al_2009, Bird_Kazantzidis_Weinberg_2012, DOnghia_et_al_2016}.

Giant molecular clouds (GMCs), on which we focus in this paper, are also an important perturber that can diffuse the stellar orbit in a galactic disc \citep[e.g.][see also Section 8.4.1 in \citealt{Binney_Tremaine_2008}]{Spitzer_Schwarzschild_1953, Schonrich_Binney_2009}. In contrast to recurrent transient spiral arms, which are thought to be a primary agent for radial migration, gravitational scattering of stars by GMCs has been thought to be a kinematic heating agent that increases the velocity dispersion of the stellar disc, especially in the vertical direction \citep[e.g.][]{Lacey_1984, Gustafsson_et_al_2016}. Numerical calculations of stellar orbits of star--GMC encounters in a small patch of a galactic disc have shown that the increasing rate of the velocity dispersion can be fitted with a simple power law of the form $\sigma \propto t^{\beta}$ and that the exponent $\beta$ is $\sim$ 0.25 in both radial and vertical directions \citep{Kokubo_Ida_1992, Hanninen_Flynn_2002}. The observed age-velocity dispersion relation shows a larger exponent ranging between 0.2--0.5 \citep[e.g.][]{Nordstrom_et_al_2004, Sharma_et_al_2014, Mackereth_et_al_2019, Sharma_et_al_2021}, and recent galaxy simulations have suggested that it originates from different heating histories among different age populations. For instance, in the early evolutional phase of the Galactic disc, the gas mass fraction is high, so the initial peculiar velocities of newborn stars are large because the internal velocity dispersion of their parent GMCs is also large \citep[e.g.][]{Kumamoto_Baba_Saitoh_2017, Bird_et_al_2021}. In the late phase, as the gas is consumed by star formation, the importance of spiral arms and the bar as a heating agent increases relative to GMCs \citep{Aumer_Binney_Schonrich_2016a, Aumer_Binney_Schonrich_2016b}. 

The discussion about the stellar scattering by GMCs and the observed age-velocity dispersion relation, as described in the previous paragraph, has been on the 10\Gyr\ scale history of the Galactic disc. However, the short-term effect of GMCs on stellar scattering immediately after the stellar birth on the Galactic plane is not well understood. \citet{Kokubo_Ida_1992} find that when the galactic shear motion dominates relative motion between a star and a GMC, the velocity dispersion evolves as $\sigma_R \propto t^{0.5}$ and $\sigma_z \propto \exp (t)$. Once the velocity dispersion of stars grows and exceeds the shear velocity, the exponents converge to around 0.25. Their results suggest that the stellar scattering by GMCs is much more efficient in the first evolutionary phase just after stellar birth and that GMCs may cause not only dynamical heating but also significant radial migration.

A numerical simulation is a powerful tool for investigating the time evolution of the Galaxy. In particular, $N$-body galactic disc simulations have been broadly used to study the dynamical effects of various kinds of physical processes that cause stellar scattering \citep[e.g.][]{Halle_et_al_2018, Kamdar_et_al_2019, Mikkola_McMillan_Hobbs_2020, Wu_et_al_2020, Wu_et_al_2022}. For instance, \citet{Aumer_Binney_Schonrich_2016a} have performed $N$-body galactic disc simulations, incorporating not only stars but also GMCs as $N$-body particles with gravitational smoothing assuming spline-kernel potentials. With large parameter space, they compare the effects of each physical process on stellar scattering, including spiral arms, the bar, and GMCs. However, as they mention, the effects of GMCs are still not properly taken into account because real GMCs do not have spherical structures but are hierarchical and substructured. To be considered GMCs properly, hydrodynamic calculations are needed. Some studies have done $N$-body + hydrodynamic simulations of a galactic disc \citep{Roskar_et_al_2012, Khoperskov_et_al_2021}. Cosmological galaxy formation simulations have also treated the evolution of dark matter, stars, and interstellar gas in a self-consistent manner, and they have also been used for studies on stellar scattering \citep{Minchev_Chiappini_Martig_2013, Agertz_et_al_2021}. However, their spatial resolutions are not high enough to fully consider the distribution of GMCs because they focus more on the 10 Gyr scale long-term evolution of the galactic disc.

High-resolution observations of the Milky Way and nearby galaxies have revealed that interstellar gas is not randomly distributed in space but has kpc-scale filamentary structures along which GMCs are located \citep[e.g.][]{Rosolowsky_et_al_2003, Koda_et_al_2009, Miville-Deschenes_et_al_2017, Sun_et_al_2018, Maeda_et_al_2020}. The global distribution of interstellar gas and GMCs are thought to be governed by self-gravity, radiative cooling, stellar feedback, and galactic scale gas flows, such as spiral arms and galactic shear. GMCs themselves also interact with each other via gravity and hydrodynamic effects \cite[e.g.][]{Jog_Ostriker_1988}. Hydrodynamics is needed to account for such physical processes of interstellar gas. Recent high-resolution hydrodynamic galaxy simulations have made remarkable progress in treating interstellar gas dynamics self-consistently and reproducing the observed spatial distributions and properties of GMCs \citep[e.g.][]{Dobbs_Bonnell_Pringle_2006, Tasker_Tan_2009, Fujimoto_et_al_2019, Jeffreson_et_al_2021}.

In this paper, we investigate the 1 Gyr scale short-term effect of GMCs on stellar scattering, especially focusing on the effect immediately after stellar birth. For that purpose, we perform hydrodynamical simulations of a galactic disc with resolutions high enough to reproduce realistic GMC distributions. To exclude the influence of other perturbers that can change the orbits of stars, such as spiral arms and the central bar, we use an axisymmetric form for the entire galactic potential. Moreover, we place massless tracer particles in the galactic disc as initial conditions to mimic newborn stars and thin and thick disc stars. Then we use them to investigate the stellar scattering by GMCs, rather than to use star particles created by sub-grid models for star formation implemented in the simulations. That is because the star formation particles can have initial peculiar velocities inherited from the internal velocity dispersion of their parent GMCs. In addition, that is also because the large mass of the particles, 300\Msun\ in our simulations, can cause gravitational interaction with each other, especially after their birth as a group in a GMC, which might result in their unrealistic peculiar motion. Therefore, in order to focus only on stellar scattering off from GMCs, we use the massless tracer particles initially placed in a galactic disc. This paper is organised as follows. In Section~\ref{sec: methods}, we present our numerical model of a galactic disc. In Section~\ref{sec: results}, we first show the time evolution of the simulated galactic disc and statistical quantities such as total star formation rate and gas mass in the disc. We also show spatial distributions and properties of GMCs, such as mass and size. After that, as our main results, we show the statistical properties of stellar scattering by GMCs; we first show the heating history of the galaxy with the time evolution of the velocity dispersion of tracer particles. We then show the time evolution of orbital properties for each particle, such as the galactic radius, specific angular momentum, peculiar velocity, and height. In Section~\ref{sec: conclusions}, we summarise our findings.

%This is a simple template for authors to write new MNRAS papers.
%See \texttt{mnras\_sample.tex} for a more complex example, and \texttt{mnras\_guide.tex}
%for a full user guide.
%
%All papers should start with an Introduction section, which sets the work
%in context, cites relevant earlier studies in the field by \citet{Fournier1901},
%and describes the problem the authors aim to solve \citep[e.g.][]{vanDijk1902}.
%Multiple citations can be joined in a simple way like \citet{deLaguarde1903, delaGuarde1904}.

\section{Methods}
\label{sec: methods}

We study the gravitational scattering of stars induced by giant molecular clouds (GMCs) in a galactic disc by performing hydrodynamical simulations of the interstellar medium (ISM) of a Milky-Way-like disc galaxy. To represent the shear motion of the ISM in a flat rotation curve and, at the same time, to exclude the effects of non-axisymmetric galactic structures such as spiral arms and the bar, the simulations use a fixed axisymmetric logarithmic potential that represents the smooth gravitational field from the old stellar disc and dark matter halo (see Section~\ref{subsec: galaxy model}). To represent the dynamical evolution of the turbulent multiphase ISM and GMCs, we include hydrodynamics, self-gravity, radiative cooling, photoelectric heating, star formation, and stellar feedback in the form of photoionization, stellar winds and supernovae (see Section~\ref{subsec: hydrodynamics}). Note that our simulation is different from an ordinary $N$-body galaxy simulation in that the gravitational field of the entire galaxy is represented by an external potential, not by $N$-body particles. We also note that there are two types of particles in our simulations; we call them `tracer particles' and `star particles', respectively. The former is a massless particle that we initially place in the galaxy, and we trace their motion in the galactic disc and investigate the effect of stellar scattering by GMCs (see Section~\ref{subsec: tracer particles}). On the other hand, the latter is a particle that is repeatedly created by the star formation process through the dynamical evolution of the ISM (see Section~\ref{subsec: star particle mass}). Further details on our numerical method are given in the following subsections.

\subsection{Galaxy model}
\label{subsec: galaxy model}

\begin{table}
	\centering
	\caption{Summary of the simulation models compared in this paper.}
	\label{tab: models}
	\begin{tabular}{lll} % four columns, alignment for each
		\hline
		Simulation & Resolution & Star particle initial mass\\
		\hline
		\textit{8pc\_0msun} (fiducial) & $\Delta x =  8$\pc & $m_{*} =   0$\Msun\\ % SM_ver016
		\textit{16pc\_0msun}           & $\Delta x = 16$\pc & $m_{*} =   0$\Msun\\ % SM_ver015
		\textit{16pc\_300msun}         & $\Delta x = 16$\pc & $m_{*} = 300$\Msun\\ % SM_ver026
		\hline
	\end{tabular}
\end{table}

The galaxy is modelled in a three-dimensional simulation box of (128\kpc)$^3$ and 128$^3$ root grids with isolated gravitational boundary conditions and periodic fluid boundaries. In addition to the static root grid, we impose five additional levels of statically refined regions enclosing the whole galactic disc of 14\kpc\ radius and 1\kpc\ height, ensuring that the galactic disc is resolved with a maximum cell size of 31.25\pc. We further refine a cell by a factor of 2 if the Jeans length, $\lambda_{\rm J} = c_{\rm s} \sqrt{\upi / (G \rho)}$, drops below 16 cell widths and repeat the process until reaching a maximum refinement level we set as an initial parameter. In addition, after star formation, we require that any cell containing the newly formed star particle whose age is less than 40\Myr\ be refined to the maximum refinement level. We run two types of simulations with different resolutions; one has six refinement levels and a minimum cell size of 15.625\pc, and the other has seven refinement levels and a minimum cell size of 7.8125\pc, as we summarise in Table~\ref{tab: models}.

The simulated galaxy is set up as an isolated disc of gas orbiting in a static background potential which represents both an old stellar disc and dark matter halo. The form of the background potential is 
\begin{equation}
    \Phi(R, z) = \frac{1}{2} v_{\rm flat}^2 \ln{\left( R^2 + R_{\rm core}^2 + \frac{z^2}{q_{\Phi}^2} \right)} + {\rm constant},
	\label{eq: logarithmic potential}
\end{equation}
where $R$ and $z$ are the radius and height in the cylindrical coordinate system, $v_{\rm flat}$ is the constant circular velocity at large radii ($R \gg R_{\rm core}$) where the rotation curve becomes nearly flat, $R_{\rm core}$ is the core radius, and $q_{\Phi}$ is the axial ratio of the equipotential surfaces. The circular velocity is set to $v_{\rm flat} = 200$\kms, roughly consistent with the current Milky Way value \citep[e.g.][]{Reid_et_al_2019, VERA_2020} and ones for normal spiral galaxies \citep[e.g.][]{Sofue_et_al_1999}. The core radius is set to $R_{\rm core} = 0.5$\kpc\ so that a main disc region of $R \gtrsim 2$\kpc\ has a flat rotation curve. The axial ratio is set to $q_{\Phi} = 0.7$, with which the corresponding density does not become negative in a main disc region of $z \lesssim 3$\kpc\ (see the equation (2.71c) in \citealt{Binney_Tremaine_2008}). The form of the circular velocity is then given by
\begin{equation}
    v_{\rm circ}(R) = \frac{v_{\rm flat}R}{\sqrt{R^2 + R_{\rm core}^2}}.
	\label{eq: circular velocity}
\end{equation}
This logarithmic potential produces a flat rotation curve at large radii. For example, $v_{\rm circ}(4\rm\kpc) =  198.5$\kms\ and $v_{\rm circ}(8\rm\kpc) = 199.6$\kms. In addition, this axisymmetric potential makes it possible to exclude the effects of the galactic spiral arm and bar and focus only on GMCs as a perturber for circular orbiting stars.

\begin{figure}
    \centering
	\includegraphics[width=\columnwidth]{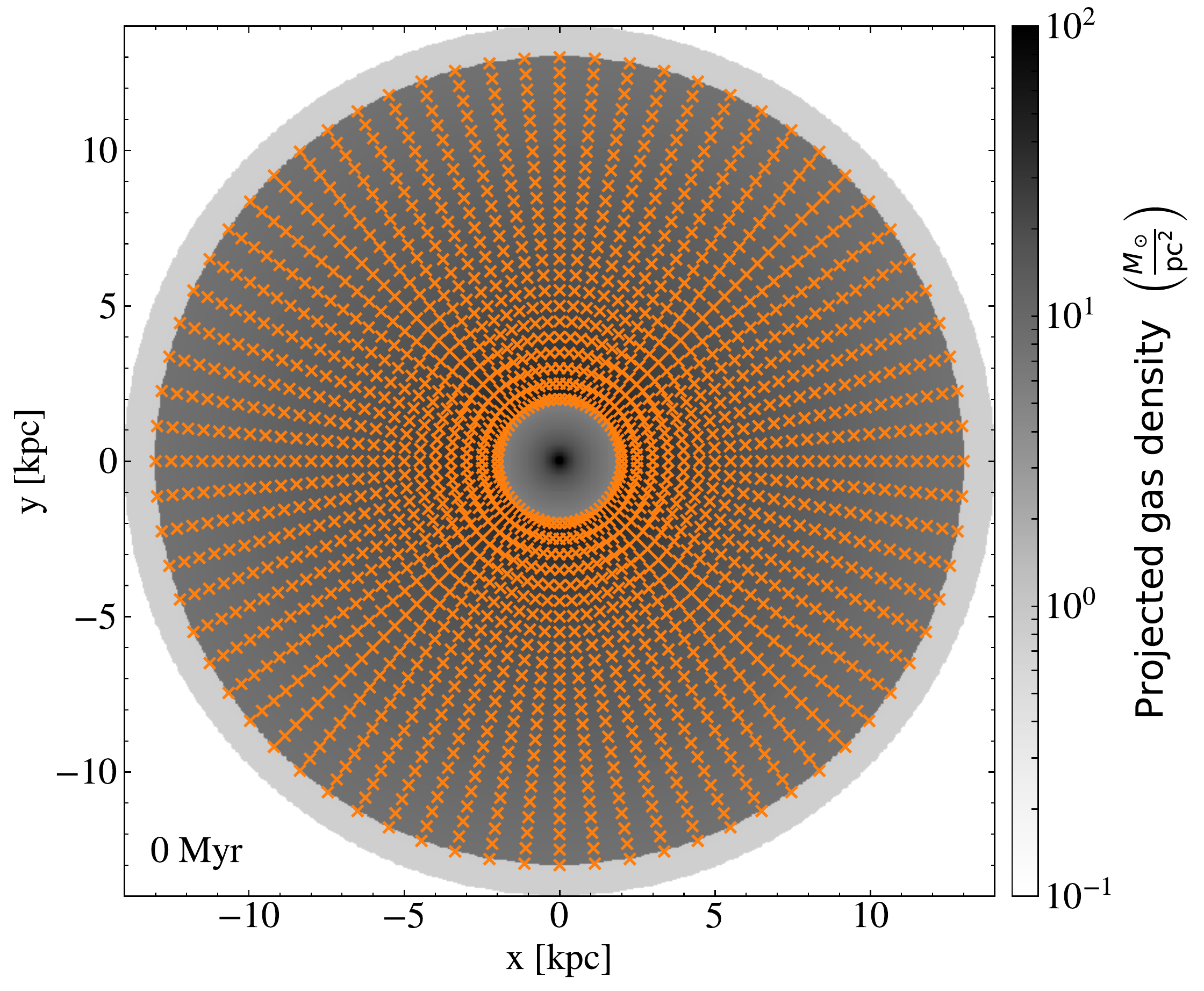} \\
    \includegraphics[width=\columnwidth]{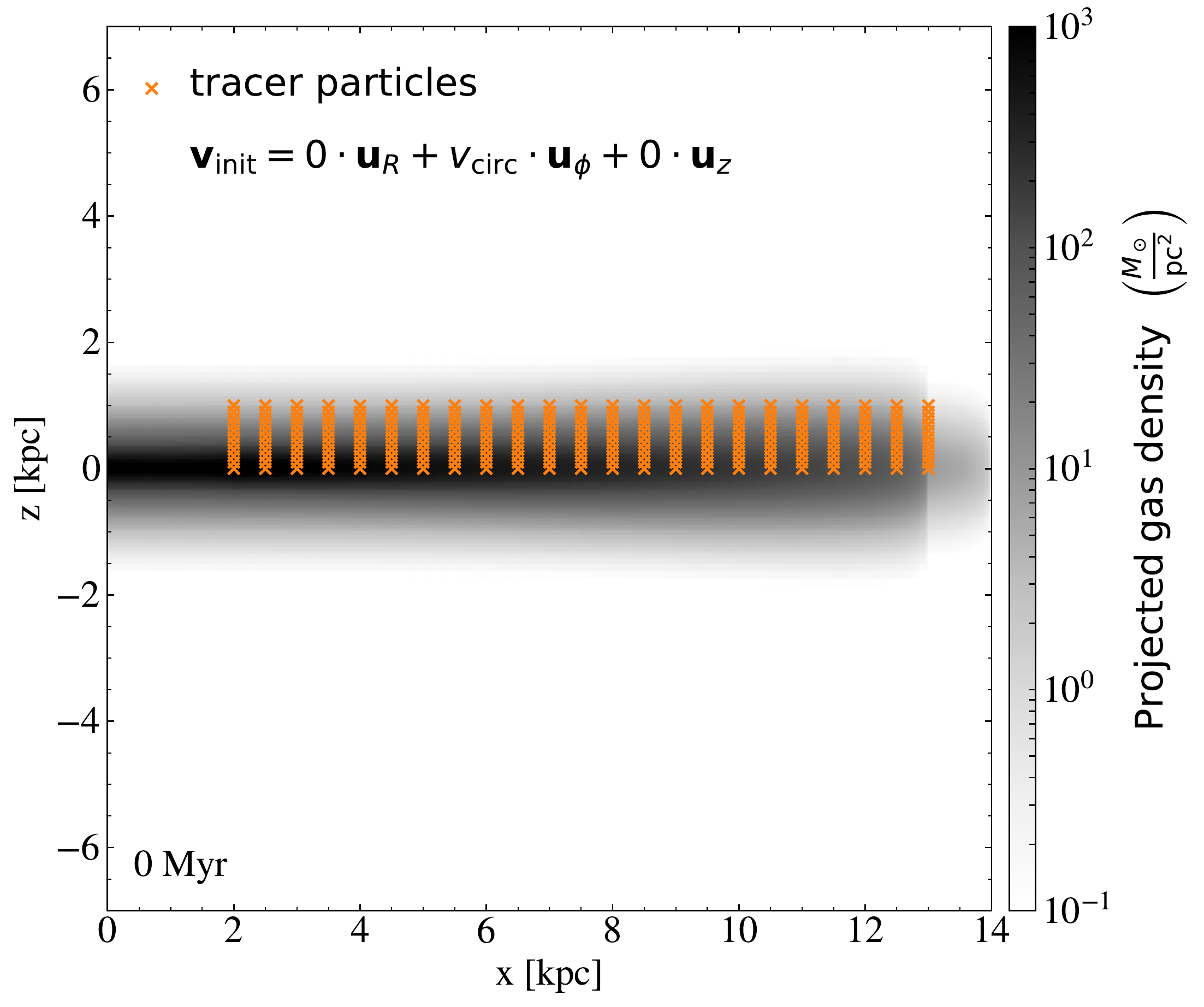}
    \caption{Initial distribution of the gas and tracer particles in the galactic disc. Panels show the face-on (top) and edge-on (bottom) disc, and the orange x marks show the positions of the tracer particles.}
    \label{fig: initial disc}
\end{figure}

The initial gas density distribution, as we show in Fig.~\ref{fig: initial disc}, is 
\begin{equation}
    \rho(R, z) = \frac{\kappa c_{\rm s}}{2 \upi G Q z_{\rm h}} {\rm sech}^2 \left( \frac{z}{z_{\rm h}} \right),
	\label{eq: initial gas distribution}
\end{equation}
where $\kappa = \sqrt{2} (v_{\rm circ} / R)$ is the epicyclic frequency for a flat rotation curve, $c_{\rm s}$ is the sound speed, here set equal to 10\kms, and $z_{\rm h}$ is the vertical scale height, which is assumed to increase with galactocentric radius following the observed radially dependent \ion{H}{i} scale height for the Milky Way presented in \citet{Binney_Merrifield_1998}. For instance, $z_{\rm h} \sim 290$\pc\ at $R = 8$\kpc. The initial gas disc profile is divided radially into three parts with the Toomre gravitational stability parameter $Q$. In the main region, between $R = 2$\kpc\ and $R = 13$\kpc, the initial gas density $\rho$ is set so that $Q = 2$. The other regions, from $R = 0$\kpc\ to $R = 2$\kpc\ and from $R = 13$\kpc\ to $R = 14$\kpc, are initialised with $Q = 20$. Beyond $R = 14$\kpc, the disc is surrounded by a very low-density gas of 10$^{-30}$\gcm. In total, the initial gas mass in the simulation box is $7.3\times10^9$\Msun, and the initial gas mass in the whole galactic disc of 14\kpc\ radius and 1\kpc\ height is $7.2\times10^9$\Msun. We do not consider gas inflow from the galactic halo.

The initial gas velocity is set in a circular motion as calculated via equation~(\ref{eq: circular velocity}). In addition to the circular motion, 10\kms\ velocity perturbation is added.

Since our goal is not to reproduce the detailed structures of the current Milky Way but to model a typical disc galaxy with a mass size similar to the Milky Way, we have not fine-tuned the parameters of the galaxy model. For instance, the circular velocity of $v_{\rm flat} = 200$\kms\ is slightly smaller than the inferred Milky Way value of $\sim 230$\kms (e.g. \citealt{Reid_et_al_2019}; \citealt{VERA_2020}; see also the review by \citealt{Sofue_2020}). Regarding the vertical frequency $\nu$, it can be estimated with the form of $\nu = \sqrt{4 \upi G \rho_{\rm total}}$ where $\rho_{\rm total}$ is the total density of stars, dark matter and the gas, and then we get $\nu \sim 50$\kmskpc at ($R$, $z$) = (8\kpc, 0\pc) in our model, which is smaller than the inferred Milky Way value of $\sim$ 80-100\kmskpc\ \citep[e.g.][]{Binney_Tremaine_2008, Li_Widrow_2021}. However, once simulations start, the gaseous disc becomes thinner due to radiative cooling and self-gravity, which raises the mid-plane density. As a result, we get $\nu \sim 90$\kmskpc, which is within the range of the inferred Milky Way values.

\subsection{Hydrodynamics}
\label{subsec: hydrodynamics}

We carry out hydrodynamical simulations of an isolated galactic disc with a similar setup used in \citet{Fujimoto_Krumholz_Tachibana_2018, Fujimoto_Krumholz_Inutsuka_2020}, except for some minor updates in parameters. We refer readers to those papers for full details of the hydrodynamical method and here summarise the most important aspects.

We use the adaptive mesh refinement (AMR) code {\sc enzo} \citep{Bryan_et_al_2014, Brummel-Smith_et_al_2019}. We select a direct-Eulerian piecewise parabolic mesh hydrodynamics method, along with a Harten–Lax–van Leer with Contact Riemann solver, to follow the motion of the gas, and we select a dual-energy formalism to follow the internal and kinetic energy of the gas. To model the thermal physics of the gas, we make use of radiative cooling down to 10\K\ implemented as tabulated cooling rates as a function of density and temperature and diffuse photoelectric heating in which electrons are ejected from dust grains via far-ultraviolet (FUV) photons implemented in as a constant heating rate. Self-gravity of the gas is also implemented. As for particle dynamics, we use a standard particle-mesh scheme implemented in {\sc enzo}; particle positions and velocities are updated according to the local gravitational acceleration using a drift-kick-drift algorithm, providing second-order accuracy. 

Star formation begins when a cell has reached a maximum refinement level and a threshold number density for star formation. The threshold number density depends on the resolution of the simulation: 23\cmcube\ for $\Delta x =16$\pc\ and 50\cmcube\ for $\Delta x =8$\pc. These densities are chosen so that they correspond to the density that is Jeans unstable at each resolution for the equilibrium temperature dictated by our heating and cooling processes. We set a star formation efficiency per free-fall time in cells above the threshold is 0.01. To avoid creating an extremely large number of star particles whose mass is insufficient to have a well-sampled stellar population, we form star particles stochastically with a particle mass of 300\Msun\ rather than spawn particles in every cell at each time step. Star particles are allowed to form in the main disc region between $R = 0.5$\kpc\ and $R = 14$\kpc. 

Within each star particle, we draw stellar population from the initial mass function (IMF) and model stellar feedback in the form of photoionization, stellar winds, and supernovae. For the photoionization, we heat the particle's host and neighbouring 26 cells to a temperature of $10^4$\K\ or less, depending on the total ionizing luminosity from each star particle. For the stellar winds, their mechanical luminosities are added to the host cell as thermal energy. For supernova feedback, we deposit a total momentum of $5 \times 10^5$\Msunkms\ for $\Delta x =16$\pc\ simulations and $8 \times 10^5$\Msunkms\ for $\Delta x =8$\pc\ simulations, directed radially outward in the 26 neighbouring cells. The total net increase in kinetic energy in the cells is then deducted from the available budget of $10^{51}$\erg, and the balance of the energy is then deposited in the host and neighbouring 26 cells as thermal energy. The stellar feedback in each star particle typically lasts up to a few tens of Myr after the particle's creation. To ensure that the stellar feedback always works on the maximum refinement level, we require that any cell containing the star particle whose age is less than 40\Myr\ is refined to the maximum refinement level. In addition to the energy and momentum feedback, we include gas mass injection from stellar winds and supernovae to each particle's host cell each time step. In this process, the mass is deducted from the star particle mass at the same time. Because of the stellar feedback, almost half of the star particles typically lose more than 50\Msun\ by 40\Myr\ after birth.

In addition to the star formation and feedback, to prevent unresolved collapse at the maximum refinement level, we employ a pressure floor such that the Jeans length is resolved by at least four minimum cells on the maximum refinement level.

\subsection{Tracer particles}
\label{subsec: tracer particles}

To study the stellar scattering in the galactic disc, we place tracer particles in the simulated galaxy as initial conditions, as shown in Fig.~\ref{fig: initial disc}. The particles are placed every 500\pc\ between $R = 2$\kpc\ and $R = 13$\kpc\ in the radial direction, every 5 degrees in the azimuthal direction, and every 100\pc\ between $z = 0$\kpc\ and $z = 1$\kpc\ in the height direction. The total number of the particles is 18216 ($= 23 \times 72 \times 11$). The particle's initial velocity $\mathbfit{v}_{\rm init}$ is set along with the galactic rotation; $\mathbfit{v}_{\rm init} = 0 \cdot \mathbf{u}_R + v_{\rm circ} \cdot \mathbf{u}_{\phi} + 0 \cdot \mathbf{u}_z$, where the $\mathbf{u}_R$ unit vector is in the radial direction, $\mathbf{u}_{\phi}$ is in the direction of rotation, and $\mathbf{u}_z$ is upward from the galactic disc. The $v_{\rm circ}$ is set with equation~(\ref{eq: circular velocity}).

The mass of the tracer particle is set to nearly zero so that its gravitational force does not affect the motions of other tracer particles, star particles, and the gas. On the other hand, they feel the gravity from the gas and star particles.

\subsection{Star particle mass}
\label{subsec: star particle mass}

As described in Section~\ref{subsec: hydrodynamics}, our star formation recipe requires us to create the star particle with a mass of 300\Msun, which represents a star cluster rather than individual stars. However, the star particles may act as compact and heavy gravitational sources and scatter the tracer particles unrealistically strongly. To investigate the possibility, we perform two types of simulations, as summarised in Table~\ref{tab: models}. One is that while using the 300\Msun\ to draw a stellar population from the IMF to model the stellar feedback, we set the initial star particle mass to nearly zero so that the particle does not gravitationally affect the motion of tracer particles. The other is that we set the initial star particle mass to 300\Msun, as originally required in the star formation recipe. In both cases, the mass loss from the stellar feedback is returned to the gas cell; in the case that the initial star particle mass is set to zero, the star particle mass keeps zero though the particle's host cell receives the feedback mass. In the case that the initial star particle mass is set to 300\Msun, the feedback mass is deducted from the star particle mass.

\section{Results}
\label{sec: results}

\subsection{Evolution of the galactic disc}
\label{subsec: evolution of the galactic disc}

\begin{figure}
    \centering
	\includegraphics[width=\columnwidth]{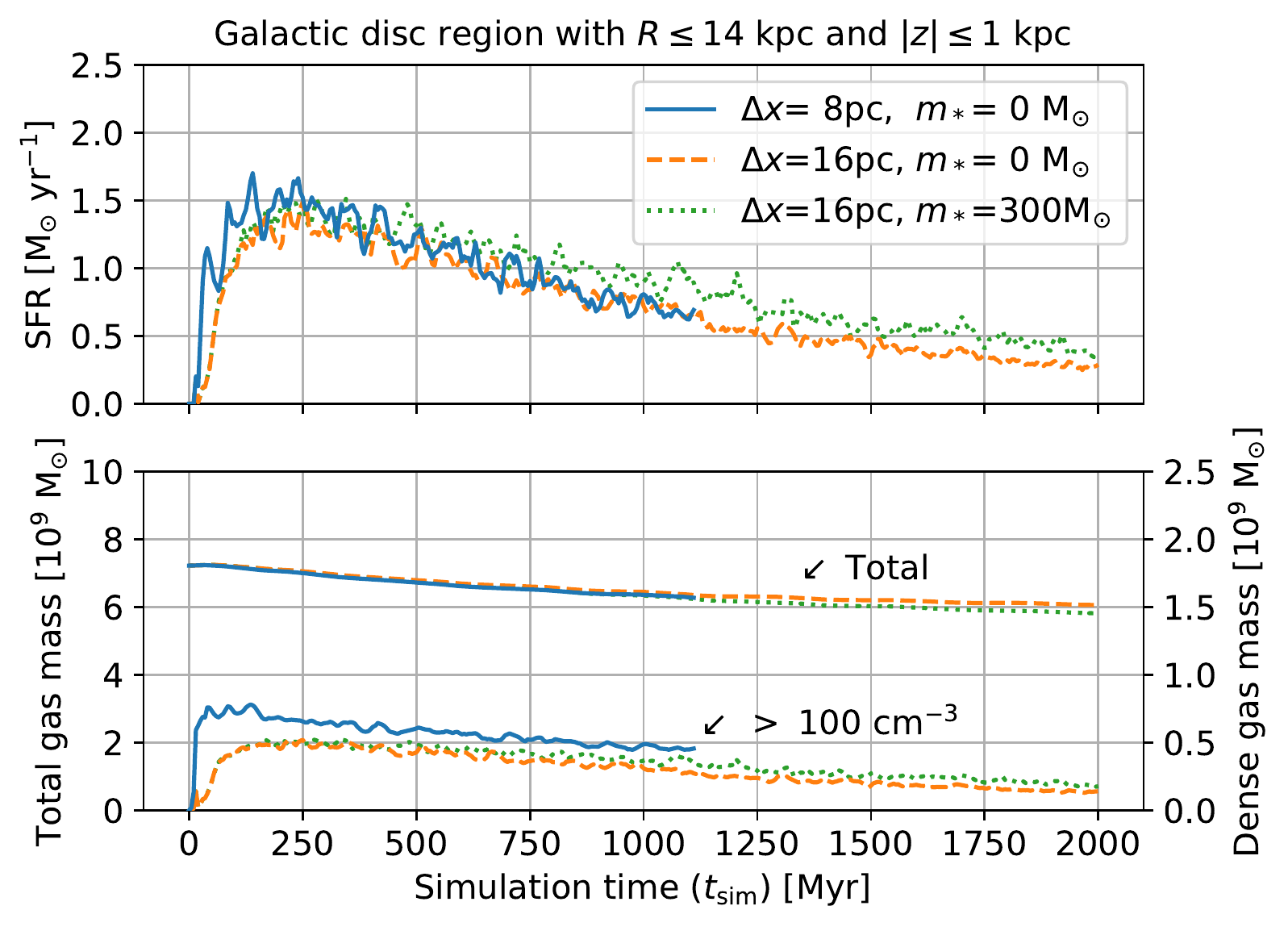}
    \caption{The time evolution of SFR (top) and masses (bottom) of the total gas and dense gas ($> 100$\cmcube) in the galactic disc with $R \leq 14$\kpc\ and $|z| \leq 1$\kpc\ for the three simulations.}
    \label{fig: SFR and gas mass evolution}
\end{figure}

\begin{figure}
    \centering
	\includegraphics[width=\columnwidth]{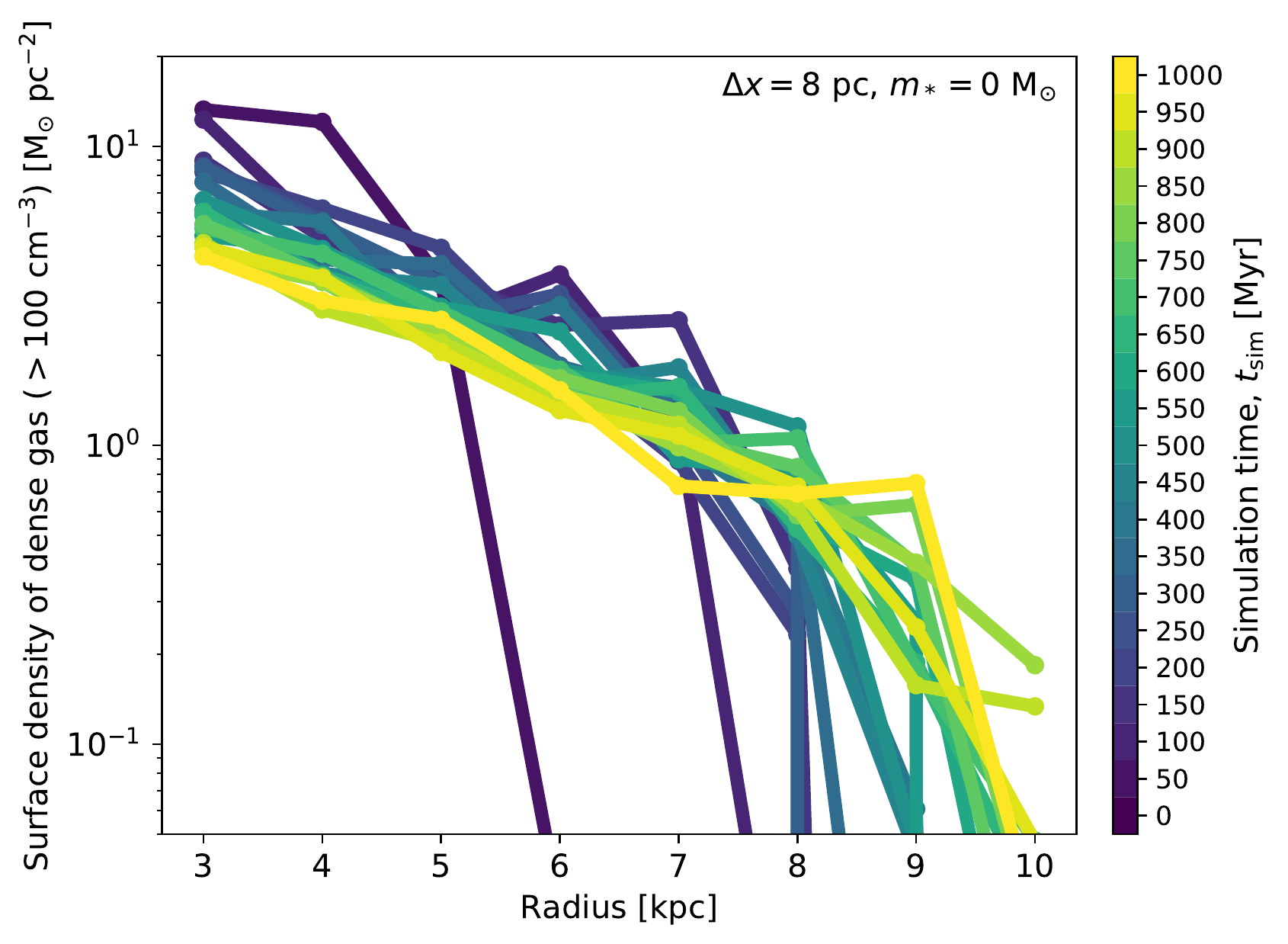}
    \caption{The time evolution of azimuthally averaged radial profiles of the surface density of dense gas ($> 100$\cmcube). The simulation model is \textit{8pc\_0msun}.}
    \label{fig: radial profile dense gas}
\end{figure}

The purpose of our simulations is not to reproduce a long-time evolution of the Galaxy over 10 Gyr but to reproduce a steady-state Galactic disc of the most recent a few Gyr. In this subsection, we check if the total quantities of the simulated galactic disc, such as the star formation rate (SFR) and gas mass, show little time variation and agree with ones of the current Milky Way. 

We run the simulations for 1.1\Gyr\ for the high resolution model ($\Delta x = 8$\pc) and 2\Gyr\ for the low resolution models ($\Delta x = 16$\pc); 1\Gyr\ corresponds to $4 \sim 8$ galactic rotation periods in the main part of the disc for our analysis ($4 < R < 8$\kpc) because one rotation periods of the galactic disc are $\sim 125$\Myr\ at $R = 4$\kpc\ and $\sim 250$\Myr\ at $R = 8$\kpc. Fig.~\ref{fig: SFR and gas mass evolution} shows the time evolution of the SFR and masses of the total gas and dense gas of $> 100$\cmcube, which corresponds to GMCs, in the galactic disc with $R \leq 14$\kpc\ and $|z| \leq 1$\kpc. Overall, there is no significant difference between the three simulations. By the time of $t_{\rm sim} \sim 200$\Myr, the SFR and masses of the total gas and dense gas stop increasing and start a gradual decline without significant changes in the quantities. Although those quantities drop to small values by the time of $t_{\rm sim} = 2$\Gyr, the galactic disc is in a nearly steady state for the first 1 Gyr, which is the time for our main analysis. In the quasi-equilibrium state, the SFR is 0.5-1.5\MsunYr for 1 Gyr. Given that star formation in the galactic centre region ($R < 0.5$\kpc) is omitted in the simulation (see Section~\ref{subsec: hydrodynamics}), the SFR is in good agreement with the inferred Milky-Way SFR of 1-2\MsunYr \citep[e.g.][]{Murray_Rahman_2010, Robitaille_Whitney_2010, Chomiuk_Povich_2011, Licquia_Newman_2015, Elia_et_al_2022}. In regard to the gas mass, the total gas mass is $\sim 6 \times 10^9$\Msun, and the dense gas mass is $\sim 5 \times 10^8$\Msun, which are also consistent with Milky Way's values; recent Galactic surveys for H\textsc{i} and CO emissions show a total gas mass of $8 \times 10^9$\Msun\ within a radius of 30\kpc\ and 11 per cent of which is H$_2$ gas \citep{Nakanishi_Sofue_2016}. The gas surface density is 10-20\MsunPc\ in the main disc region of $4 < R < 8$\kpc, consistent with the Milky Way's value \citep[e.g.][]{Wolfire_et_al_2003, Yin_et_al_2009, McKee_Parravano_Hollenbach_2015}.

Fig.~\ref{fig: radial profile dense gas} shows the time evolution of azimuthally averaged radial profiles of the surface density of dense gas ($> 100$\cmcube). We see that by the time of $t_{\rm sim} \sim 150$\Myr, the production of dense gas has almost been completed in the main part of the disc with $R < 8$\kpc. After that, there is no significant change in the radial profile, except for a slight decrease of the surface densities with time, which is consistent with the gradual decrease of the total dense gas mass in the disc, as shown in Fig.~\ref{fig: SFR and gas mass evolution}. On the other hand, in the outer part with $R > 8$\kpc, the dense gas production is slow, and it has taken $\sim 500$\Myr\ for completion (See also Fig.~\ref{fig: spatial distribution of tracer particles}). We also see a radial gradient in the dense gas surface densities: for example, $\sim$ 3.7, 1.7, and 0.56 \MsunPc\ at $R =$ 4, 6, and 8\kpc, respectively, at $t_{\rm sim} = 600$\Myr. That is consistent with observations; the radial gradient of a dense gas such as $\mathrm{H_2}$ has been seen in a main disc region outside the Galactic bar ($R > 3$\kpc) in the Milky Way \citep[e.g.][]{Nakanishi_Sofue_2016, Miville-Deschenes_et_al_2017} and nearby galaxies \citep[e.g.][]{Bigiel_et_al_2008}.

\subsection{Distribution of giant molecular clouds}
\label{subsec: distribution of giant molecular clouds}

In this subsection, we first see spatial distributions of GMCs formed in the simulated galactic disc and then see their properties, such as mass and size. We also check if they are consistent with observations. 

\subsubsection{Spatial distribution}
\label{subsubsec: spatial distribution}

\begin{figure}
    \centering
	\includegraphics[width=\columnwidth]{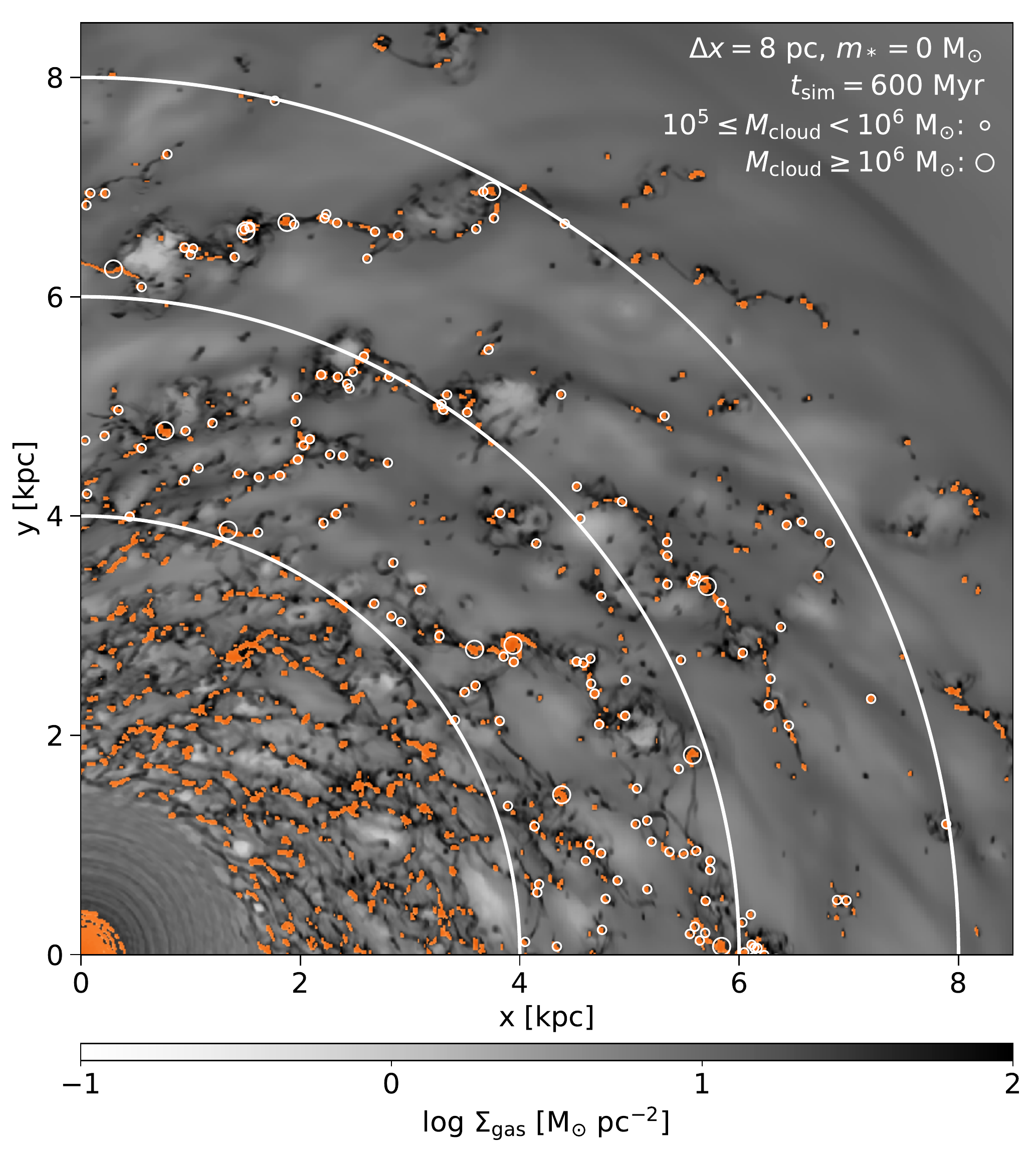}
    \caption{The gas distribution in the first quadrant of the galactic disc viewed face-on, shown in greyscale, with dense gas ($> 100$\cmcube) distribution overlaid with orange. The identified GMCs in a region between 4\kpc\ $< R <$ 8\kpc\ are also shown in white circles: small ones indicate intermediate clouds ($10^5\ \mathrm{M}_{\odot} \leq M_{\rm cloud} < 10^6\ \mathrm{M}_{\odot}$), and large ones indicate massive clouds ($M_{\rm cloud} > 10^6\ \mathrm{M}_{\odot}$). The simulation model is \textit{8pc\_0msun}, and the panel shows the surface densities at $t_{\rm sim} = 600$\Myr. The galactic centre is located at the lower-left corner, and the galactic disc rotates anticlockwise. The three white arcs show galactocentric radii $R$ of 4, 6, and 8\kpc.}
    \label{fig: spatial distribution of clouds}
\end{figure}

\begin{figure}
    \centering
	\includegraphics[width=\columnwidth]{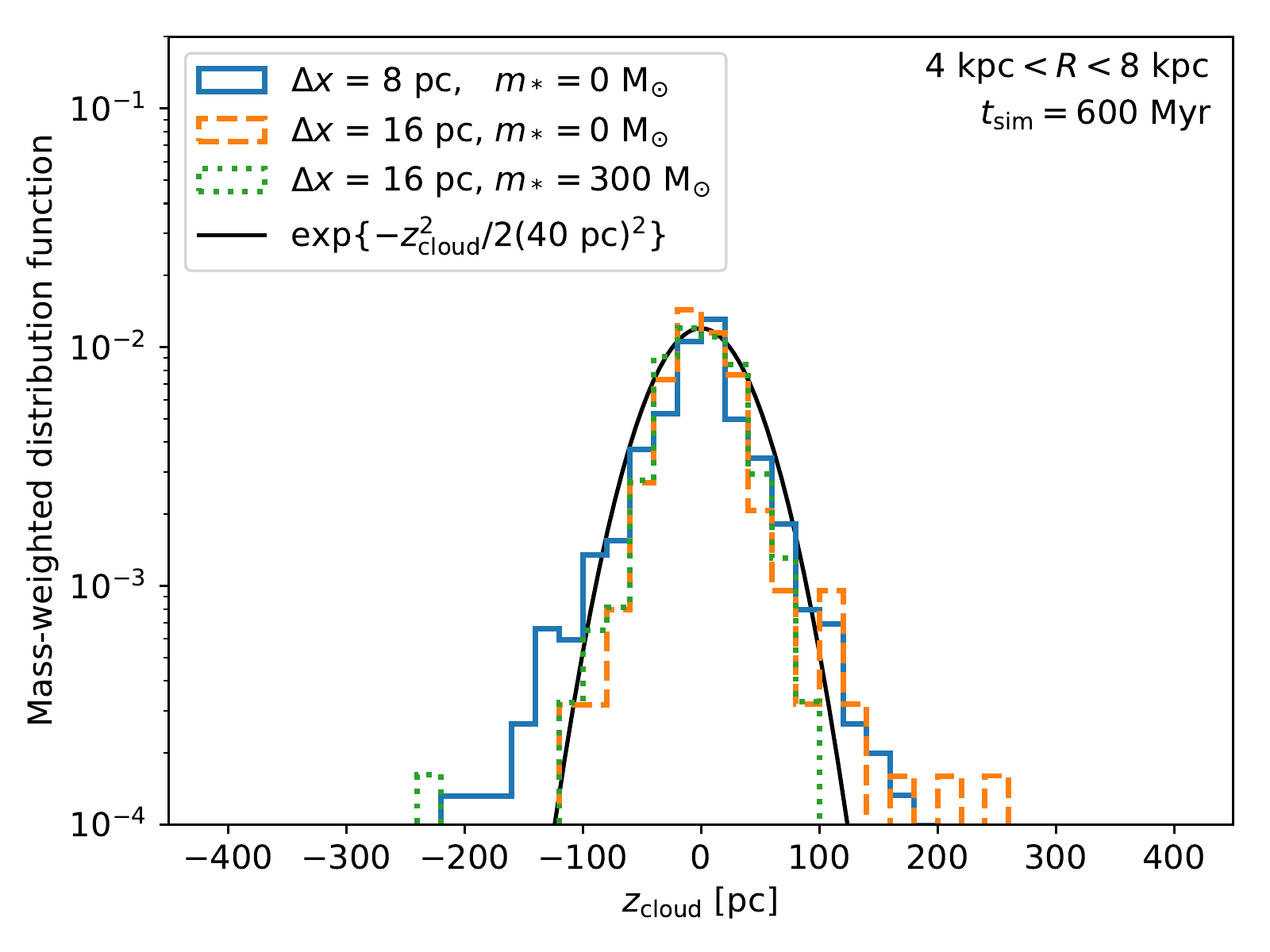}
    \caption{Cloud mass-weighted normalised distribution functions of the cloud's vertical position, $z_{\rm cloud}$, to the galactic disc. The sample clouds are taken from a region between 4\kpc\ $< R <$ 8\kpc\ at $t_{\rm sim} = 600$\Myr. The black solid line is a Gaussian distribution with a standard deviation of 40\pc.}
    \label{fig: vertical distribution of clouds}
\end{figure}

Once the disc has been in the quasi-equilibrium state after $t_{\rm sim} \sim 200$\Myr, the galactic disc is fully fragmented and produces dense gas clouds, the so-called giant molecular clouds (GMCs). Fig.~\ref{fig: spatial distribution of clouds} shows the gas distribution in the first quadrant of the galactic disc viewed face-on at $t_{\rm sim} = 600$\Myr, overlaid with the dense gas ($> 100$\cmcube) distribution. There is no large difference in the disc's morphology among the three simulations, so we only show the result of the model \textit{8pc\_0msun} in the figure. 

To quantitatively examine the spatial distribution and properties of the dense gas clouds formed in the simulated disc, we define a GMC as a connected structure contained within a contour at a threshold number density. We use 100\cmcube\ as the threshold number density because it is similar to the mean volume densities of typical observed GMCs. Note that since the formation of molecules is not being followed in our simulations, the gas is purely atomic. 

The spatial distribution of the identified GMCs is shown in Fig.~\ref{fig: spatial distribution of clouds} with white circles; small ones indicate intermediate clouds ($10^5\ \mathrm{M}_{\odot} \leq M_{\rm cloud} < 10^6\ \mathrm{M}_{\odot}$), and large ones indicate massive clouds ($M_{\rm cloud} > 10^6\ \mathrm{M}_{\odot}$). Low-mass clouds ($M_{\rm cloud} < 10^5\ \mathrm{M}_{\odot}$) are not shown in the figure. We see that GMCs are distributed in kpc-scale filamentary structures rather than being distributed randomly in the galactic disc and that some GMCs form clusters in the filament, forming hundreds of pc-scale dense gas associations, the so-called giant molecular cloud associations (GMAs). These hierarchical and clustered distributions of GMCs have been observed in nearby galaxies \citep[e.g.][]{Rosolowsky_et_al_2003, Koda_et_al_2009, Sun_et_al_2018}.

As there is a radial gradient in the surface densities of dense gas (Fig.~\ref{fig: radial profile dense gas}), there is also a radial dependence in the spatial distribution of GMCs. The azimuthally averaged surface number densities of intermediate to massive GMCs larger than $10^5$\Msun\ are $\sim$ 8, 4, and 2 \kpc$^{-2}$ at $R =$ 4, 6, and 8\kpc, respectively, at $t_{\rm sim} = 600$\Myr\ for the simulation model of \textit{8pc\_0msun}.

In addition to the GMC distribution projected onto the galactic mid-plane, the vertical distribution of GMCs to the galactic disc may also be an important element that affects stellar scattering. Fig.~\ref{fig: vertical distribution of clouds} shows cloud mass-weighted normalised distribution functions of the cloud's vertical position. We see that most clouds are located within 100\pc\ of the galactic disc and that the scale height of the vertical distribution is $\sim 40$\pc. That is consistent with observations that the scale height of the molecular gas disc in the Milky Way is 30-50\pc\ between radii $3 < R < 8$\kpc\ \citep{Malhotra_et_al_1994, Nakanishi_Sofue_2006} and that the scale heights in edge-on disc galaxies with similar gas surface densities to the Milky Way are $\sim 40$\pc\ \citep{Yim_et_al_2014}. In addition, we find no model dependence of the vertical distribution between the three simulation models.

\subsubsection{Mass and size distribution}
\label{subsubsec: mass and size distribution}

\begin{figure*}
    \centering
	\includegraphics[width=\columnwidth]{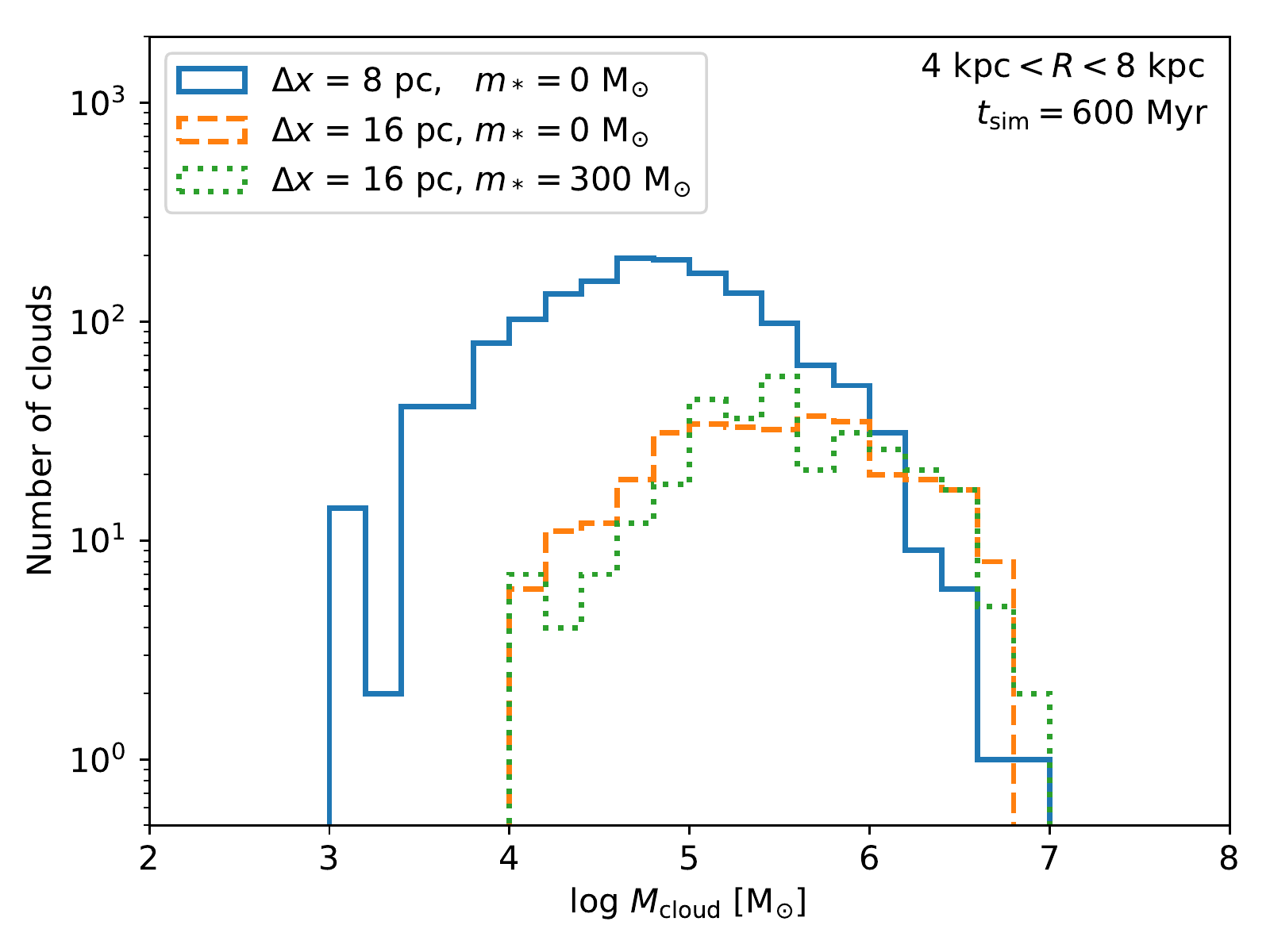}
	\includegraphics[width=\columnwidth]{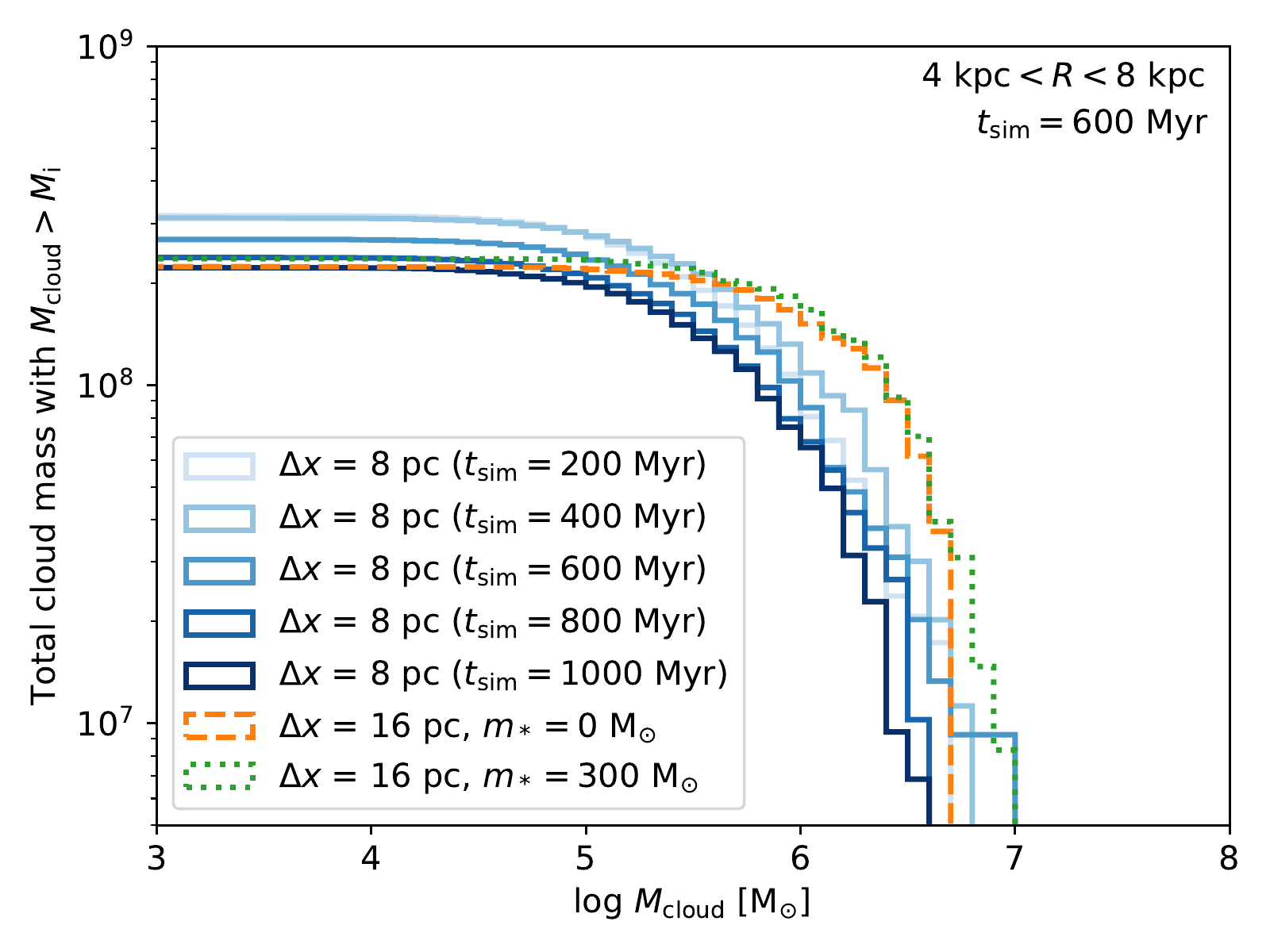}
	\includegraphics[width=\columnwidth]{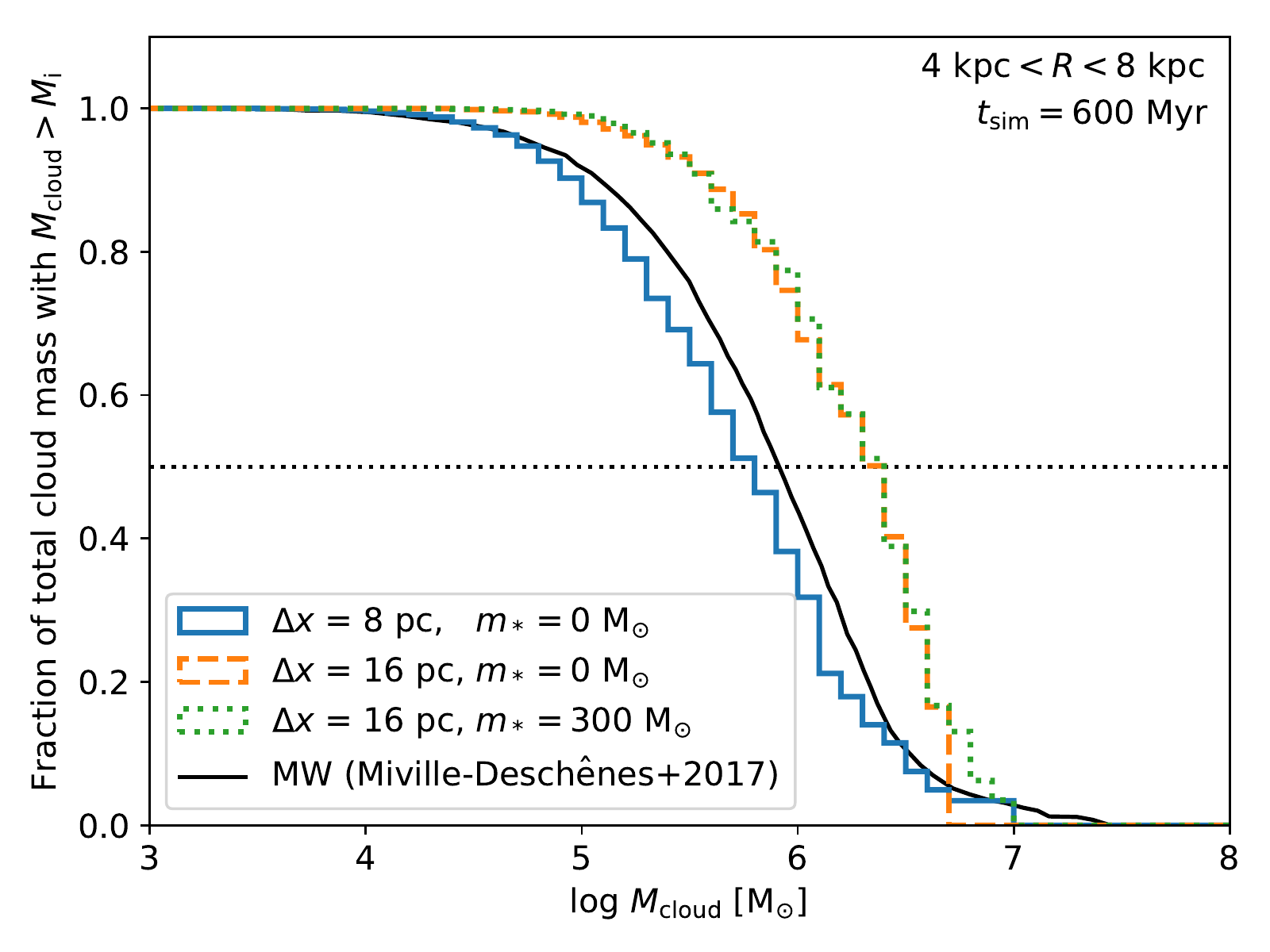}
	\includegraphics[width=\columnwidth]{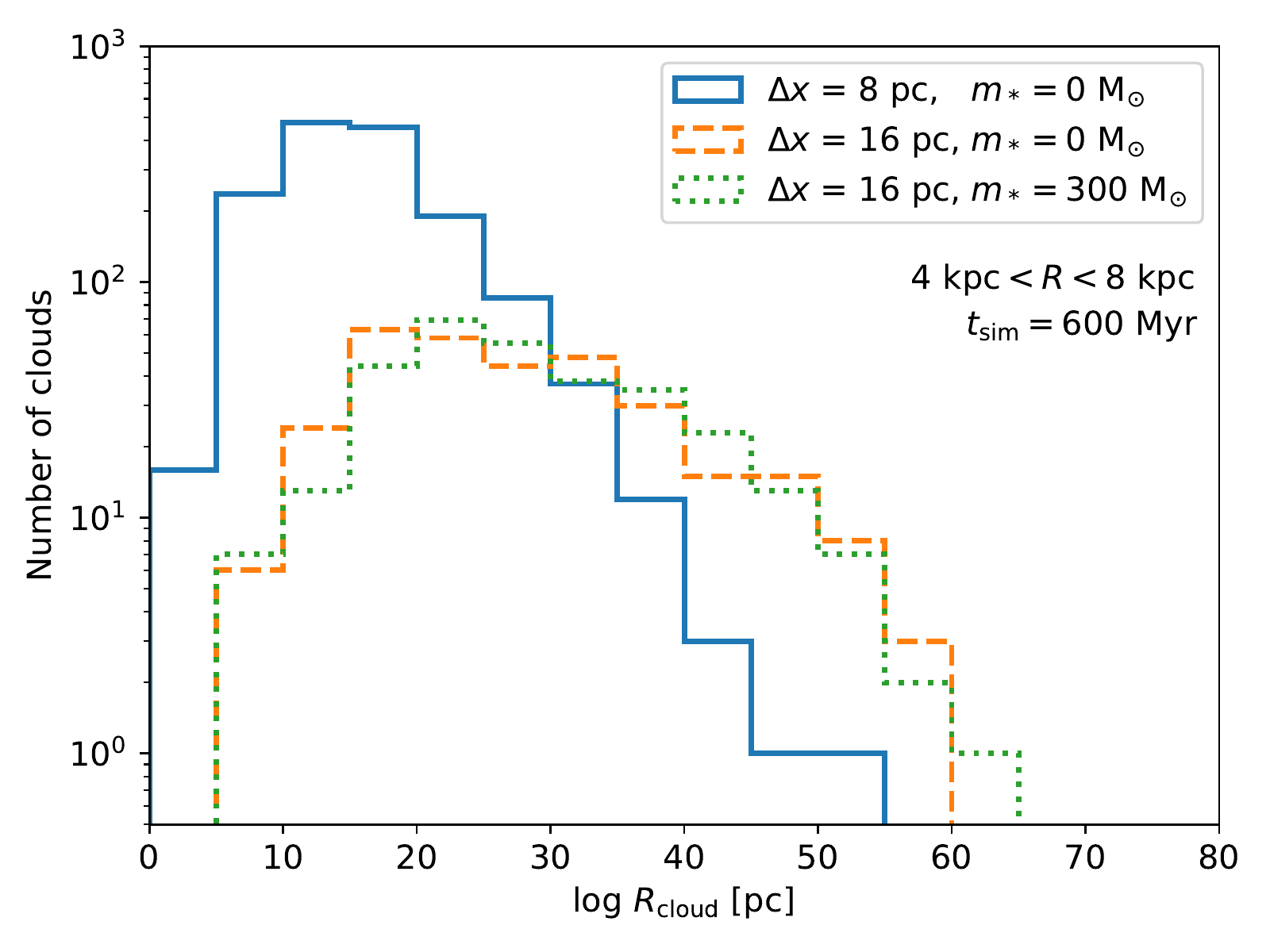}
    \caption{Mass and size distributions of GMC: histograms of cloud mass (top left), cumulative distribution functions of cloud mass (top right), normalised cumulative distribution functions of cloud mass (bottom left), and histograms of cloud radius (bottom right). The observed Milky-Way's GMC mass distribution \citep{Miville-Deschenes_et_al_2017} is shown with a black line. The sample clouds are taken from a region between 4\kpc\ $< R <$ 8\kpc. Unless stated otherwise, we show the distributions at $t_{\rm sim} = 600$\Myr.}
    \label{fig: distribution of cloud mass and size}
\end{figure*}

In this sub-subsection, we show the GMC's properties. The top left panel of Fig.~\ref{fig: distribution of cloud mass and size} shows the histogram of the mass of clouds located between 4\kpc\ $< R <$ 8\kpc\ at $t_{\rm sim} = 600$\Myr. There is a resolution dependence for low to intermediate-mass clouds with $M_{\rm cloud} < 10^6$\Msun. As for massive clouds with $M_{\rm cloud} > 10^6$\Msun, the distributions seem almost consistent among the three models, although the numbers are slightly low in the high-resolution model. This tendency might come from the lack of resolution because, in the low-resolution models, low to intermediate-mass clouds can be identified as a part of neighbouring massive clouds. The top right panel of Fig.~\ref{fig: distribution of cloud mass and size} provide a supplementary explanation. This panel shows cumulative distribution functions of the cloud mass; the $y$-axis is the total cloud mass with $M_{\rm cloud} > M_i$. There is little difference in the mass range of $M_i < 10^5$\Msun, indicating little difference in the total cloud mass among simulation models. This result is consistent with the lower panel of Fig.~\ref{fig: SFR and gas mass evolution}, which shows little model dependence of the total dense gas mass in the galactic disc. On the other hand, the total mass of clouds with the range of $M_i > 5 \times 10^5$\Msun\ in the high-resolution model is smaller than those in the low-resolution models, suggesting that some massive clouds are split into smaller clouds in the high-resolution model. The figure also shows little dependence of the cloud mass distribution on the simulation time.

We next compare the mass distributions with observations. The bottom left panel of Fig.~\ref{fig: distribution of cloud mass and size} shows the normalised cumulative distribution functions of the cloud mass, together with the observed Milky-Way's GMC mass distribution \citep{Miville-Deschenes_et_al_2017}. Although there is a resolution dependence, as discussed in the previous paragraph, the simulated distributions, especially in the high-resolution model, are in good agreement with the observed one. 

We also examine the size distributions. The bottom right panel of Fig.~\ref{fig: distribution of cloud mass and size} shows the histogram of cloud radius. We define the cloud radius as $R_{\rm cloud} \equiv (3V_{\rm cloud}/4\upi)^{1/3}$, where $V_{\rm cloud}$ is the cloud volume which calculates the sum of the cell volumes in the cloud. Note that this definition assumes that clouds are spherical, but it is only sometimes accurate because clouds often have an elongated structure. However, we use this definition to simplify the comparison of the size distribution between different simulation models. We find that the cloud radius in the high-resolution model tends to be small, as we have discussed with the mass distributions. 

With regard to comparison with observations, \citet{Miville-Deschenes_et_al_2017} have shown that the size of the Milky Way's GMCs ranges from 1\pc\ to 150\pc, with the most probable value of $\sim 30$\pc. The typical value is almost consistent with ours of $\sim 20$\pc. However, our simulations do not show large clouds whose radius is over 100\pc. That might be because we underestimate the cloud size due to the assumption that clouds are spherical. It is also possible that the threshold number density for cloud finding is so high that a diffuse ambient gas is not regarded as a part of the cloud. Therefore, we do not consider the slight disagreement in the cloud size distributions as a failure of our simulation. Since the mass distribution is consistent with observations, we argue that our simulations successfully reproduce a realistic GMC distribution to investigate stellar scattering by GMCs.

\subsection{Stellar scattering by GMCs}
\label{subsec: stellar scattering}

\begin{figure*}
    \centering
	\includegraphics[width=\textwidth]{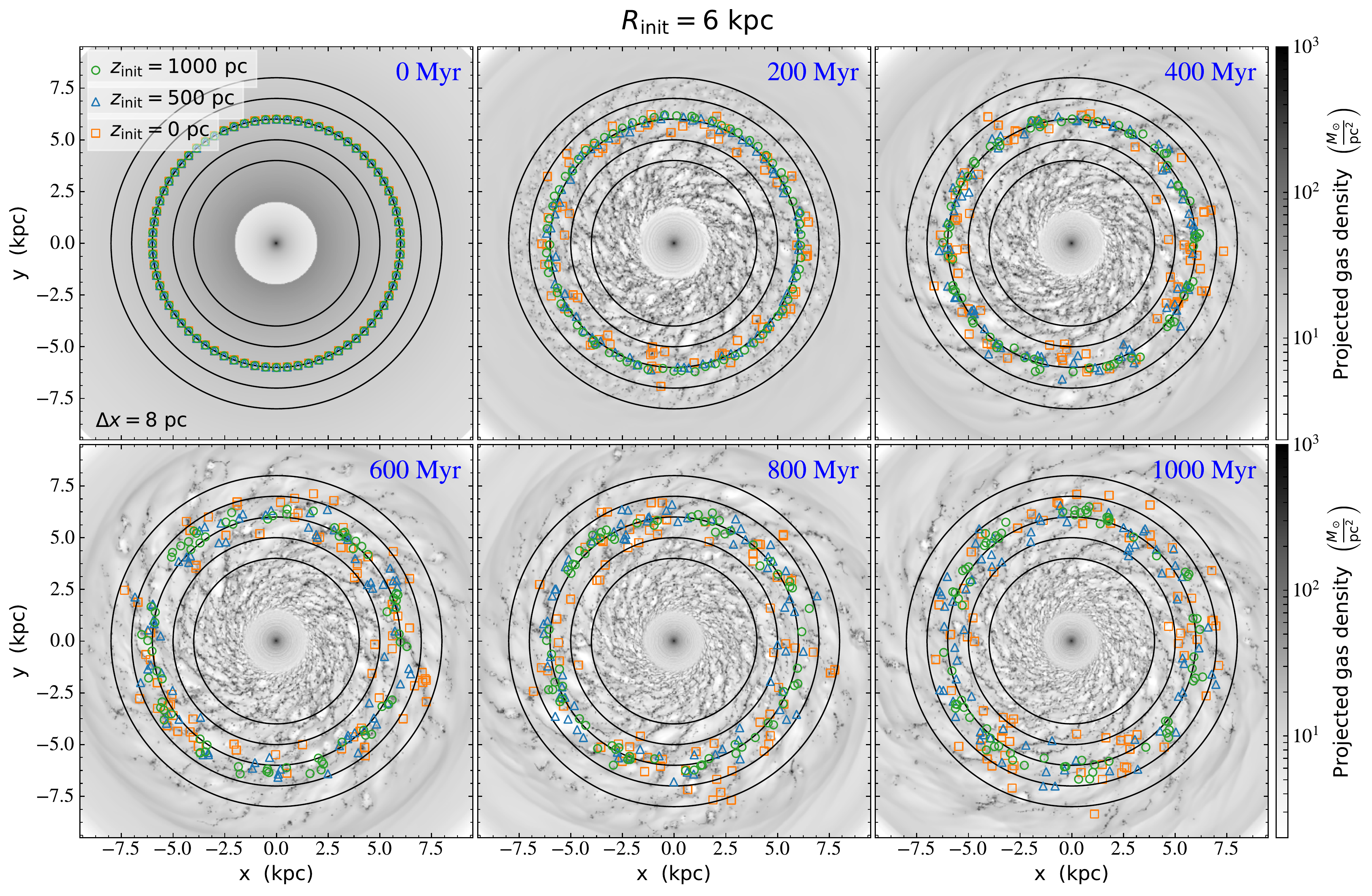}
    \caption{Spatial distribution of tracer particles and their time evolution. Particles whose initial radial position is $R_{\rm init} = 6$\kpc\ are shown, and their initial heights are $z_{\rm init} = $ 0, 500, and 1000\pc, marked with orange squares, blue triangles, and green circles. The number of particles at a given initial radius and a given initial height is 72, so the total number of particles shown in each panel is 216 ($= 72 \times 3$). The background greyscale shows the surface densities of the total gas, and the black circles indicate galactocentric radii of 4, 5, 6, 7, and 8\kpc. Each panel shows time evolution: $t_{\rm sim} = $ 0, 200, 400, 600, 800, and 1000\Myr, respectively. The galactic disc rotates anticlockwise. The simulation model is \textit{8pc\_0msun}.}
    \label{fig: spatial distribution of tracer particles}
\end{figure*}

In this subsection, we will look into the effects of GMCs on stellar scattering, which is the main topic of this paper. For that, we analyse the motions of tracer particles, which are massless and initially placed in the galactic disc. We focus on particles whose initial heights are $z_{\rm init} = $ 0, 500, and 1000\pc. That is because star formation occurs within 100\pc\ of the galactic plane or because the observed scale heights of the Galactic thin and thick stars are $\sim$ 300 and 900\pc\ \citep[][]{Gilmore_Reid_1983, Juric_et_al_2008}, which allows us to consider those tracer particles as newborn stars, thin disc stars, and thick disc stars, respectively. We also mainly focus on particles whose initial radius is $R_{\rm init} = 6$\kpc\ because the Solar system is thought to form in the inner part of the Galaxy, not around the current radius of 8\kpc\ \citep[e.g.][]{Wielen_Fuchs_Dettbarn_1996}. Regarding the dependence on simulation models, we mainly show results of the high-resolution model of \textit{8pc\_0Msun} as our fiducial model, in which the resolution is $\Delta x = 8$\pc\ and the star particle's initial mass is $m_* = 0$\Msun. That is because we find no significant difference in the particle motion between models, as shown in the following sections.

We show the spatial distribution of tracer particles and their time evolution in Fig.~\ref{fig: spatial distribution of tracer particles}, in which particles whose initial radial position is $R_{\rm init} = 6$\kpc\ and whose initial heights are $z_{\rm init} = $ 0, 500, and 1000\pc\ are shown. We see that most particles deviate from their initial radial positions of $R_{\rm init} = 6$\kpc\ by the time of $t_{\rm sim} = 1000$\Myr, and some move more than 1\kpc, showing effective stellar scattering by GMCs. We also see that particles whose initial height is closer to the midplane of the galactic disc, such as ones with $z_{\rm init} = 0$\pc, move more in the radial direction, suggesting that the scattering is more effective for newborn stars than old disc stars. However, in this figure, it is unclear which physical process for stellar scattering is occurring: dynamical disc heating due to increased velocity dispersion, the so-called blurring, or radial migration due to changes in angular momentum, the so-called churning.

In the following sub-subsections, we examine a more quantitative analysis of disc heating and radial migration by investigating tracer particles' velocity dispersion and angular momentum.

\subsubsection{Heating history of the galactic disc}
\label{subsubsec: heating history}

\begin{figure*}
    \centering
	\includegraphics[width=\columnwidth]{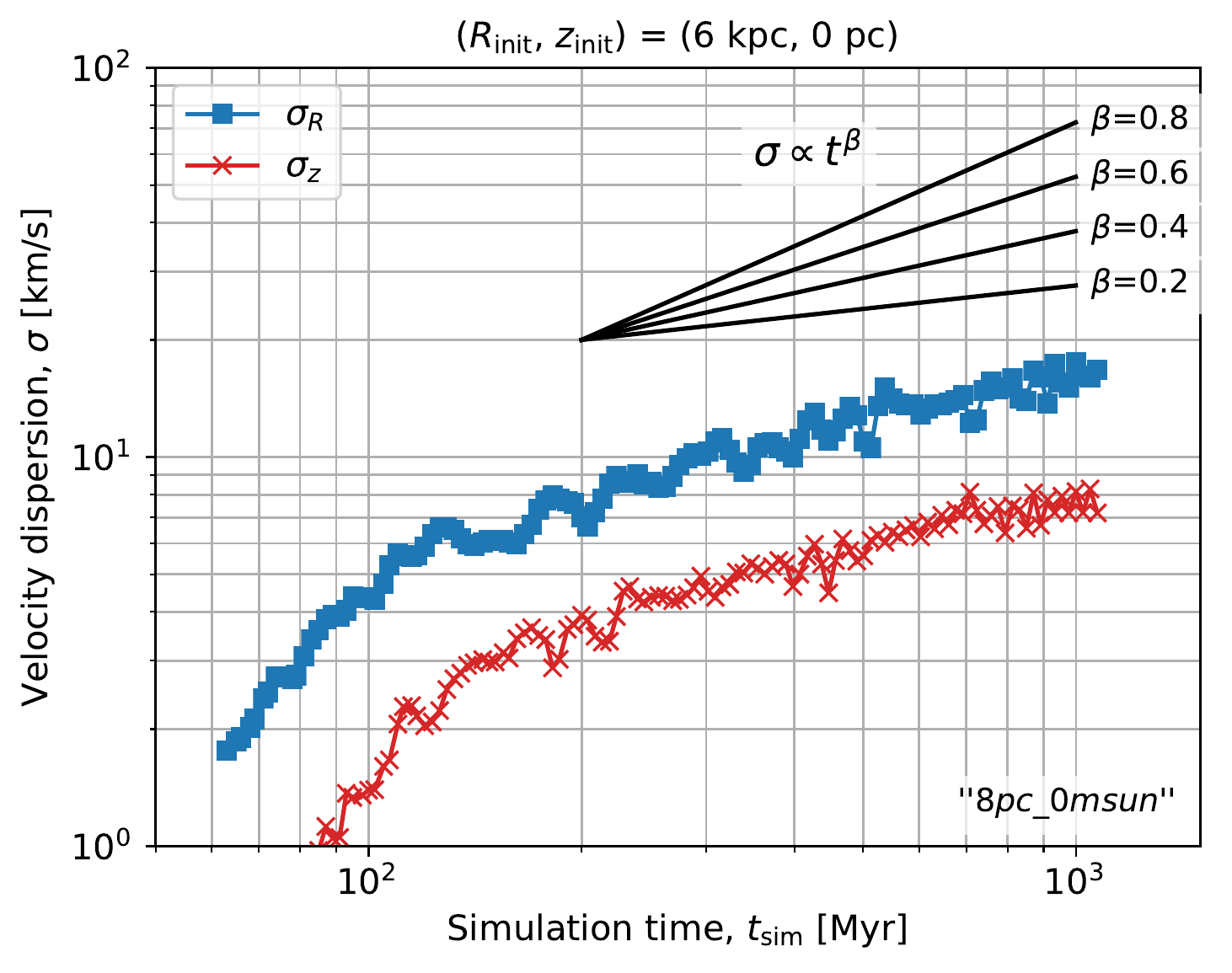}
	\includegraphics[width=\columnwidth]{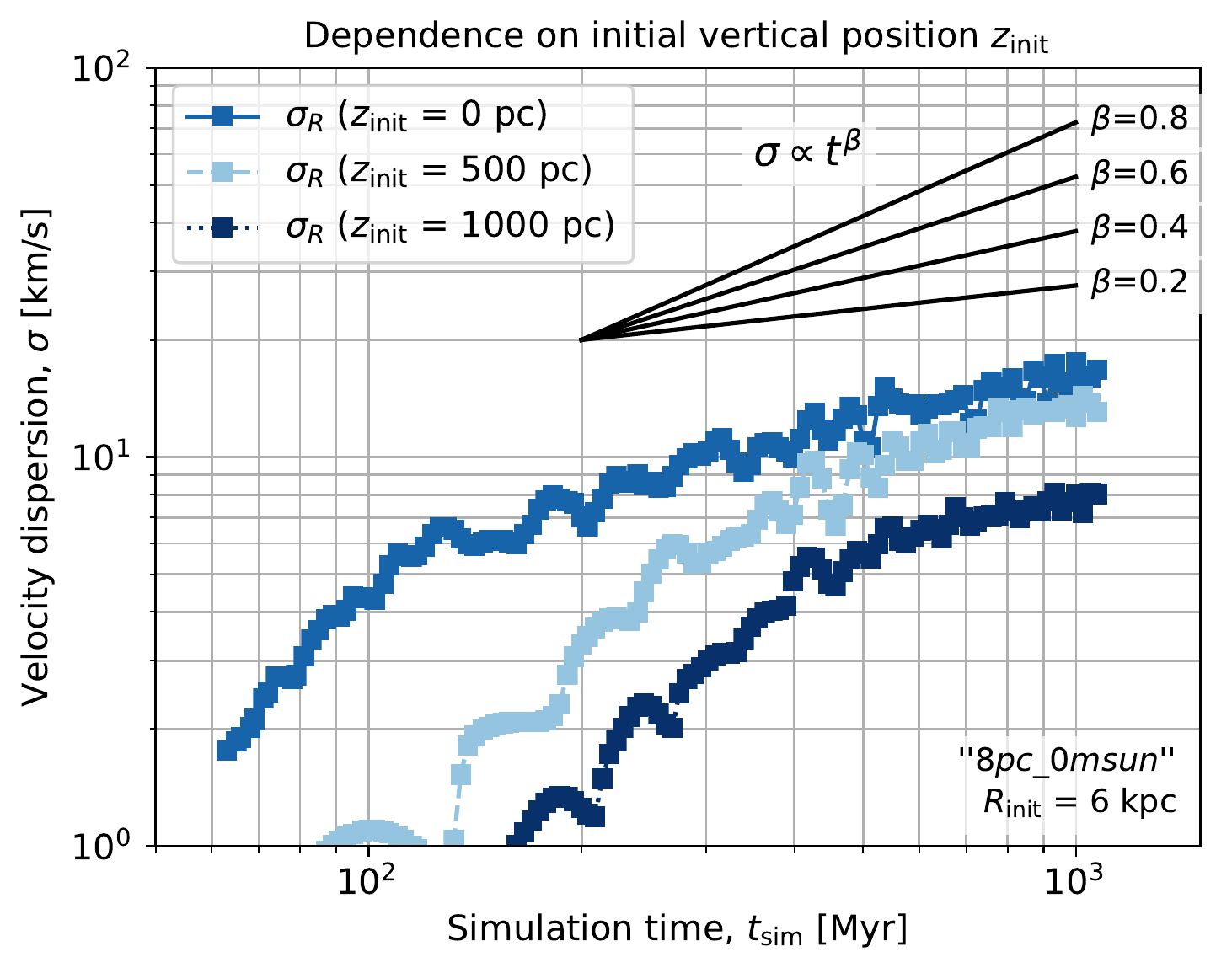}
	\includegraphics[width=\columnwidth]{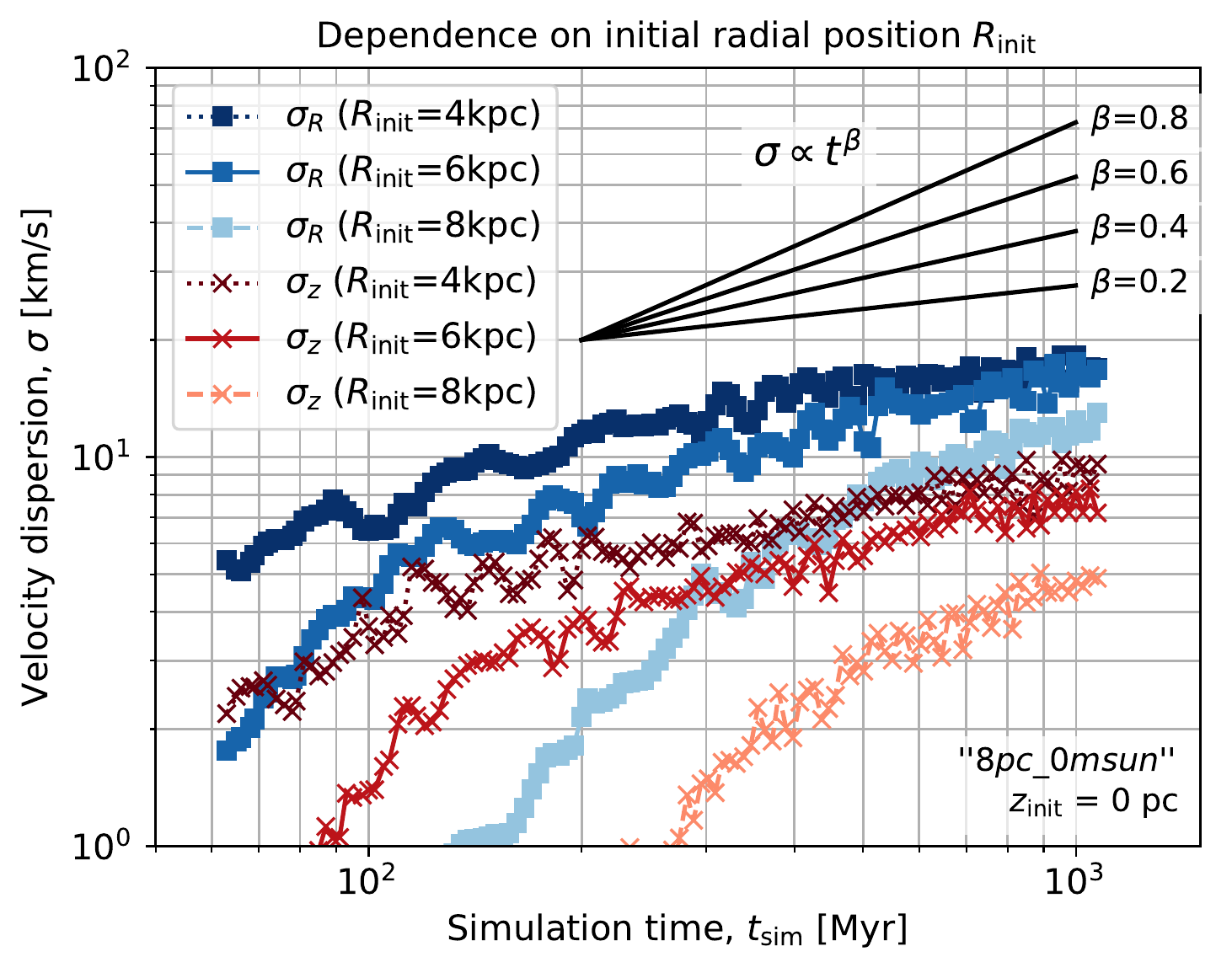}
	\includegraphics[width=\columnwidth]{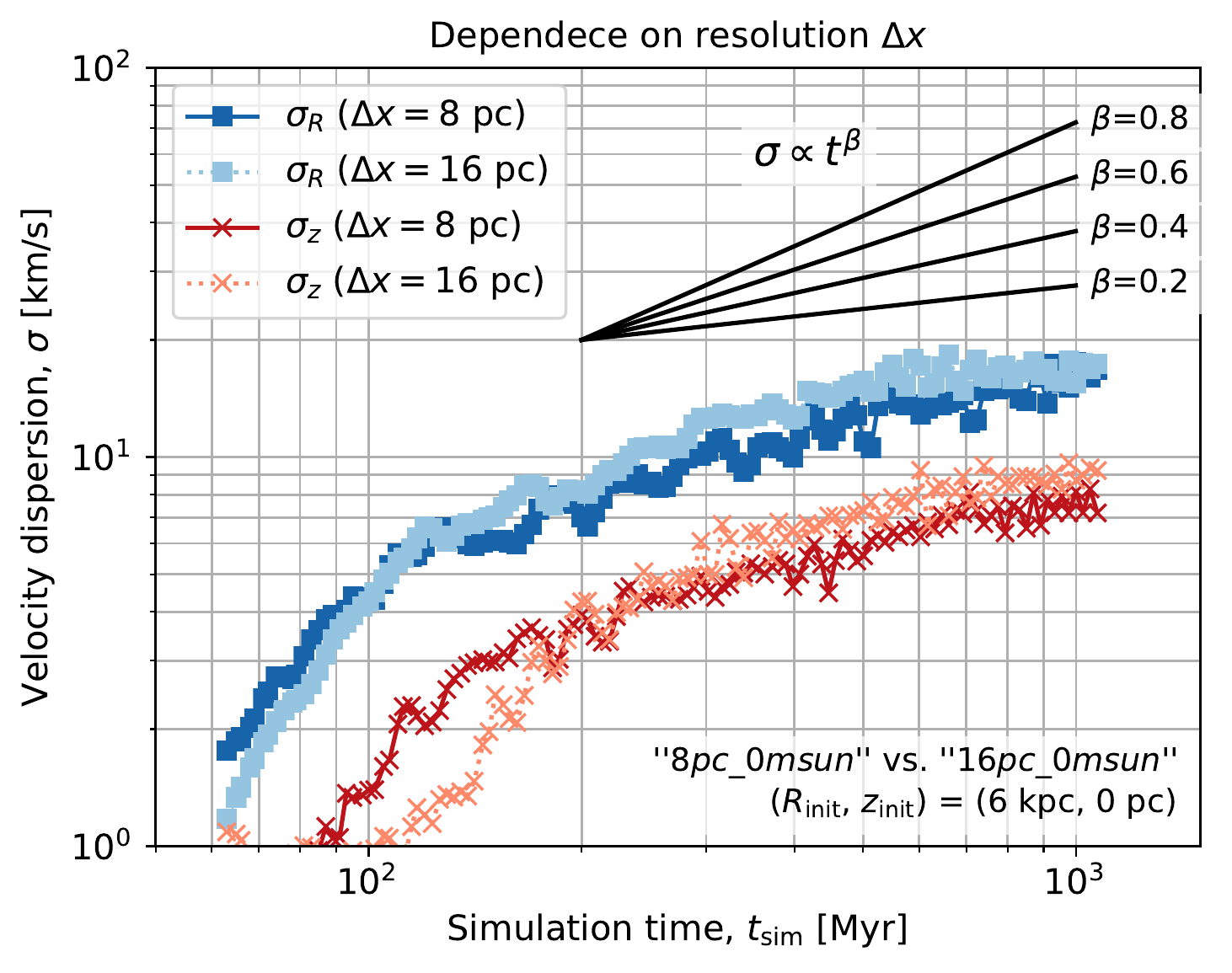}
	\includegraphics[width=\columnwidth]{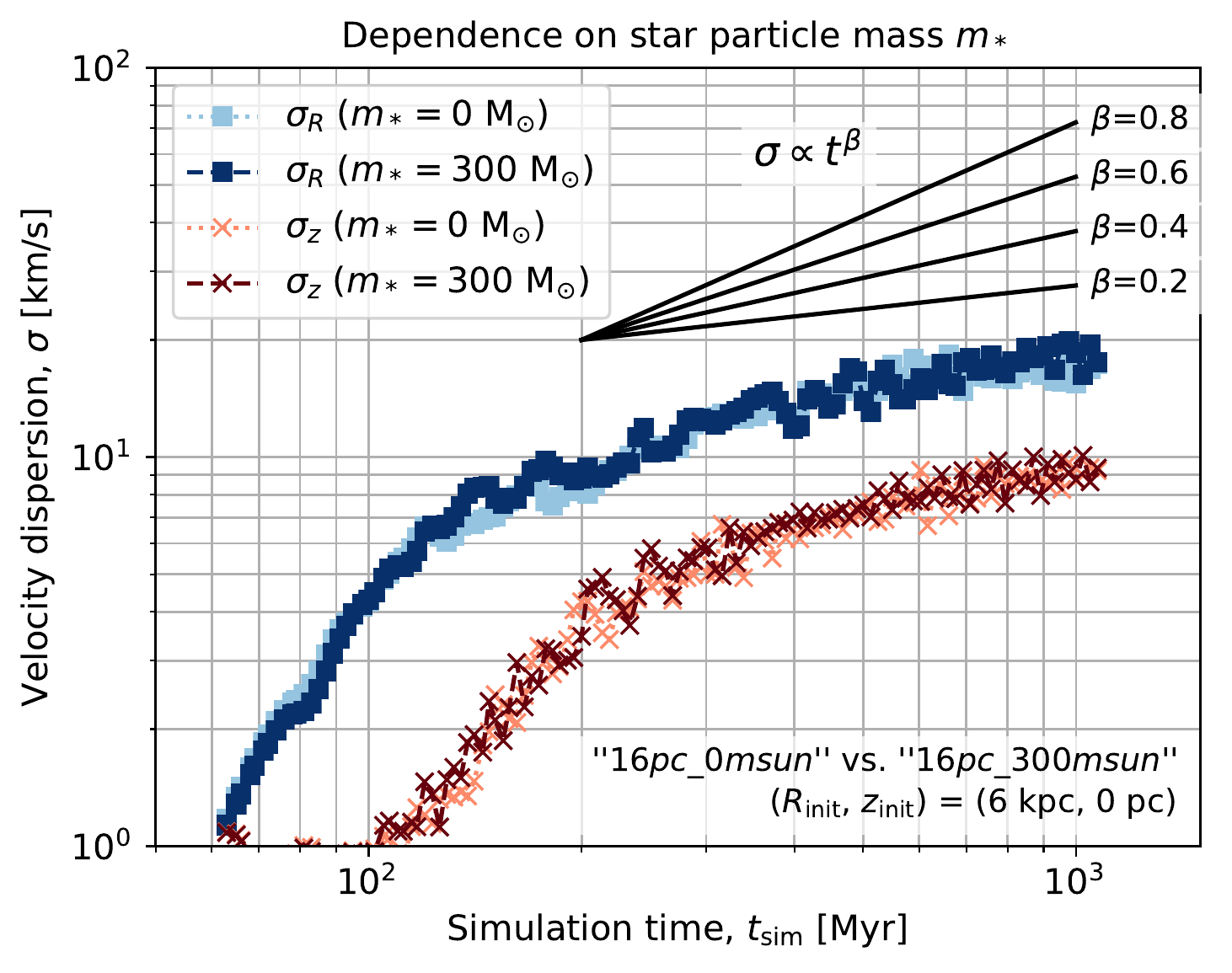}
	\includegraphics[width=\columnwidth]{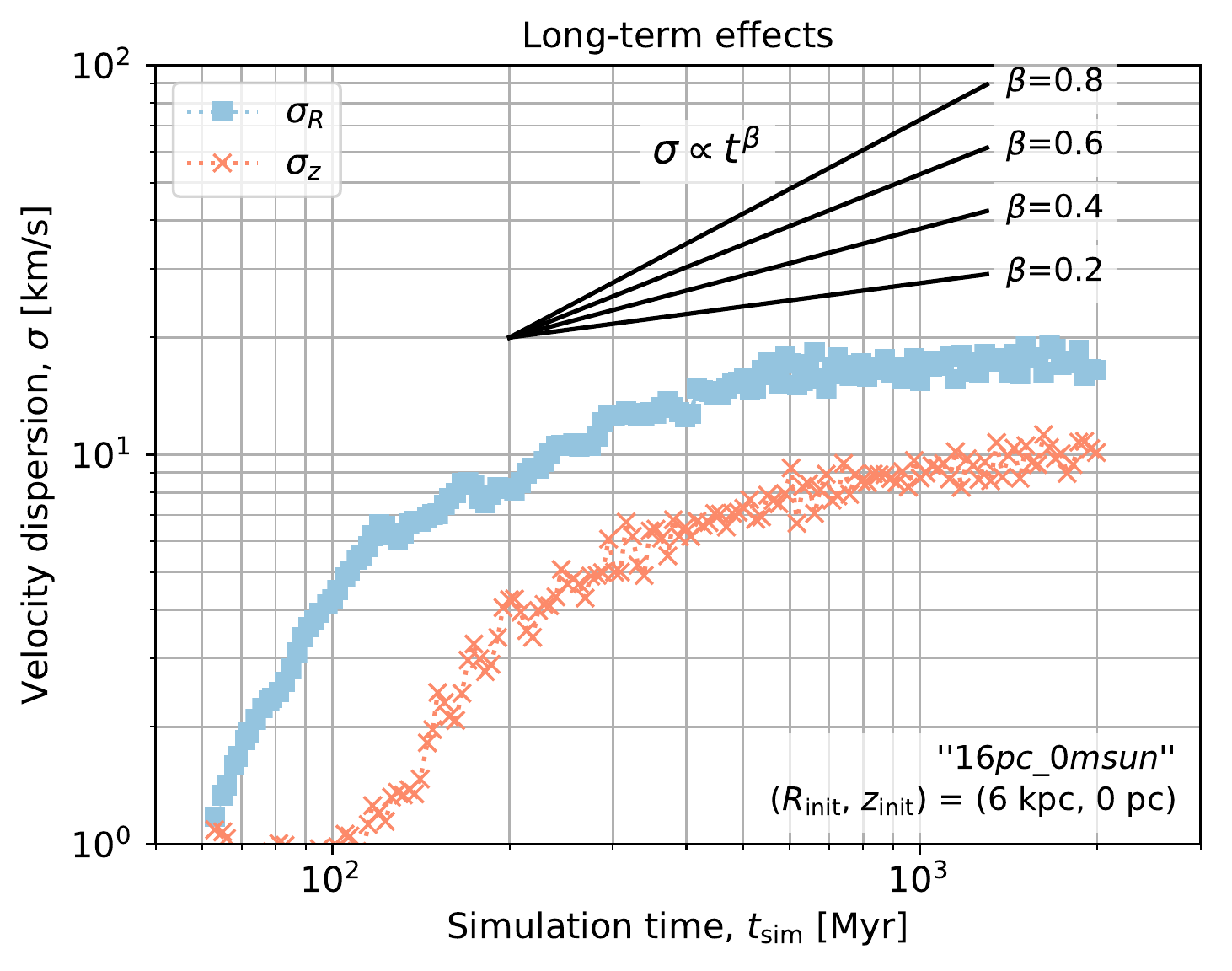}
    \caption{Heating history: time evolution of the radial and vertical velocity dispersion ($\sigma_R$ and $\sigma_z$). The velocity dispersion is defined as the standard deviation of radial or vertical velocities for sample particles. Top left: tracer particles with $R_{\rm init} = 6$\kpc\ and $z_{\rm init} = $ 0\pc\ from the simulation model of \textit{8pc\_0msun}. Top right: dependence on the initial vertical position ($z_{\rm init} =$ 0, 500, and 1000\pc). Middle left: dependence on the initial radial position ($R_{\rm init} =$ 4, 6, and 8\kpc). Middle right: dependence on the resolution ($\Delta x =$ 8 and 16\pc). Bottom left: dependence on star particle mass ($m_* =$ 0 and 300\Msun). Bottom right: long-term evolution of up to 2\Gyr\ (simulation model of \textit{16pc\_0msun}). Black solid lines show power laws of the form $\sigma \propto t^{\beta}$ with $\beta =$ 0.2, 0.4, 0.6, and 0.8.}
    \label{fig: time evolution of sigma}
\end{figure*}

Fig.~\ref{fig: time evolution of sigma} shows the time evolution of the radial and vertical velocity dispersion, $\sigma_R$ and $\sigma_z$, therefore, the heating history of the disc. The velocity dispersion $\sigma$ is defined as the standard deviation of velocities for sample particles. The top left panel of Fig.~\ref{fig: time evolution of sigma} shows the case of particles that have $R_{\rm init} = 6$\kpc\ and $z_{\rm init} = $ 0\pc, in which we regard the particles as newborn stars. We see a rapid increase of velocity dispersion in both directions in the first 100-200\Myr. After that, the increases slow down, and the slopes become shallow; if we fit them with a simple power law of the form $\sigma \propto t^{\beta}$, the exponents $\beta$ decrease with time from $\beta > 1$ to $\beta \sim 0.4$ by the time of $t_{\rm sim} = 1.1$\Gyr. Therefore, our result suggests that newborn stars formed close to the galactic mid-plane can be efficiently scattered by GMCs within $\sim$ 1\Gyr, especially within the first few hundred Myr, and that the young disc is heated kinematically with a large heating exponent $\beta$.

At first glance, the large exponents of $\beta \geq 0.4$ look inconsistent with previous analytical or numerical works, in which the heating exponent is often referred to as $\beta \sim 0.25$ \citep[][]{Lacey_1984, Hanninen_Flynn_2002, Aumer_Binney_Schonrich_2016b}. However, those previous works discuss the 10-Gyr scale long-term history of the disc. On the other hand, our study focuses on the 1-Gyr scale short-term history of the disc. In fact, our result is qualitatively consistent with previous work done by \citet{Kokubo_Ida_1992}, who perform numerical calculations of stellar orbits of star-GMC encounters in a differentially rotating disc with an epicycle approximation. They show that in an early phase, when the galactic shear motion dominates relative motion between a star and a GMC, the heating process is efficient, and the velocity dispersion rapidly increases with time as  $\sigma_R \propto t^{0.5}$ and $\sigma_z \propto \exp (t)$. In the later phase when the velocity dispersion exceeds the shear velocity, the heating rate becomes small as $\sigma \propto t^{0.25}$ in both directions, in which the exponent is consistent with those in other previous works discussing 10-Gyr scale heating history \citep[][]{Hanninen_Flynn_2002, Aumer_Binney_Schonrich_2016b}. The large exponent of $\beta \geq 0.4$ in our simulation suggests that GMCs gravitationally scatter newborn stars efficiently for at least 1\Gyr\ and that it takes more than 1\Gyr\ for the heating exponent to decrease to 0.25.

We next compare with observations. Our simulation shows that in 1\Gyr, the velocity dispersions increase from 0\kms\ to $\sigma_R \sim 17$\kms\ and $\sigma_z \sim 8$\kms. On the other hand, observations of stars in the solar neighbourhood show that young stars whose ages are around 1-2\Gyr\ have velocity dispersions of $\sigma_R \sim 25$\kms\ and $\sigma_z \sim 10$\kms\ \citep[e.g.][]{Nordstrom_et_al_2004, Sharma_et_al_2021}, which might be slightly larger than those in our simulation. The further increases of those values can be expected by the additional physical processes that are not considered in our simulations, such as the inheritance of the internal velocity dispersion of parent GMCs, clustering effects of embedded star formation in a GMC \citep{Kroupa_2002}, or other perturbers, such as spiral arms. Regarding the ratio of radial and vertical velocity dispersion, our simulation shows $\sigma_z/\sigma_R \sim 0.5$ at $t_{\rm sim} = 1$\Gyr, which is consistent with observations and theoretical works, in which the ratio ranges around 0.4-0.6 \citep[e.g.][]{Kokubo_Ida_1992, Hanninen_Flynn_2002, Nordstrom_et_al_2004, Aumer_Binney_Schonrich_2016b, Sharma_et_al_2021}. Although the observed age-velocity dispersion relation and its exponent have also been discussed in the context of the heating history of the Galaxy \citep[e.g.][]{Aumer_Binney_Schonrich_2016b, Kumamoto_Baba_Saitoh_2017}, we do not discuss them. The first reason is that tracer particles in our simulations are mono-age populations, so we can not draw age-velocity dispersion relations from them. The second reason is that the age-velocity dispersion relation has been discussed in the context of the 10-Gyr scale evolution of the Galactic disc, but our study focuses on the 1-Gyr scale short-term heating history.

We next look at the dependence of the heating history on the initial vertical position of particles. The top right panel of Fig.~\ref{fig: time evolution of sigma} shows the heating history of particles whose initial heights are $z_{\rm init} =$ 0, 500, and 1000\pc; we regard those particles as newborn stars, thin disc stars, and thick disc stars, respectively. Note that we only show the radial velocity dispersion $\sigma_R$, not the vertical one. The reason is that although the stellar disc with finite scale height generally has finite vertical velocity dispersion, those for particles with $z_{\rm init} =$ 500 and 1000\pc\ in our simulations become zero. The zero velocity dispersion is an outcome of the same phases of the vertical oscillation for particles that have the same initial heights. 

The figure shows a clear dependence; the larger the initial height, the later the timing at which the velocity dispersion begins to increase. This dependence might come from differences in the time spent in the galactic disc where GMCs as perturbers exist; the vertical velocity near the galactic mid-plane is $\sim$ 30 and 50\kms\ for particles with $z_{\rm init} =$ 500 and 1000\pc, respectively, and those correspond to $\sim$ 7 and 4\Myr\ duration within 100\pc\ from the galactic mid-plane. In addition, one period of vertical oscillation is $\sim$ 100 and 120\Myr\ for particles with $z_{\rm init} =$ 500 and 1000\pc, respectively, so the number of times they pass through the galactic disc in 400\Myr\ is $\sim$ 8 and 7 times. Therefore, the total duration in the galactic disc in 400\Myr\ is $\sim$ 60 and 30\Myr\ for particles with $z_{\rm init} =$ 500 and 1000\pc, respectively, while particles with $z_{\rm init} = 0$\pc\ stay in the galactic disc until the amplitudes of their vertical oscillation become large due to gravitational scattering by GMCs. As a result, the final velocity dispersion at $t_{\rm sim} = 1.1$\Gyr\ is $\sim$ 17, 13, and 8\kms\ for 0, 500, and 1000\pc\ initial height, respectively. Our results suggest that thin and thick disc stars can also be heated by GMCs in a radial direction depending on their heights, although not as much as newborn stars.

The middle left panel of Fig.~\ref{fig: time evolution of sigma} shows the dependence on the initial radial position $R_{\rm init}$, comparing particles with $R_{\rm init} =$ 4, 6, and 8\kpc. We see that the smaller the initial radius, the earlier the timing at which the velocity dispersion begins to increase. That might be because there is a radial gradient in surface densities of dense gas, as shown in Fig.~\ref{fig: radial profile dense gas}; the inner part of the galaxy has a denser environment where the number surface density of GMCs is high. We also see that particles with $R_{\rm init} = 8$\kpc\ are particularly slow to start to be heated. That might be because the dense gas production in the outer part beyond a radius of 8\kpc\ is particularly slow, as we have shown in Fig.~\ref{fig: radial profile dense gas}.

We next check a model dependence of the heating history; in Fig.~\ref{fig: time evolution of sigma}, the middle right panel shows the dependence on resolution $\Delta x$, and the bottom left panel shows the dependence on star particle mass $m_*$. There is a slight difference between models with different resolutions; in the case of the low resolution, the radial velocity dispersion $\sigma_R$ is slightly larger than in the high-resolution model between $200 < t_{\rm sim} < 800$\Myr. After that, however, the heating rate becomes small, and the velocity dispersion at $t_{\rm sim} \sim 1$\Gyr\ is similar in both models. The resolution dependence might come from a slightly different distribution of GMCs between models with different resolutions, as we have shown in subsection~\ref{subsec: distribution of giant molecular clouds}. Except for the point above, there is no significant model dependence. The no significant difference between the models with and without 300\Msun\ star particle mass, shown in the bottom left panel of Fig.~\ref{fig: time evolution of sigma}, suggests two important things. One is that in our galaxy simulations, which do not have enough resolution to resolve individual stars, the 300\Msun\ star particles required for the star formation recipe do not have a significant effect on the stellar scattering compared with the effects of GMCs. Therefore, the possibility we state in subsection~\ref{subsec: star particle mass} that the massive star particles act as compact and heavy gravitational sources and scatter the tracer particles unrealistically strongly is ruled out. The other is that young stellar clusters born in GMCs do not significantly affect the stellar scattering compared with the effects of GMCs. This is not surprising because the star formation efficiency per free-fall time is 0.01, so the mass contribution from newborn clusters as gravitational sources for stellar scattering is negligible compared with that from GMCs.

Lastly, we look at the longer-term effects of GMCs on the heating history. The bottom right panel of Fig.~\ref{fig: time evolution of sigma} shows the simulation model of \textit{16pc\_0msun}, which is a low-resolution model but runs for 2\Gyr, twice as the high-resolution run. We see that after $t_{\rm sim} \sim 1$\Gyr, the heating rate becomes small, and the velocity dispersion scarcely increases.

\subsubsection{The motion of individual particles}
\label{subsubsec: individual particles}

\begin{figure*}
    \centering
	\includegraphics[width=\columnwidth]{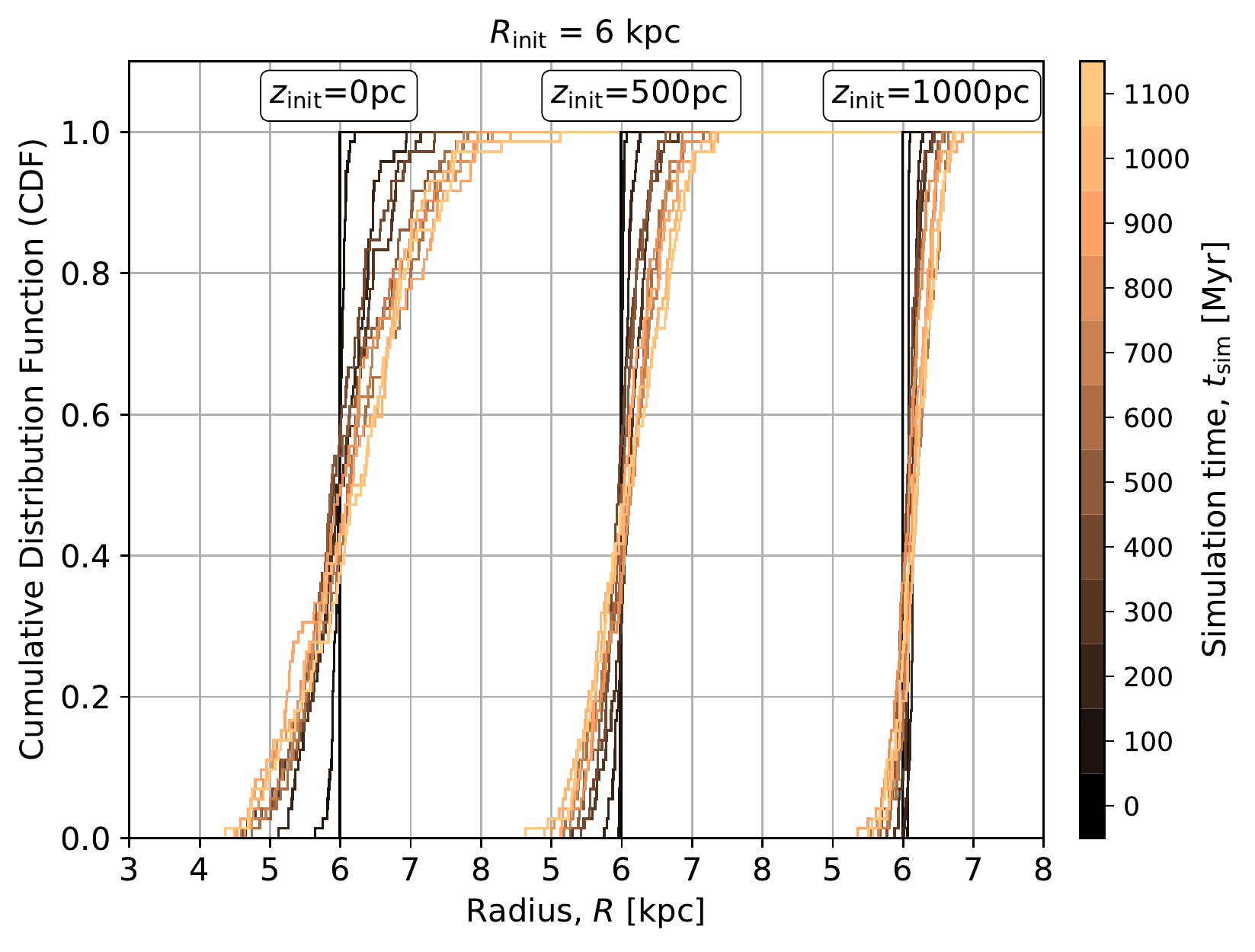}
	\includegraphics[width=\columnwidth]{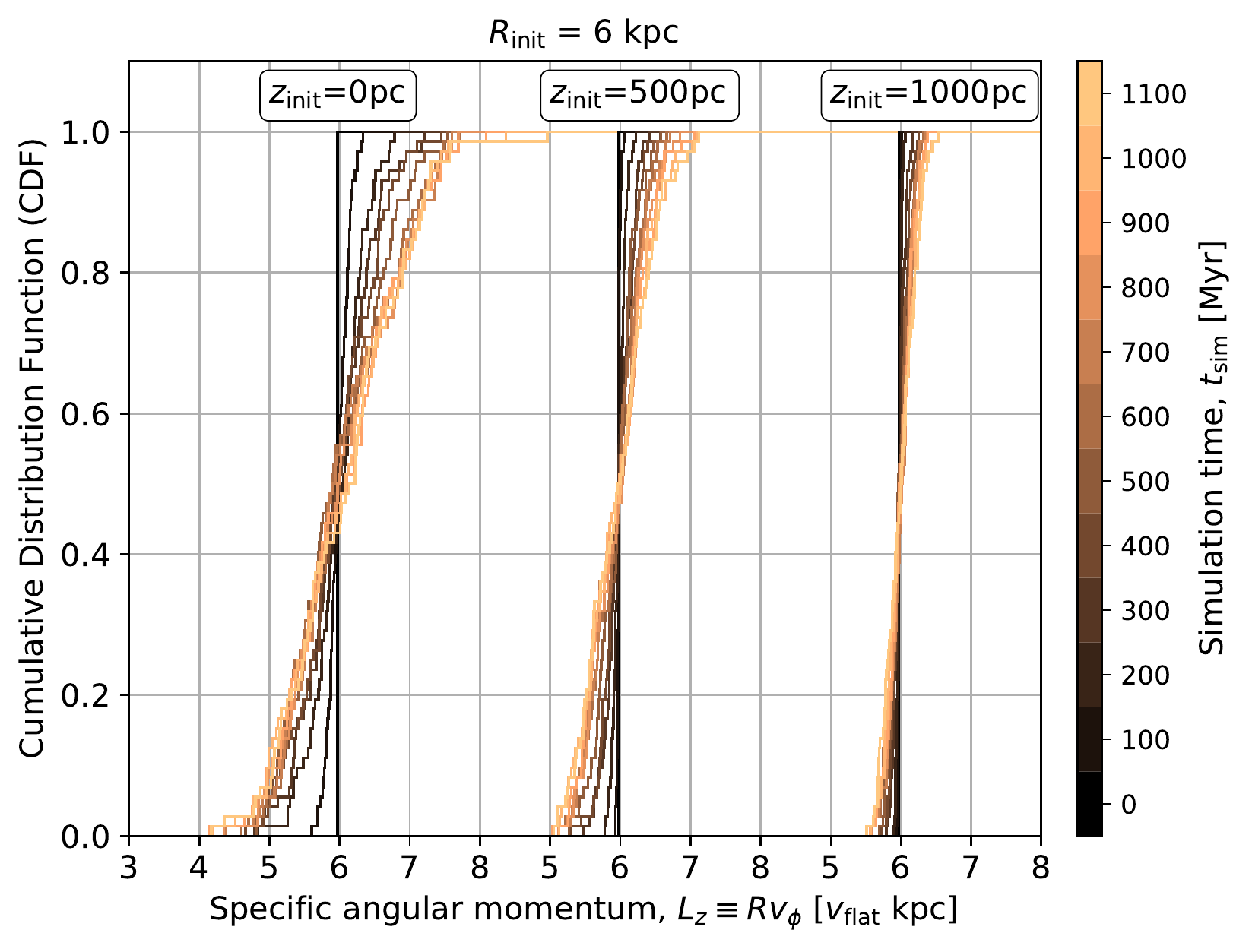} \\
	\includegraphics[width=\columnwidth]{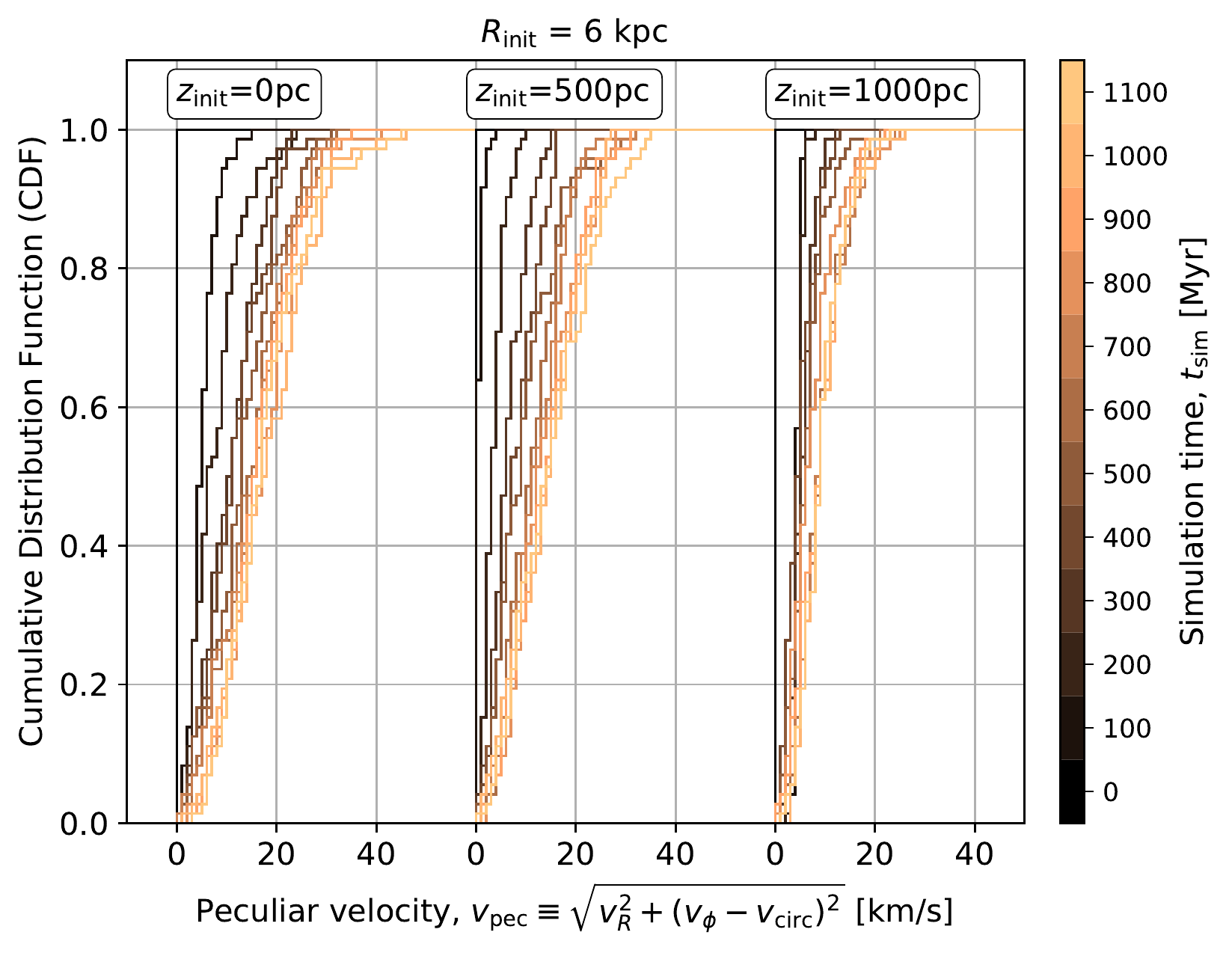}
	\includegraphics[width=\columnwidth]{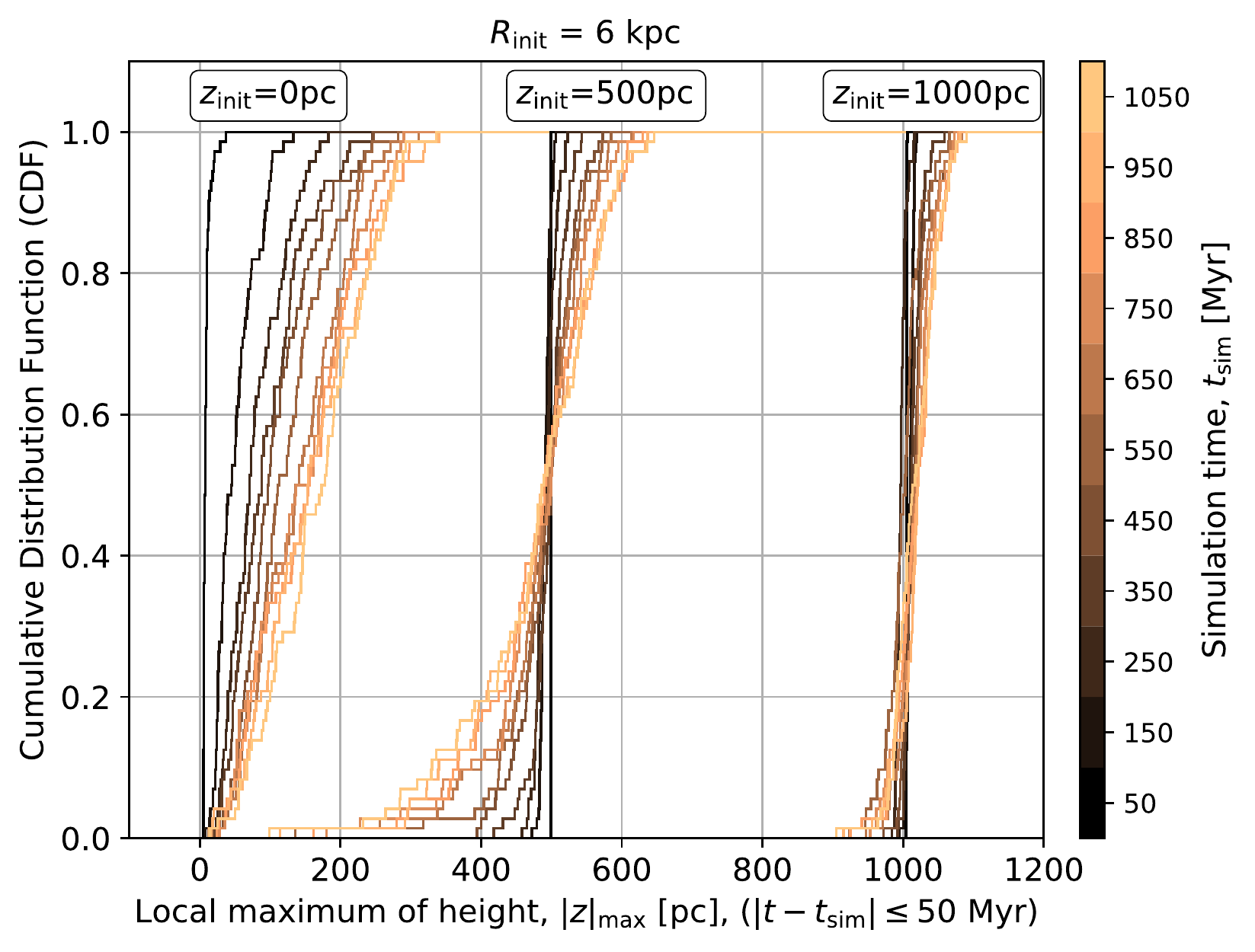}
    \caption{Time evolution of normalised cumulative distribution functions of the instantaneous radial position (top left), specific angular momentum (top right), peculiar velocity (bottom left), and a local maximum of height (bottom left) for tracer particles whose initial radius is $R_{\rm init} = 6$\kpc\ and initial heights are $z_{\rm init} = $ 0, 500, and 1000\pc. The number of samples (= particles) at each initial height is 72. The simulation model is \textit{8pc\_0msun}. For easy comparison with the radius $R$, the unit of angular momentum $L_z$ is normalized by the flat rotation speed of $v_{\rm flat} = 200$\kms.}
    \label{fig: cumulative distribution functions}
\end{figure*}

\begin{figure}
    \centering
	\includegraphics[width=\columnwidth]{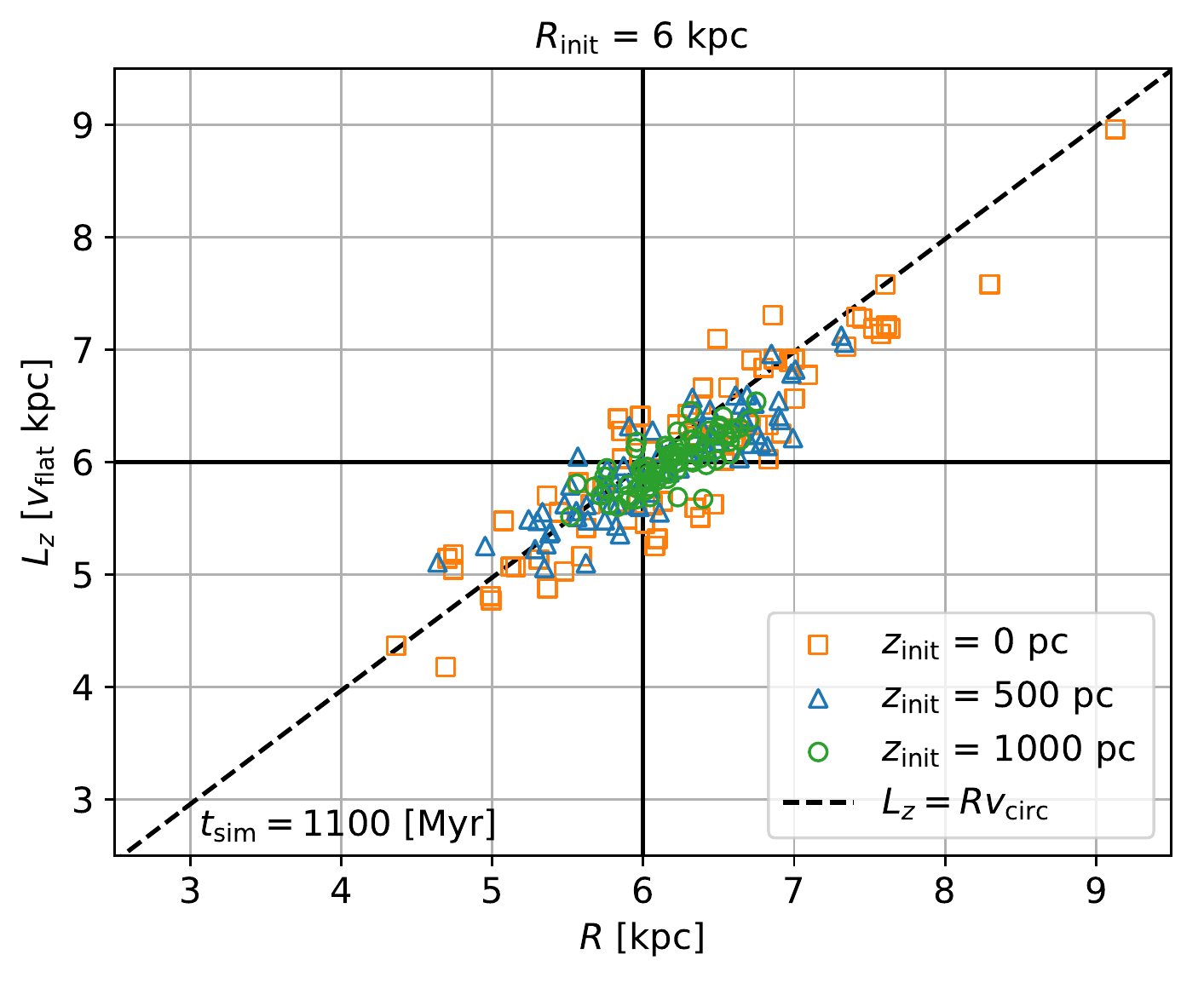}
    \caption{Scatter plot of $L_z$ versus $R$ at $t_{\rm sim} = 1100$\Myr. The dashed line indicates pure circular orbits of $L_z = R v_{\rm circ}$, where $v_{\rm circ}$ is from equation~(\ref{eq: circular velocity}).}
    \label{fig: R vs Lz}
\end{figure}

\begin{figure}
    \centering
	\includegraphics[width=\columnwidth]{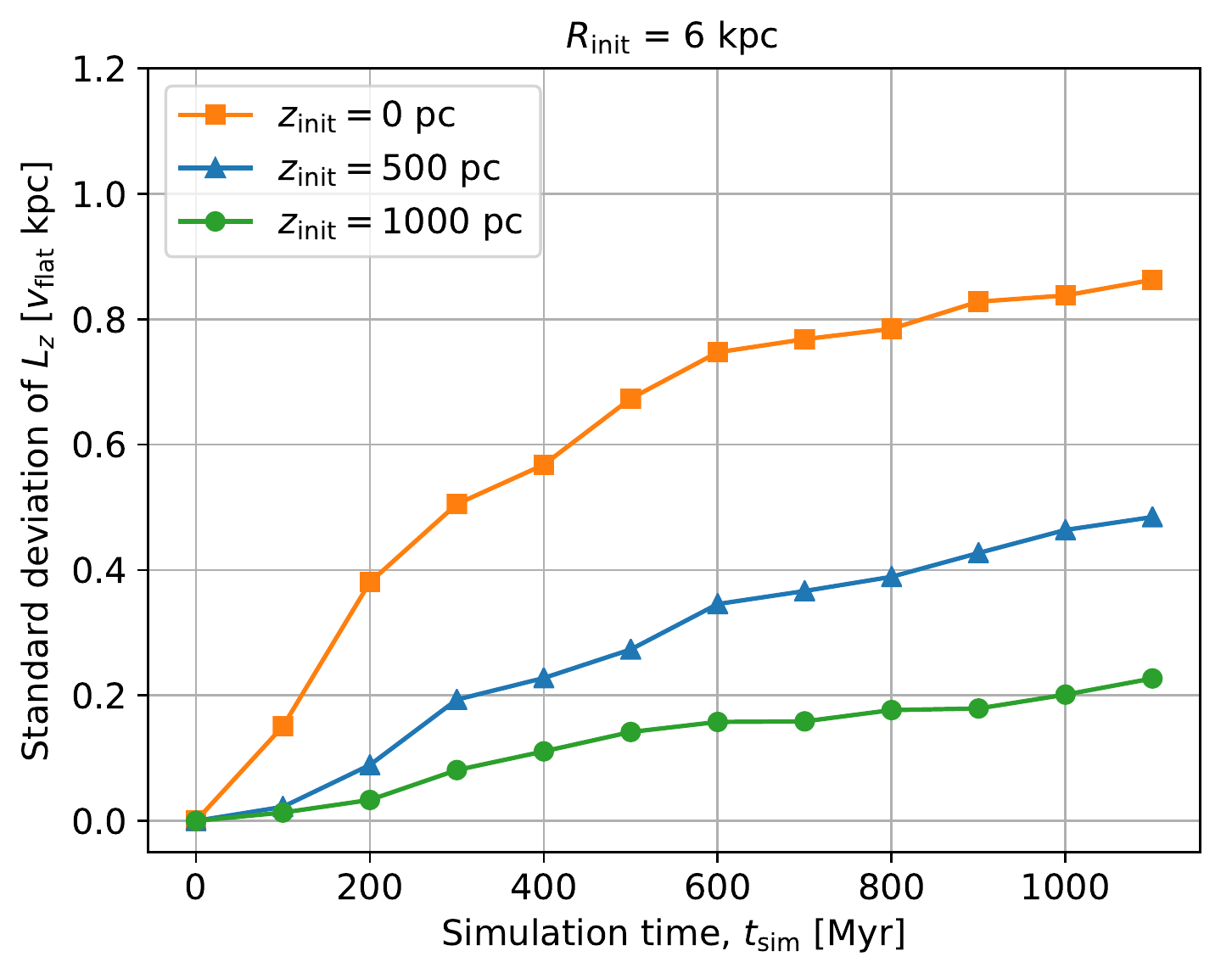}
    \caption{Time evolution of the standard deviation of the distribution of the specific angular momentum shown in the top right panel of Fig.~\ref{fig: cumulative distribution functions}.}
    \label{fig: time evolution of deviation}
\end{figure}

In this sub-subsection, we investigate radial mixing and vertical heating caused by GMCs' gravitational scattering, analysing each particle's motion. For that purpose, we measure an instantaneous radial position $R$, specific angular momentum $L_z$, peculiar velocity $v_{\rm pec}$, and a local maximum of the height $|z|_{\rm max}$ for each tracer particle. The specific angular momentum is defined as $L_z \equiv Rv_{\phi}$, where $v_{\phi}$ is the particle's velocity in the rotational direction. The peculiar velocity is the deviation of the projected velocity onto the galactic plane from the rotation velocity of the galactic disc, $v_{\rm circ}$, as given by equation~(\ref{eq: circular velocity}). Therefore, it is defined as $v_{\rm pec} \equiv \left\{v_R^2 + (v_{\phi} - v_{\rm circ})^2\right\}^{1/2}$, where $v_{R}$ is the particle's velocity in the radial direction. For the last quantity of $|z|_{\rm max}$, we measure the maximum height from the galactic disc within 50 Myr before and after a given time rather than measure an instantaneous height at the time. 

Fig.~\ref{fig: cumulative distribution functions} shows the time evolution of cumulative distribution functions of $R$, $L_z$, $v_{\rm pec}$, and $|z|_{\rm max}$ for particles with $R_{\rm init} = 6$\kpc\ and $z_{\rm init} = $ 0, 500 and 1000\pc. Looking at the distributions of the instantaneous radial position $R$ in the top left panel of Fig.~\ref{fig: cumulative distribution functions}, we see that particles deviate from their initial radii over time and that the trend is more pronounced when the initial height is small. For instance, in the case of $z_{\rm init} = 0$\pc, 20 per cent of particles move outward to $R \sim 7$\kpc\ or more, and 10 per cent move inward to $R \sim 5$\kpc\ or less by the time of 1\Gyr. This result shows that radial mixing of churning or blurring occurs more within the galactic disc. To determine whether radial migration occurs, we look at the distributions of the specific angular momentum $L_z$ in the top right panel of Fig.~\ref{fig: cumulative distribution functions}. We see a similar trend with the radial position, indicating that churning and radial migration occur more within the galactic disc. For instance, given that the galactic circular velocity of the flat rotation curve is 200\kms, 30 per cent of particles with $z_{\rm init} = 0$\pc\ change their guiding-centre radius more than 1\kpc\ to either direction inward or outward\footnote{This suggests that when we look at an individual star, it can migrate inward or outward with almost the same probability. On the other hand, when we sample stars at a given galactic radius bin, the number of stars that come from a small radius should dominate because the stellar density increases with a smaller galactic radius.}. In contrast, in the case of $z_{\rm init} = 1000$\pc, the change in the angular momentum is not as large as the one in the radial position, indicating that radial migration is small. 

Fig.~\ref{fig: R vs Lz} is a scatter plot of angular momentum $L_z$ versus instantaneous radial position $R$ at $t_{\rm sim} = 1100$\Myr, and the dashed line indicates pure circular orbits of $L_z = R v_{\rm circ}$, where $v_{\rm circ}$ is the galactic rotation velocity from equation~(\ref{eq: circular velocity}). 
This plot shows two important things. One is that the data points mostly align with the line of pure circular orbits (dashed line); particles that have largely changed their radial position from the initial radius of 6\kpc, such as ones with $R > 7$\kpc\ or $R < 5$\kpc, have largely changed their angular momentum at the same time. It indicates that strong radial migration, or churning, occurs. The other is that the distribution of the data points shows a small deviation from the pure circular orbit's line; the deviation is mostly less than 1\kpc\ in the horizontal direction. Dynamical disc heating, or blurring, is the process that increases the peculiar motion from the circular orbit and the epicycle amplitude. Therefore, the small deviation (less than 1\kpc) compared to the amount of change in radius (up to 2-3\kpc) indicates that churning occurs more efficiently than blurring for such particles. We also note that those trends are more significant when the initial height is small.

We quantify the blurring by examining the peculiar velocity and a local maximum of the height. The bottom panels of Fig.~\ref{fig: cumulative distribution functions} show the time evolution of cumulative distribution functions of $v_{\rm pec}$ and $|z|_{\rm max}$. Similar to the radial position and angular momentum, the peculiar velocity increases with time, and the medians are almost consistent with the radial velocity dispersion shown in the top right panel in Fig.~\ref{fig: time evolution of sigma}. Again, the increase with time is more pronounced when the initial height is small, although the difference between the case of $z_{\rm init} = $ 0\pc\ and 500\pc\ looks insignificant compared to the one in the angular momentum. Not only in-plane motion but also vertical motion is affected by GMCs; for particles with zero initial height, the maximum height increase with time and 40 per cent reach over 200\pc\ from the galactic mid-plane. For particles with non-zero initial heights, such as $z_{\rm init} = $ 500\pc\ and 1000\pc, they show deviation from the initial heights; some increase their vertical motion, and others decrease.

The churning and radial migration drastically occur in the first several hundred Myr, especially when the initial height is small. Fig.~\ref{fig: time evolution of deviation} shows the time evolution of the standard deviation of the specific angular momentum. When $z_{\rm init} = 0$\pc, the deviation quickly increases in an early phase, and then the growth slows down after around $t_{\rm sim} \sim 600$\Myr. Fig.~\ref{fig: long term effects} in Appendix~\ref{appendix: the long-term effects} examines the time evolution of cumulative distribution functions with the simulation model of \textit{16pc\_0msun} over a longer time period, up to 2\Gyr, and shows a similar trend. These trends are likely because particles that are initially on or close to the galactic mid-plane begin to oscillate vertically across the galactic plane with time, reducing the time spent in the galactic disc, where most GMCs are located within 100\pc\ of the galactic plane (see Fig.~\ref{fig: vertical distribution of clouds}). In fact, the larger the initial height is, the smaller the increase in the standard deviation is. Our result is consistent with recent observations of open clusters; \citet{Chen_Zhao_2020} investigate radial migration rates of open clusters using Gaia DR2 data and find that young clusters with ages of less than 1\Gyr\ tend to migrate in both inward and outward directions and to have larger migration rates than old clusters \citep[see also][]{Zhang_Chen_Zhao_2021}.

We also note that the duration of the increase seems slightly different between the angular momentum and velocity dispersion; Fig.~\ref{fig: time evolution of deviation} shows that the increase of angular momentum occurs in the first $\sim 600$\Myr, and, on the other hand, Fig.~\ref{fig: time evolution of sigma} shows that the increase of velocity dispersion occurs in the first 100-200\Myr. This is not surprising because churning and blurring are different physical processes, so they need not correlate. We will discuss this further in the next Section~\ref{subsubsec: radial migration and disc heating}.

At the end of this sub-subsection, we briefly check the dependence on other parameters. The features of the cumulative distribution functions of $R$, $L_z$, $v_{\rm pec}$, and $|z|_{\rm max}$ for particles with $R_{\rm init} = 6$\kpc\ discussed so far can also be seen in the case with different initial radial positions (see Fig.~\ref{fig: radial dependence} in Appendix~\ref{appendix: the dependence on the initial radial position}), indicating no significant dependence of the stellar scattering on the initial radial position. Regarding the dependence on resolution $\Delta x$, we see slightly stronger effects of GMCs on stellar scattering in the lower-resolution case, which might come from a slightly different distribution of GMCs with different resolutions. However, qualitative features are almost the same in both models (see Fig.~\ref{fig: resolution dependence} in Appendix~\ref{appendix: the dependence on simulation models}). Regarding the dependence on star particle initial mass $m_*$, there is almost no difference (see Fig.~\ref{fig: initial star particle mass dependence} in Appendix~\ref{appendix: the dependence on simulation models}).

\subsubsection{Relationship between radial migration and disc heating}
\label{subsubsec: radial migration and disc heating}

\begin{figure*}
    \centering
	\includegraphics[width=\columnwidth]{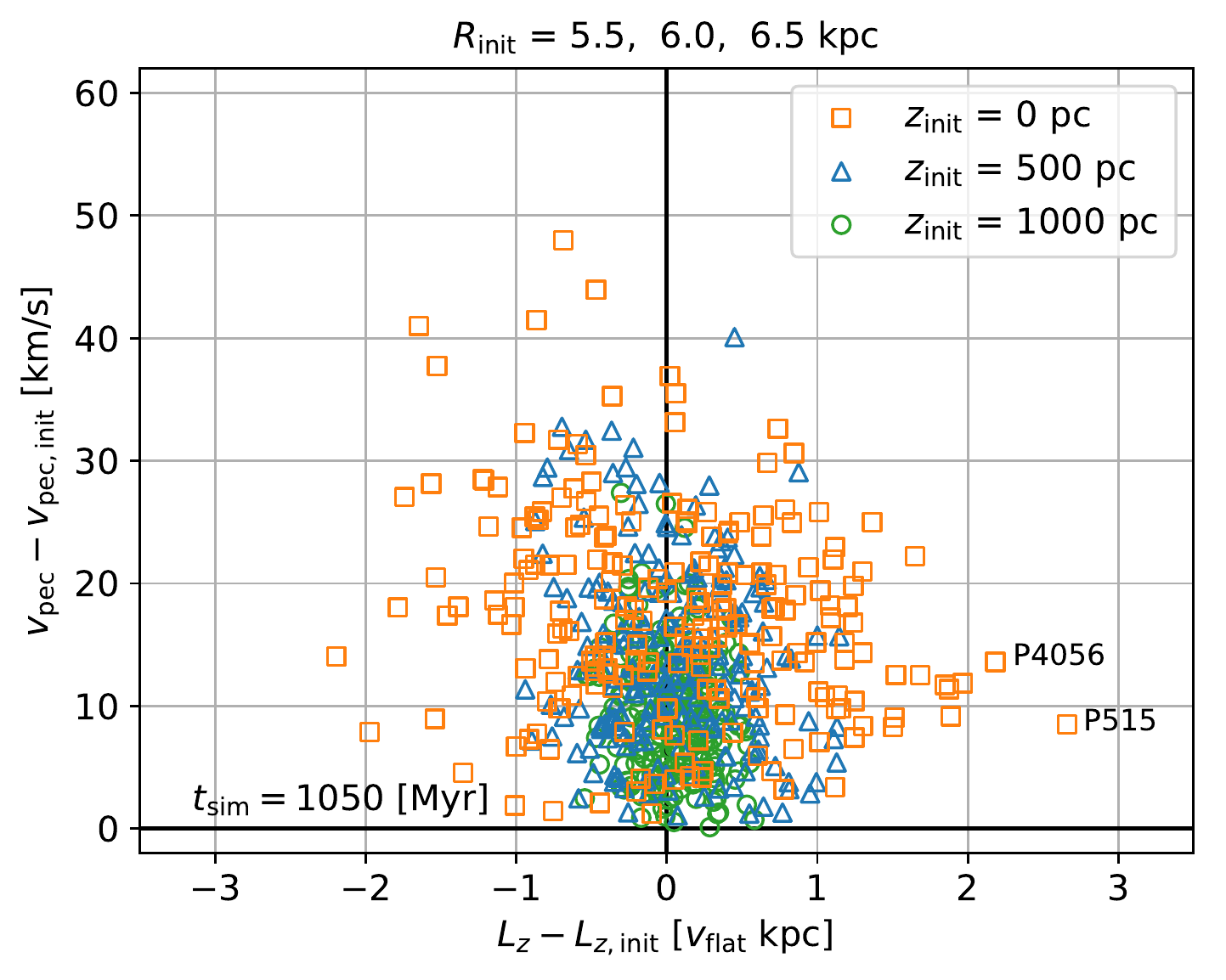}
	\includegraphics[width=\columnwidth]{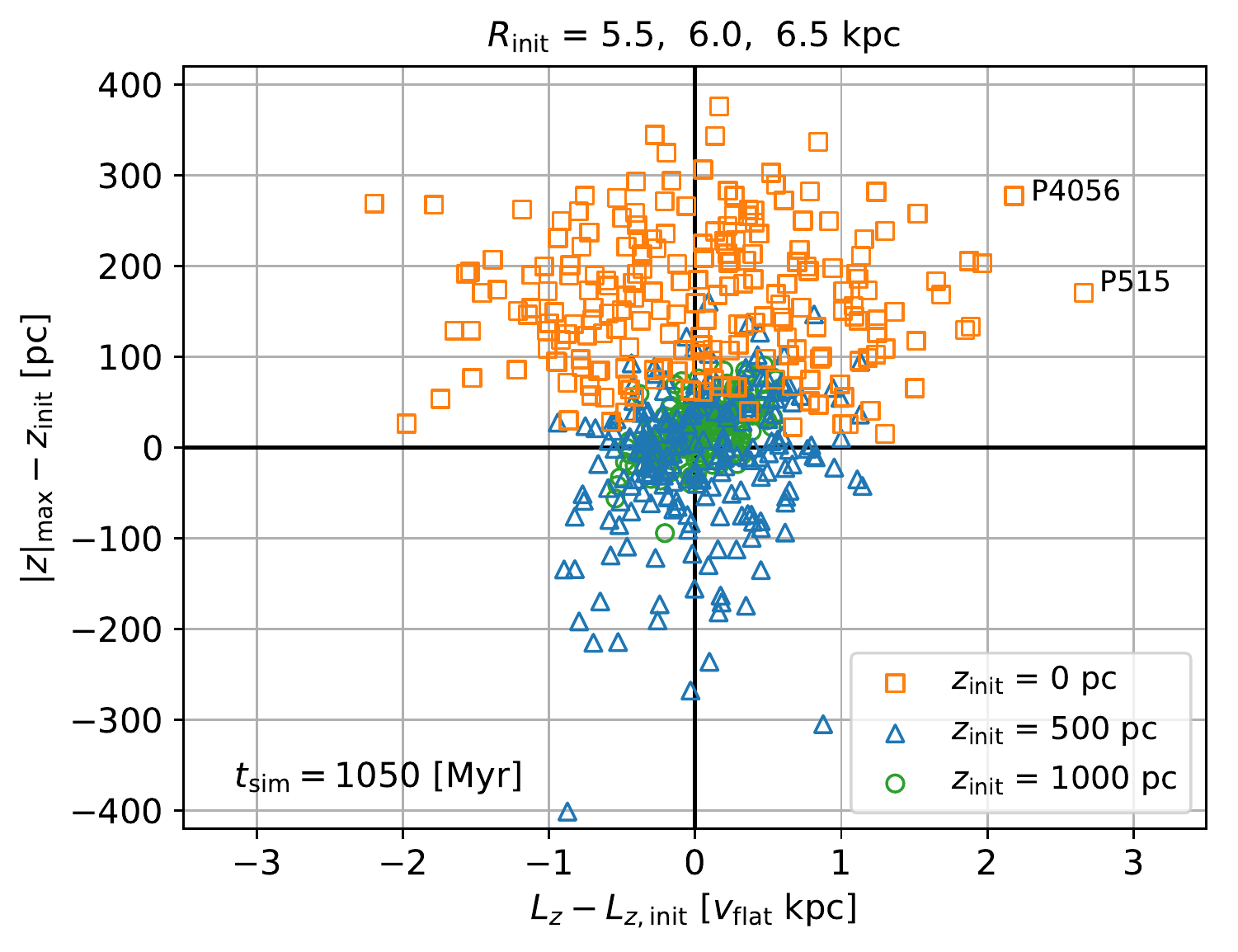}
    \caption{Scatter plots: peculiar velocity versus specific angular momentum (left) and the local maximum of height versus specific angular momentum (right). All these physical quantities are shown as differences between the initial value and the value at $t_{\rm sim} = 1050$\Myr. The samples are tracer particles whose initial radial positions are $R_{\rm init} =$ 5.5, 6.0, and 6.5\kpc\ and heights are $z_{\rm init} =$ 0, 500, and 1000\pc. The total number of sample particles at a given initial height is 216 ($= 72 \times 3$). The simulation model is \textit{8pc\_0msun}. The numbers shown in the panels indicate the particle IDs, and the time evolution of $R$, $L_z$, $v_{\rm pec}$, and $z$ for those particles is shown in Fig.~\ref{fig: time evolution each particle}.}
    \label{fig: scatter plot}
\end{figure*}

To investigate a relationship between churning and blurring, we examine a correlation between the following three physical quantities: specific angular momentum, peculiar velocity, and maximum height. Fig.~\ref{fig: scatter plot} shows the scatter plots for the three quantities, which are described as  the amount of change from their initial values. To ensure a sufficient number of samples, not only particles with an initial radius of 6\kpc\ but also those with 5.5 and 6.5\kpc\ are plotted. First, looking at the left panel of Fig.~\ref{fig: scatter plot}, which shows a dependence between angular momentum and peculiar velocity, we do not see a clear correlation; instead, the distribution seems almost random. Of course, the distribution is not perfectly random because particles that have obtained their angular momentum significantly without obtaining peculiar velocities look particularly rare. However, other than that, data points look distributed broadly. That means there is no strong correlation between churning and blurring, at least on the galactic plane, and each particle does not necessarily change the two physical quantities simultaneously. Regarding the dependence on initial heights, the spatial distribution in the scatter plot is quite different, as we expect from the results so far (e.g. Fig.~\ref{fig: cumulative distribution functions}); the smaller the initial height, the broader the data points are distributed.

We next investigate a correlation between churning and blurring in the vertical direction. The right panel of Fig.~\ref{fig: scatter plot} shows the scatter plot between specific angular momentum and the maximum height. When $z_{\rm init} = 0$\pc\ the data points are distributed only in the upper half of the plot because the maximum height minus the initial height (= 0) cannot be negative ($|z|_{\rm max} - z_{\rm init} = |z|_{\rm max} > 0$). When $z_{\rm init} = 0$\pc, although there is a slight trend that particles that have changed significant angular momentum have obtained non-zero peculiar velocities, there is no clear correlation between the two quantities, similar to the scatter plot of angular momentum and peculiar velocity discussed in the previous paragraph. This trend indicates no strong correlation between churning and blurring also in the vertical direction. In addition, the larger the initial height, the smaller the data points are distributed, as we expect. When the initial height is large, especially when $z_{\rm init} = 1000$\pc, we see a slight correlation between two quantities. It might be because the depth of the galactic potential in the mid-plane slightly increases with the galactic radius. This trend cannot be seen in the case of $z_{\rm init} = 0$\pc, maybe because the effect of the scattering off from GMCs is much stronger. We also find that the data points are distributed slightly more on the lower side of the plot when $z_{\rm init} = 500$\pc. It indicates that the interaction with GMCs tends to decrease the vertical motion of particles rather than increase it, particularly when particles have such intermediate height. This feature can also be seen in the lower right panel of Fig.~\ref{fig: cumulative distribution functions}. 

\subsubsection{Can GMCs explain the Solar system's migration?}
\label{subsubsec: Solar migration}

Lastly, we look at particles that obtain (not lose) specific angular momentum of more than $1 \times 200$\kpckms, which corresponds to 1\kpc\ or more of outward radial migration, given the circular velocity of 200\kms. Such particles can be good examples for discussing the migration of the Solar system because the Solar birth radius is thought to be 1\kpc\ or more inner than the current radius. The top left panel of Fig.~\ref{fig: scatter plot} shows that when $z_{\rm init} = 0$\pc, 35 out of 216 particles obtain such large angular momentum and that their peculiar velocities are broad, ranging from a few to $\sim$ 25\kms; Quantitatively, 19 out of 35 (54 per cent) particles obtain a peculiar velocity of $10 < v_{\rm pec} < 20$\kms\ and 6 out of 35 (17 per cent) particles obtain a peculiar velocity of $v_{\rm pec} > 20$\kms. The number of particles that obtain a small peculiar velocity of $v_{\rm pec} < 10$\kms\ is 10 out of 35 (29 per cent), and they can be a good example of the Solar system because the current Solar peculiar velocity with respect to the local standard of rest is about 15\kms\ \citep{Bland-Hawthorn_Gerhard_2016}\footnote{We do not provide a detailed comparison between our results and the Solar value for a few reasons. First, we do not calculate the full 4.6 Gyr-scale time evolution, so our simulated disc has not been heated at a similar level to the Solar age stars. Second, some physical processes, such as spiral arms, are not included in our simulations, which results in slight small velocity dispersion, as discussed in Section~\ref{subsubsec: heating history}. Third, our simulations are not fine-turned to either the current state of the Milky Way or the state it was in 4.6\Gyr\ ago. For example, there is a $\sim 30$\kms\ difference in the circular velocity between the inferred Milky Way's present-day value and the one used in our model.}, which is relatively small compared to the radial velocity dispersion of 35\kms\ for nearby stars with a similar age to the Sun \citep{Nordstrom_et_al_2004, Sharma_et_al_2021}. Notably, we see that two particles obtain significant specific angular momentum of more than $2 \times 200$\kpckms, which corresponds to 2\kpc\ or more outward radial migration, and that they obtain moderate peculiar velocities around 10\kms. The probability is not so high; it is only $\sim$ 1 per cent (2 out of 216). However, it suggests that even without other perturbers, such as spiral arms, GMC's stellar scattering alone can be responsible for most of the radial migration of the Solar system. Regarding the maximum height, the two particles have a few times larger values than the Solar value, which is estimated as 100 pc or less \citep{Bahcall_Bahcall_1985, McMillan2017}. This might be because our galactic potential has a slightly small vertical frequency compared to the Milky Way (see Section~\ref{subsec: galaxy model}); the smaller the vertical frequency of the disc, the more the stars can be launched vertically.

In Fig.~\ref{fig: time evolution each particle} in Appendix~\ref{appendix: time evolution of Sun-like motion particles}, we show the time evolution of $R$, $L_z$, $v_{\rm pec}$, and $z$ for the two particles. The angular momentum shows a gradual change, and on the other hand, the peculiar velocity shows multiple sudden changes over the 1\Gyr. This means that once particles obtain significant angular momentum, they hardly return to the initial value quickly. On the other hand, particles can have small peculiar velocities, like the Solar system, even after having large peculiar motions. This result supports the idea that the GMC's scattering can be responsible for Solar radial migration.

%\section{Maths}
%
%Simple mathematics can be inserted into the flow of the text e.g. $2\times3=6$
%or $v=220$\,km\,s$^{-1}$, but more complicated expressions should be entered
%as a numbered equation:
%
%\begin{equation}
%    x=\frac{-b\pm\sqrt{b^2-4ac}}{2a}.
%	\label{eq:quadratic}
%\end{equation}
%
%Refer back to them as e.g. equation~(\ref{eq:quadratic}).

\section{Conclusions}
\label{sec: conclusions}

We have performed hydrodynamical simulations of an isolated galactic gaseous disc. The galaxy model is an axisymmetric external potential to consider only GMCs and to exclude other perturbers, such as spiral arms. The resolutions are high enough to reproduce realistic GMC distributions. We have analysed tracer particles that are initially placed in the galaxy to mimic newborn stars and thin and thick disc old stars. We have investigated how much orbits of the tracer particles are displaced from their circular orbits due to gravitational interactions with GMCs. We have particularly focused on the 1 Gyr scale short-term effect of GMCs on stellar scattering immediately after stellar birth. Our main results are as follows.

\begin{enumerate}
    \item GMCs effectively heat the galactic disc; the velocity dispersion $\sigma$ rapidly increases in the first few hundred Myr, and the heating exponent $\beta$ becomes much larger than 1. After that, although the heating exponent decreases, it keeps large values of $0.3 \sim 0.6$ for 1\Gyr\ (Fig.~\ref{fig: time evolution of sigma}). 
    \item The efficient and rapid heating of the disc is pronounced for particles with a smaller initial height, especially ones with $z_{\rm init} = 0$\pc, which we regard as newly formed stars close to the galactic mid-plane. On the other hand, the thin and thick disc stars ($z_{\rm init} = $ 500 and 1000\pc) are moderately heated by GMCs (Fig.~\ref{fig: time evolution of sigma}).
    \item Not only disc heating but also radial migration occur; 30 per cent of newborn stars ($z_{\rm init} = 0$\pc) show a change in their angular momentum and hence, a change in the guiding-centre radius of more than 1\kpc\ to either inward or outward direction (Fig.~\ref{fig: cumulative distribution functions} and Fig.~\ref{fig: R vs Lz}).
    \item The radial migration drastically occurs in the first several hundred Myr. After that, particles start oscillating vertically across the galactic plane, and the change in angular momentum slows down (Fig.~\ref{fig: cumulative distribution functions} and Fig.~\ref{fig: time evolution of deviation}).
    \item There is no significant correlation between radial migration and disc heating; particles that change their angular momentum do not necessarily change their peculiar motion simultaneously, and vice versa (Fig.~\ref{fig: scatter plot}). 
    \item Some particles move more than 2\kpc\ radially outward without obtaining significantly large peculiar velocities (Fig.~\ref{fig: scatter plot} and Fig.~\ref{fig: time evolution each particle}).
\end{enumerate}

In summary, we find that GMCs efficiently scatter newborn stars in the first several hundred Myr immediately after the stellar birth. The GMC's stellar scattering is a random process, so it independently changes each star's angular momentum and peculiar motion. Those results suggest that our Solar system might experience significant radial migration due to gravitational interaction with nearby GMCs shortly after its birth, 4.6\Gyr\ ago. 

\section*{Acknowledgements}
YF is supported by JSPS KAKENHI Grant Number 22K20387. SI is supported by JSPS KAKENHI Grant Numbers 18H05436 and 18H05437. JB is supported by JSPS KAKENHI Grant Numbers 21H00054, 21K03633 and 22H01259. Simulations were carried out on Cray XC50 at Center for Computational Astrophysics, National Astronomical Observatory of Japan, Oakforest-PACS provided by Multidisciplinary Cooperative Research Program in Center for Computational Sciences, University of Tsukuba, and Oakbridge-CX provided by Information Technology Center, University of Tokyo, through the HPCI System Research Project (Project ID: hp210015). Computations described in this work were performed using the publicly available \textsc{enzo} code (\citealt{Bryan_et_al_2014}; \url{http://enzo-project.org}), which is the product of a collaborative effort of many independent scientists from numerous institutions around the world. Their commitment to open science has helped make this work possible. We acknowledge extensive use of the \textsc{yt} package (\citealt{Turk_et_al_2011}; \url{http://yt-project.org}) in analysing these results and the authors would like to thank the \textsc{yt} development team for their generous help.

%The Acknowledgements section is not numbered. Here you can thank helpful
%colleagues, acknowledge funding agencies, telescopes and facilities used etc.

%%%%%%%%%%%%%%%%%%%%%%%%%%%%%%%%%%%%%%%%%%%%%%%%%%
\section*{Data Availability}
 
%The inclusion of a Data Availability Statement is a requirement for articles published in MNRAS. Data Availability Statements provide a standardised format for readers to understand the availability of data underlying the research results described in the article. The statement may refer to original data generated in the course of the study or to third-party data analysed in the article. The statement should describe and provide means of access, where possible, by linking to the data or providing the required accession numbers for the relevant databases or DOIs.

The data underlying this article will be shared on reasonable request to the corresponding author.



\bibliographystyle{mnras}
\bibliography{example, reference} % if your bibtex file is called example.bib

%\begin{thebibliography}{99}
%\end{thebibliography}

%%%%%%%%%%%%%%%%%%%%%%%%%%%%%%%%%%%%%%%%%%%%%%%%%%

%%%%%%%%%%%%%%%%% APPENDICES %%%%%%%%%%%%%%%%%%%%%

\appendix

\section{Long-term effects}
\label{appendix: the long-term effects}

\begin{figure*}
    \centering
    	\includegraphics[width=\columnwidth]{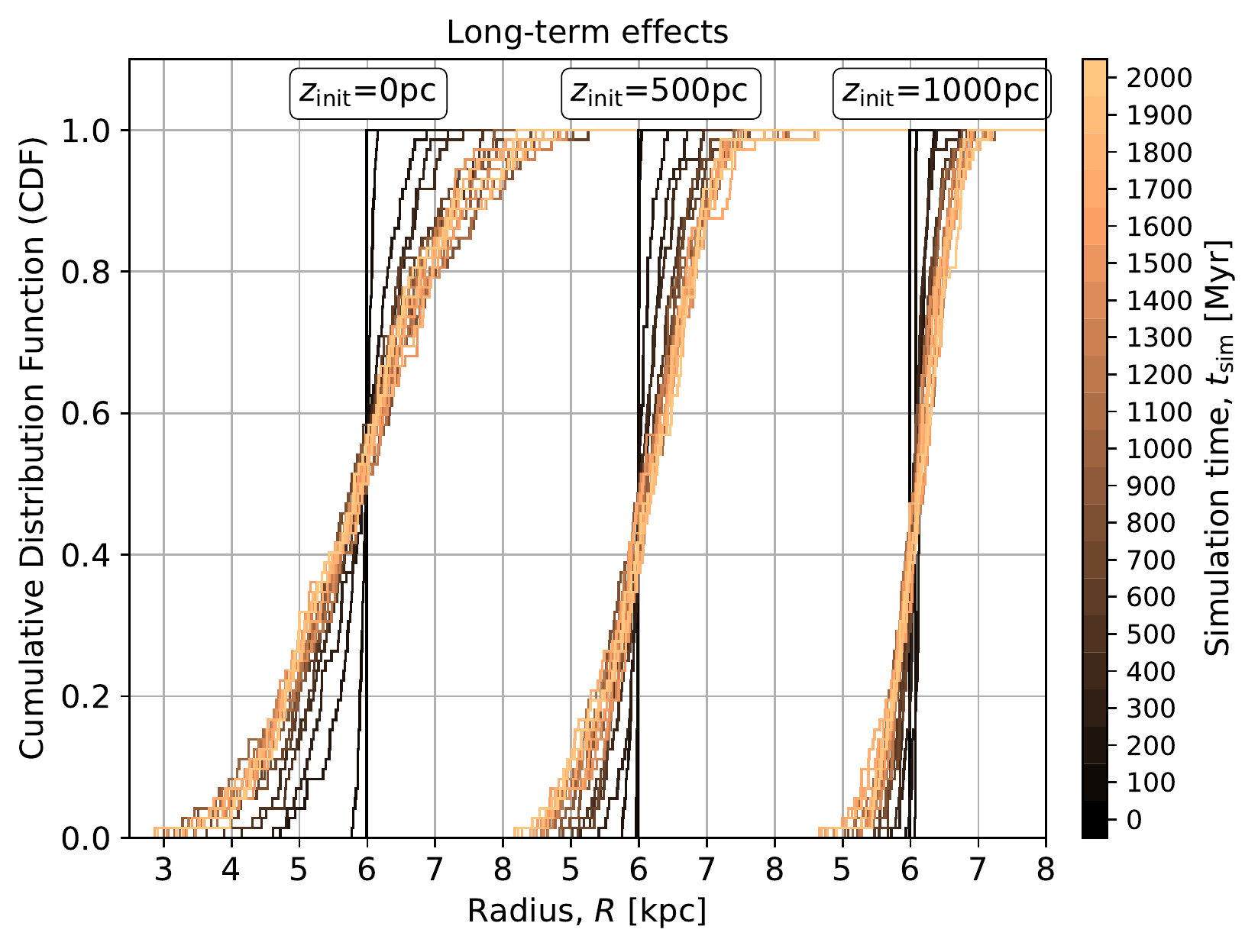}
    	\includegraphics[width=\columnwidth]{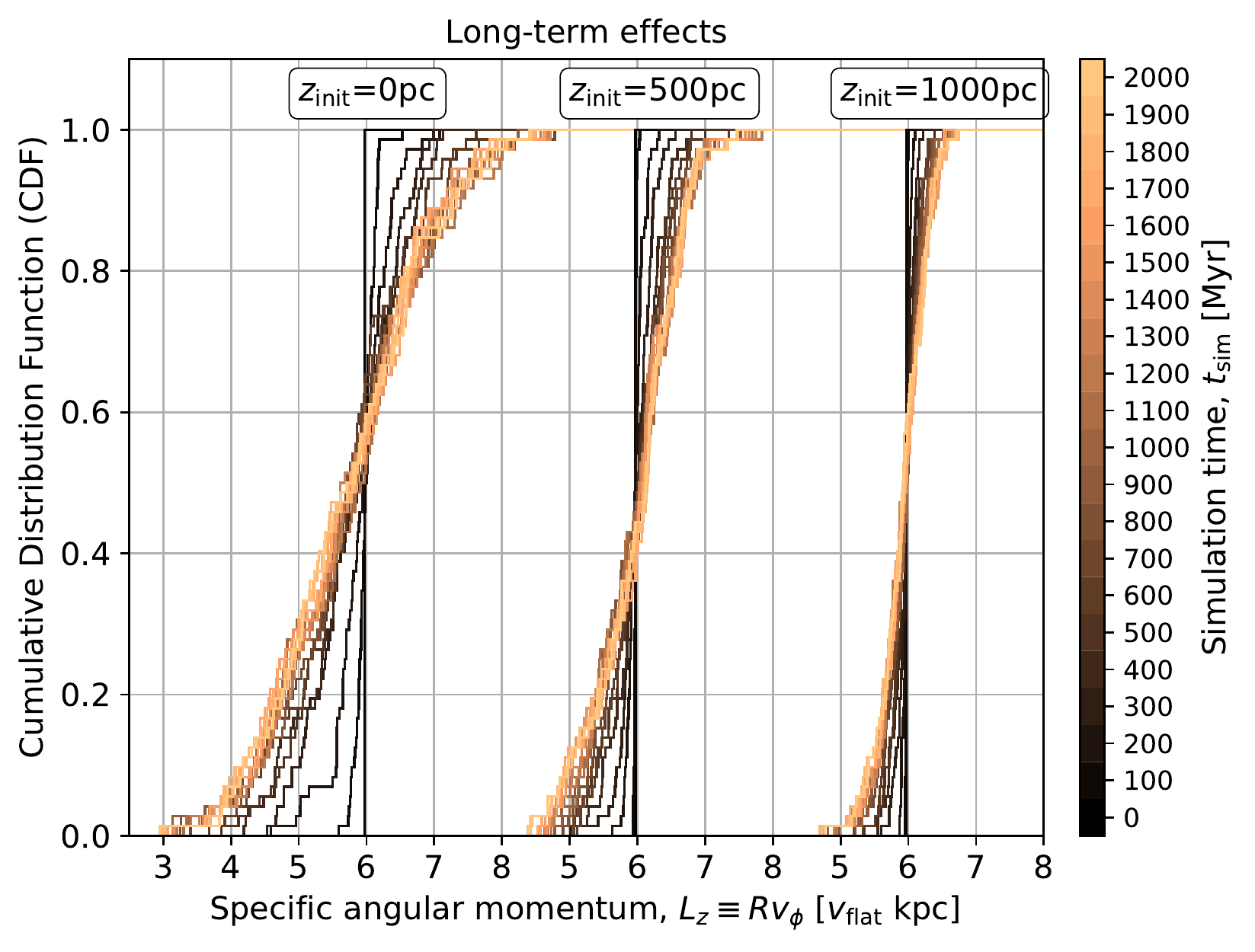}
    	\includegraphics[width=\columnwidth]{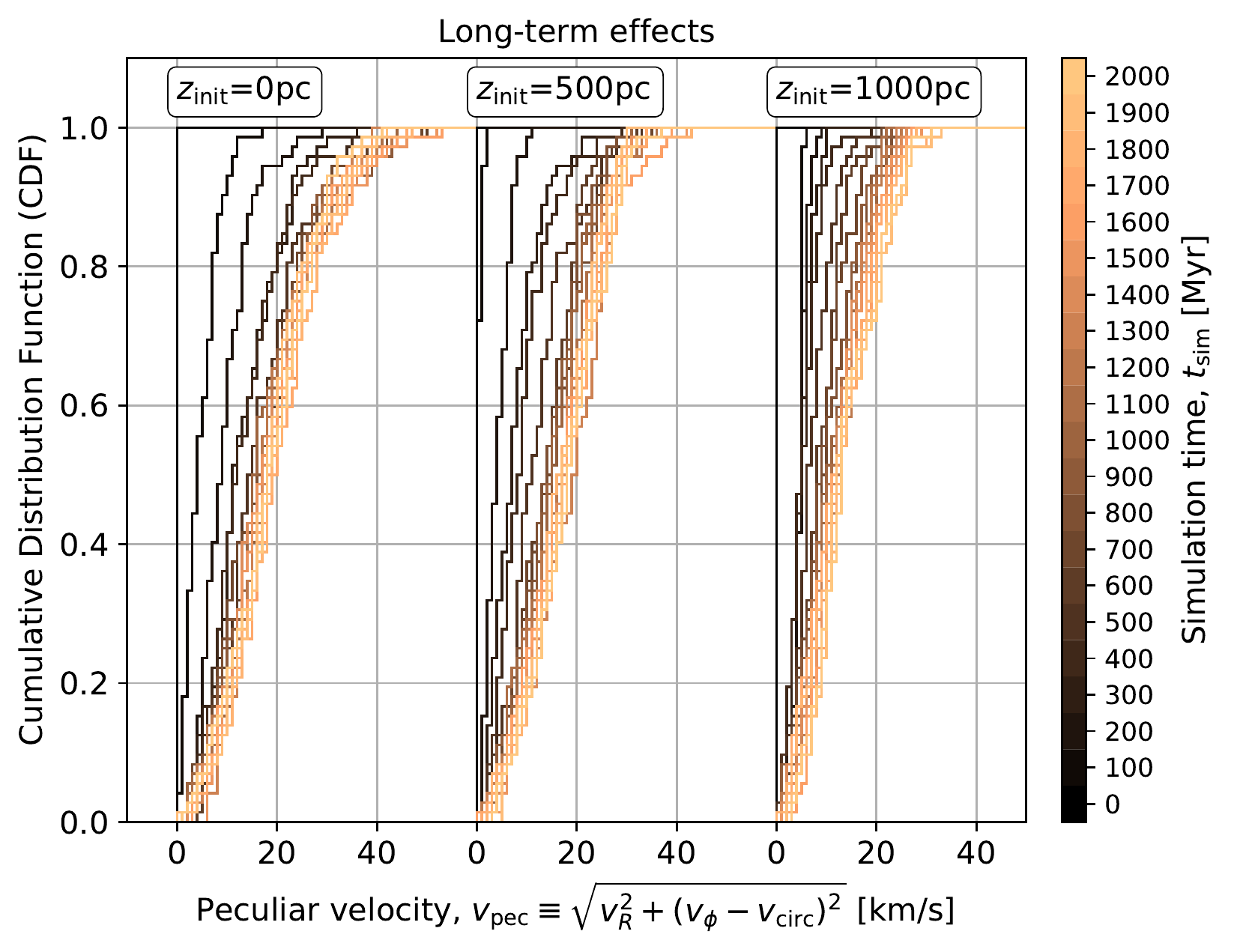}
    	\includegraphics[width=\columnwidth]{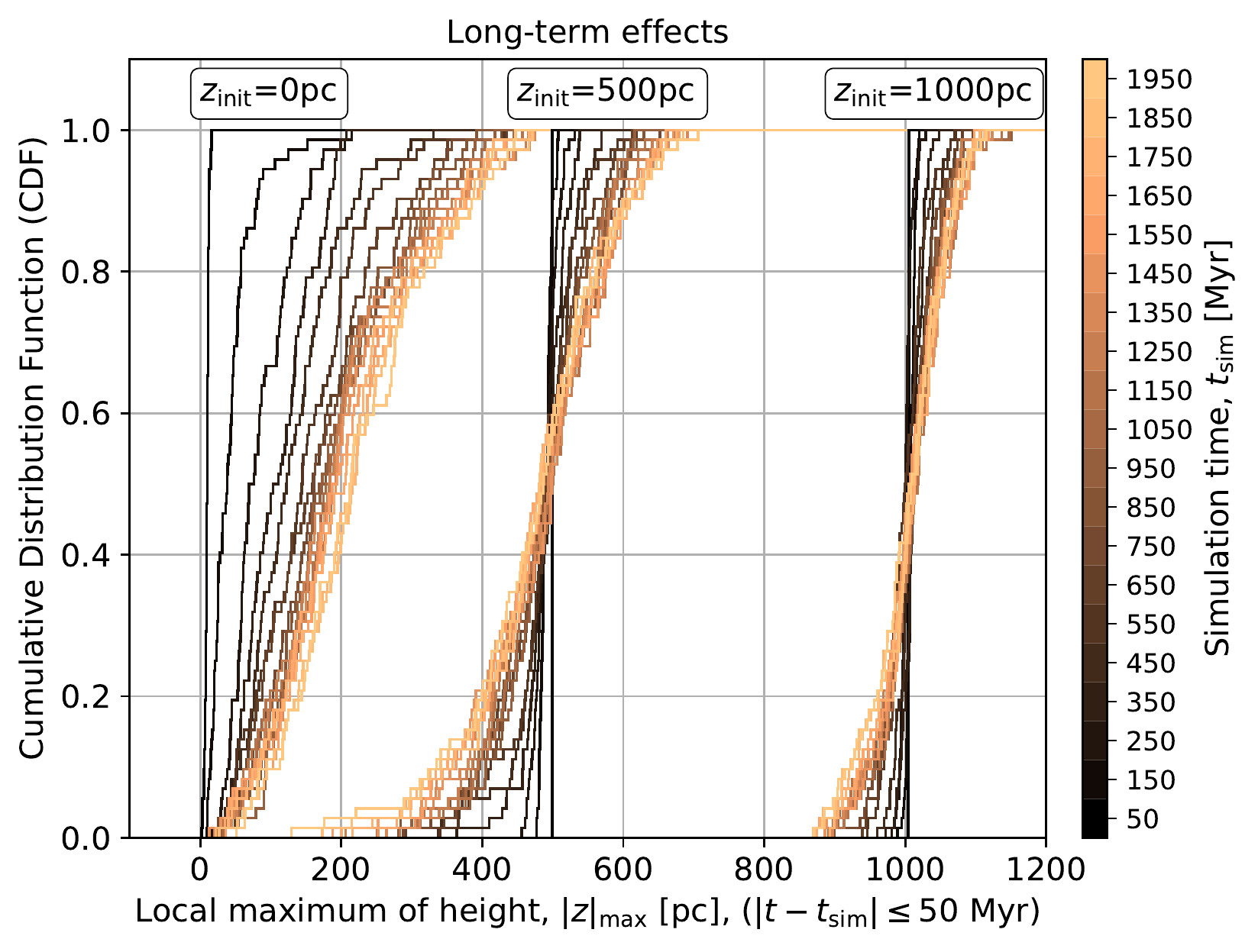}
    \caption{Long-term effects: time evolution of normalised cumulative distribution functions of $R$, $L_z$, $v_{\rm pec}$, and $|z|_{\rm max}$, showing a time evolution up to $t_{\rm sim} = 2000$\Myr. The initial radius is $R_{\rm init} = 6$\kpc\, and the initial heights are $z_{\rm init} = $ 0, 500, and 1000\pc. The simulation model is \textit{16pc\_0msun}.}
    \label{fig: long term effects}
\end{figure*}

Fig.~\ref{fig: long term effects} shows the cumulative distribution functions for the simulation model of \textit{16pc\_0msun}. This figure shows time evolution up to 2000\Myr, which is almost twice as long as 1100\Myr\ used in other plots as default. As discussed in Section~\ref{subsubsec: individual particles}, the distribution functions drastically change in the first several hundred Myr, but no significant change will appear after around 600\Myr.

\section{dependence on initial radial position}
\label{appendix: the dependence on the initial radial position}

\begin{figure*}
    \centering
	\includegraphics[width=\columnwidth]{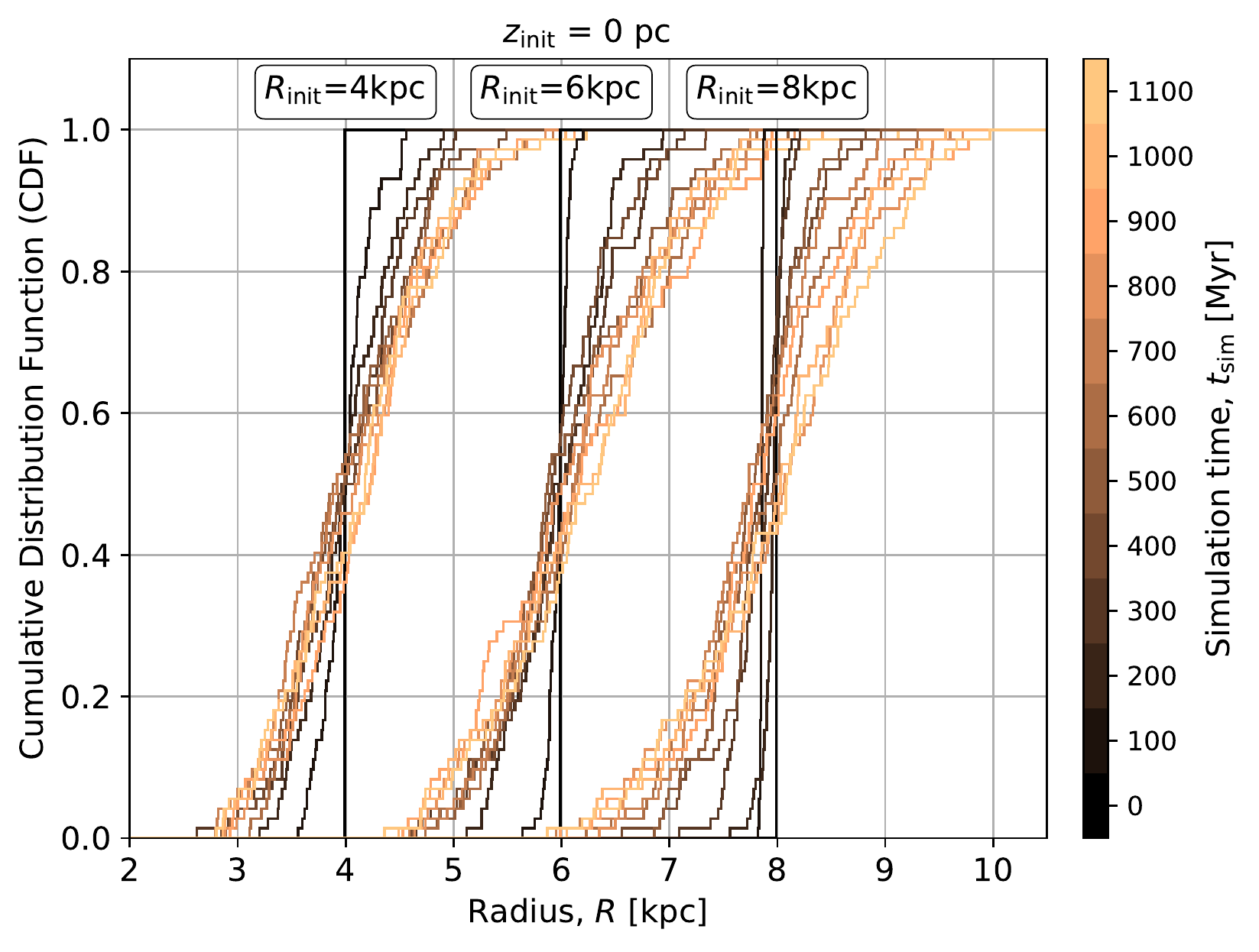}
	\includegraphics[width=\columnwidth]{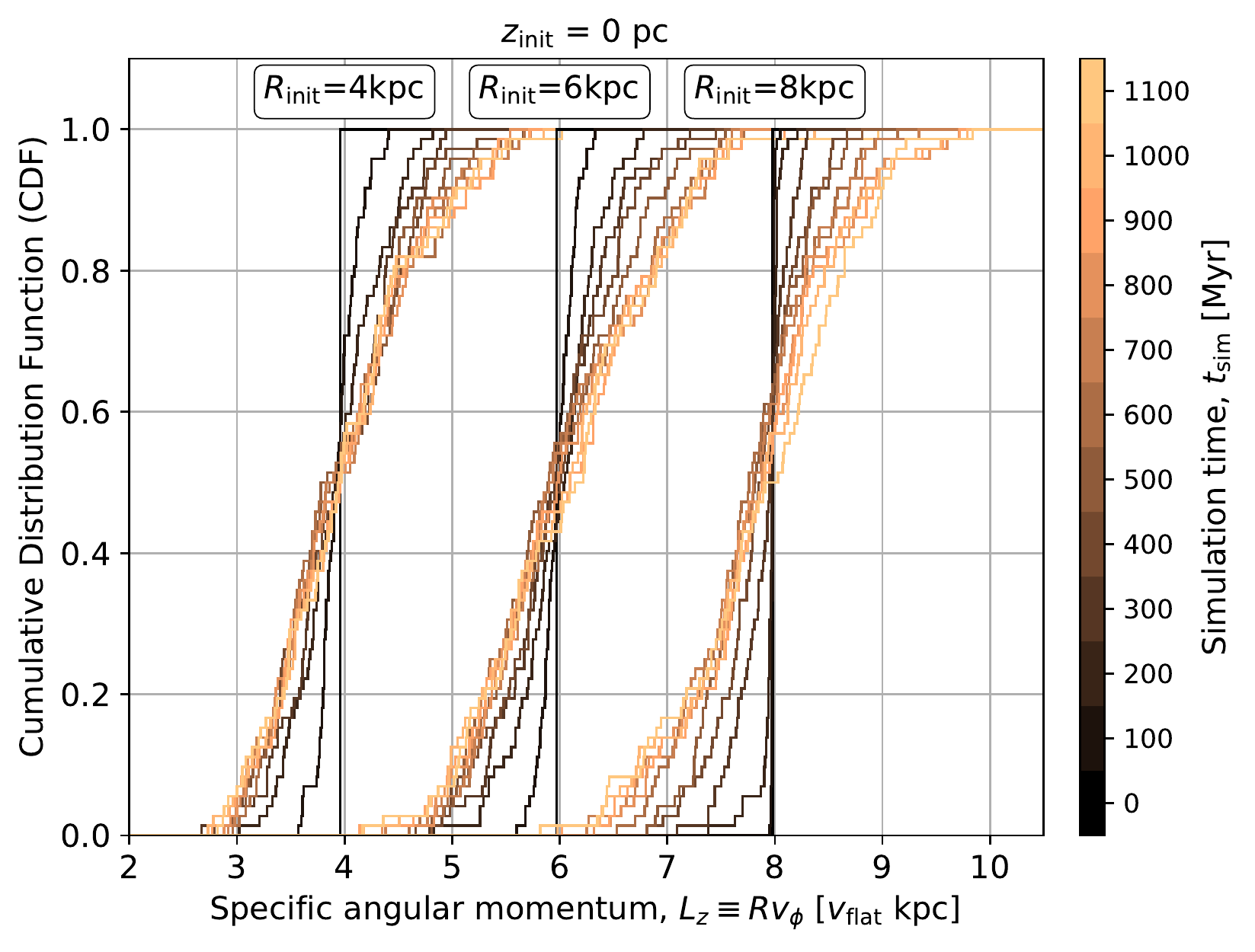}
	\includegraphics[width=\columnwidth]{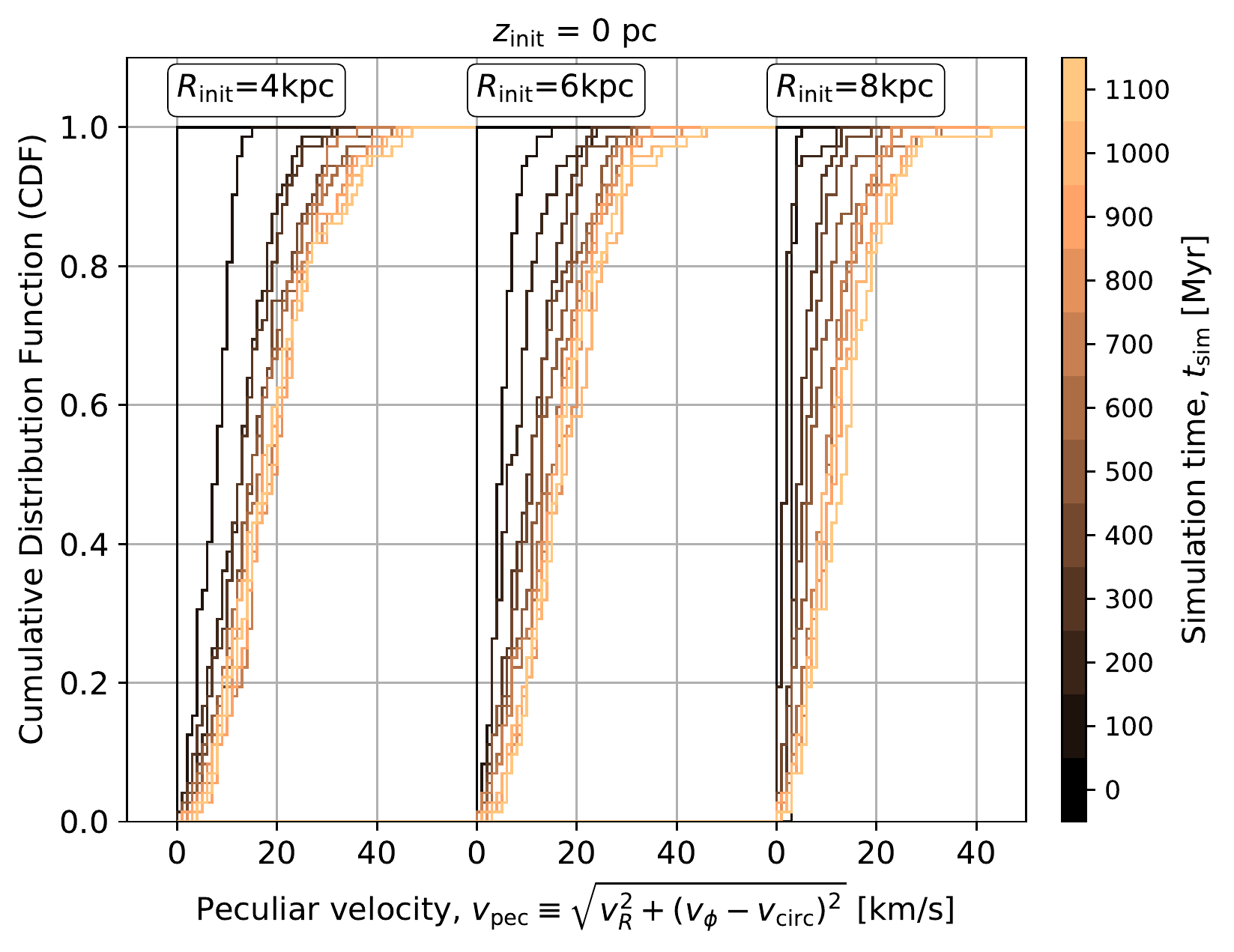}
	\includegraphics[width=\columnwidth]{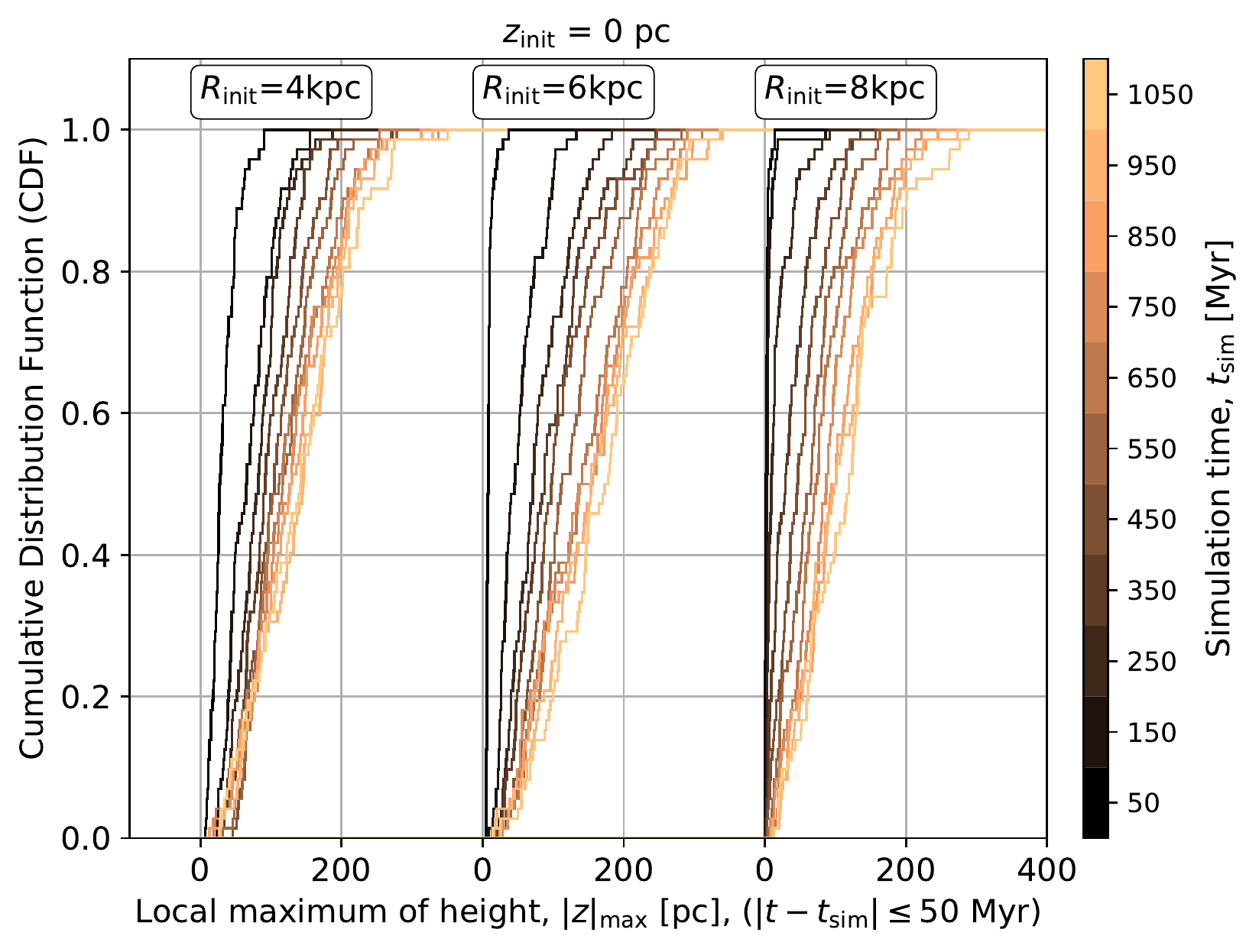}
    \caption{Dependence on the initial radial position $R_{\rm init}$: time evolution of normalised cumulative distribution functions of $R$, $L_z$, $v_{\rm pec}$, and $|z|_{\rm max}$ for tracer particles with the initial radii of $R_{\rm init} = $ 4, 6, and 8\kpc\ and the initial height of $z_{\rm init} = 0$\pc. The simulation model is \textit{8pc\_0msun}.}
    \label{fig: radial dependence}
\end{figure*}

Fig.~\ref{fig: radial dependence} shows the cumulative distribution functions for particles with initial radii of $R_{\rm init} = $ 4, 6, and 8\kpc\ and the initial height of $z_{\rm init} = 0$\pc. This figure shows no significant dependence of the stellar scattering on the initial radial position.

\section{Dependence on simulation models}
\label{appendix: the dependence on simulation models}

\begin{figure*}
    \centering
	\includegraphics[width=\columnwidth]{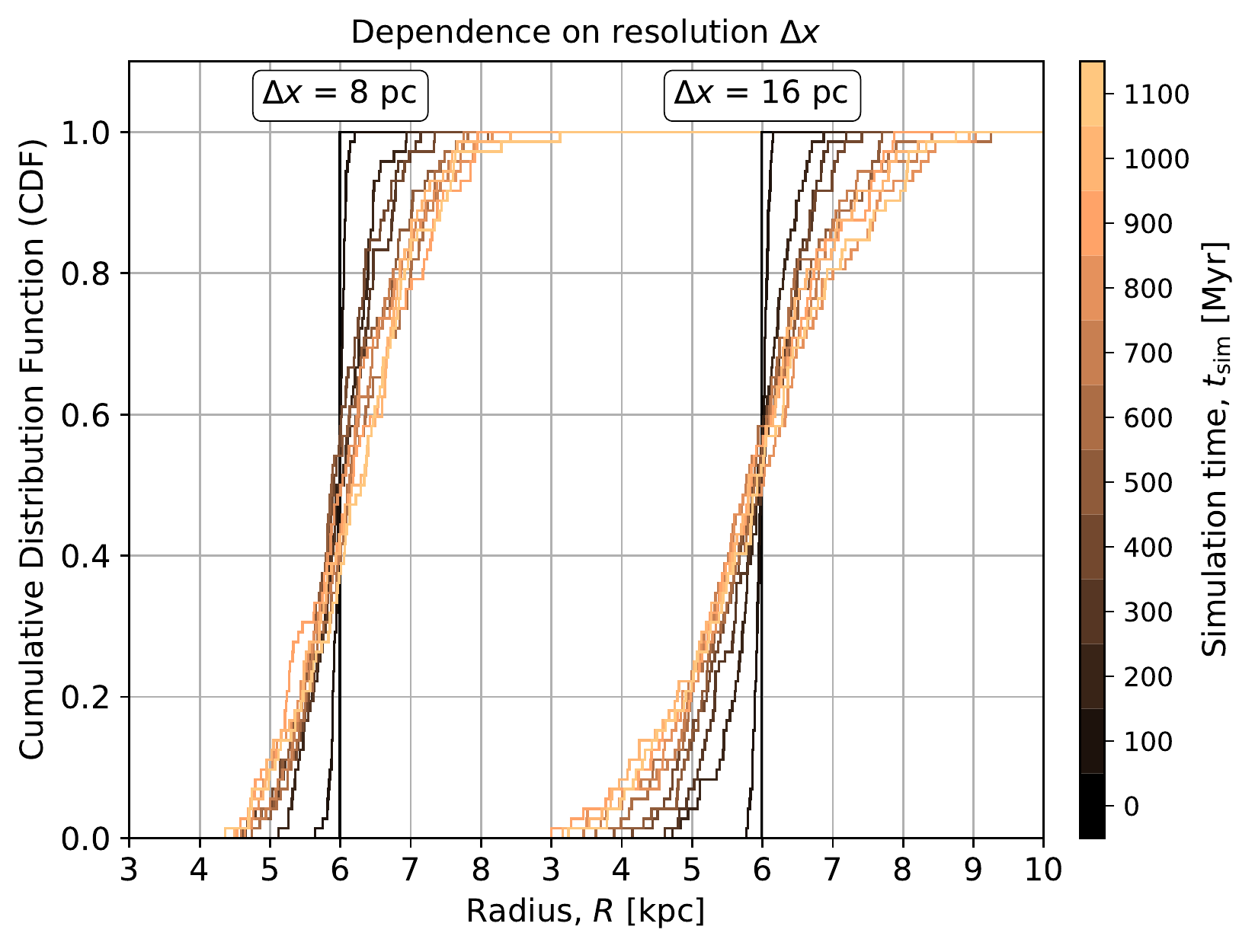}
	\includegraphics[width=\columnwidth]{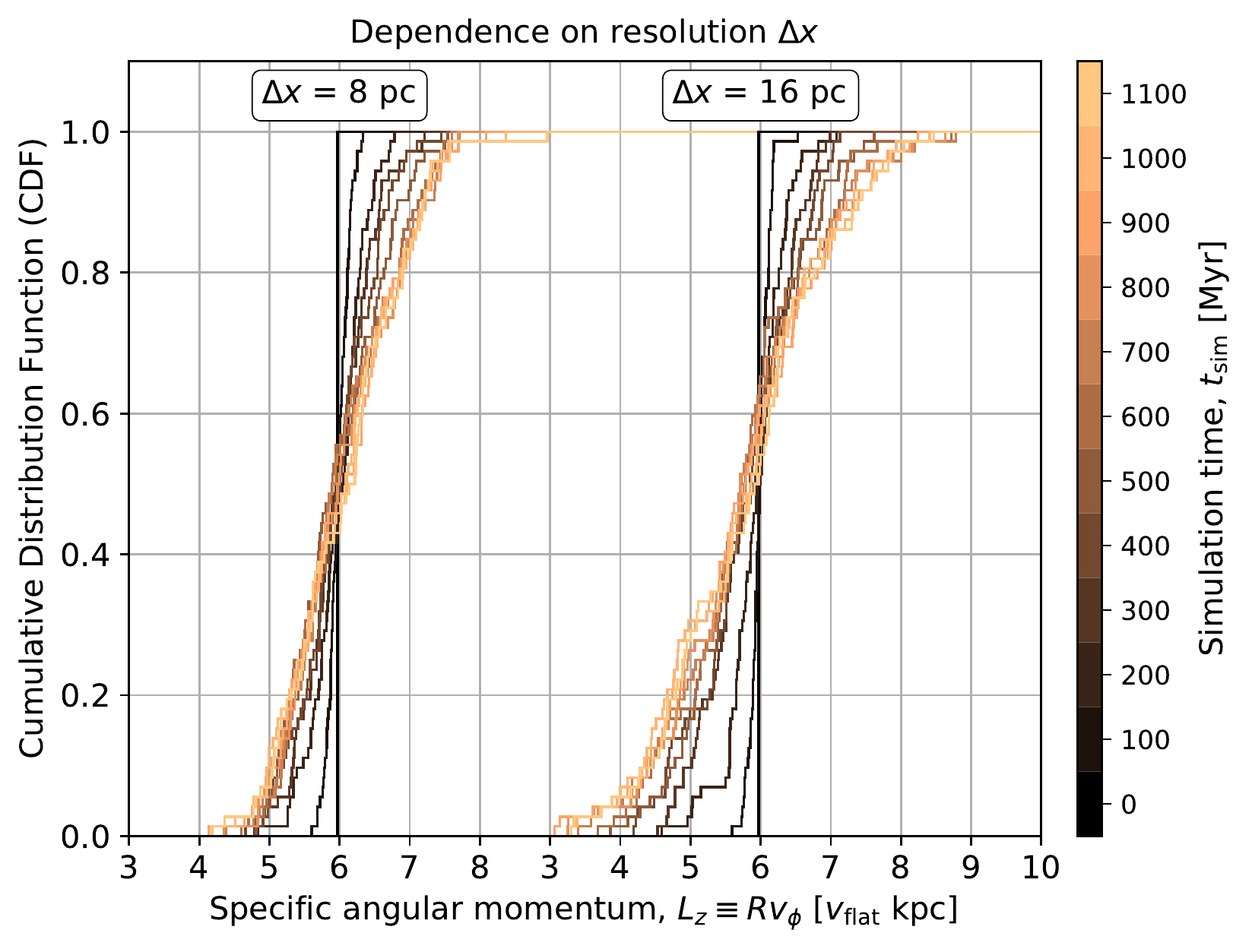}
	\includegraphics[width=\columnwidth]{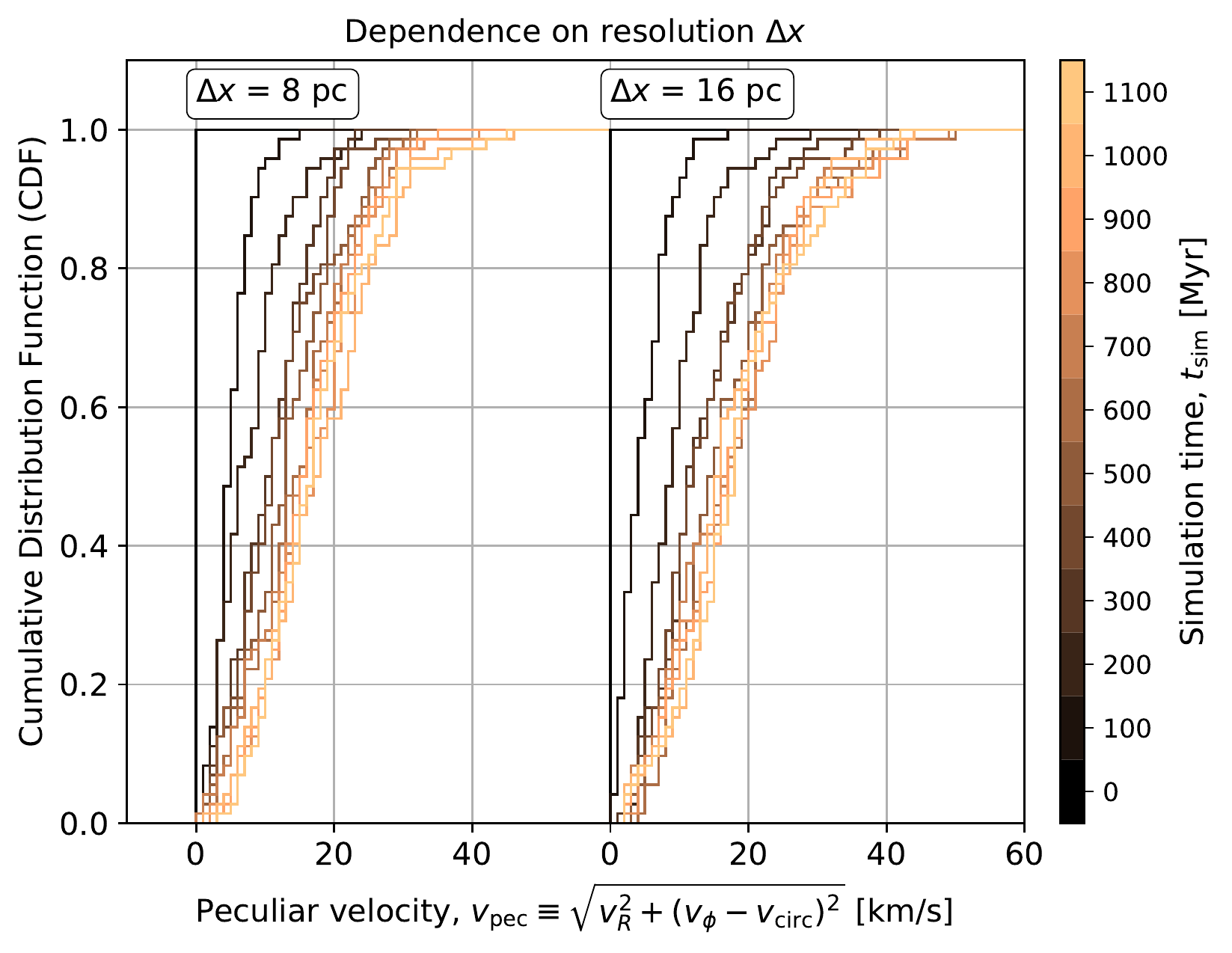}
	\includegraphics[width=\columnwidth]{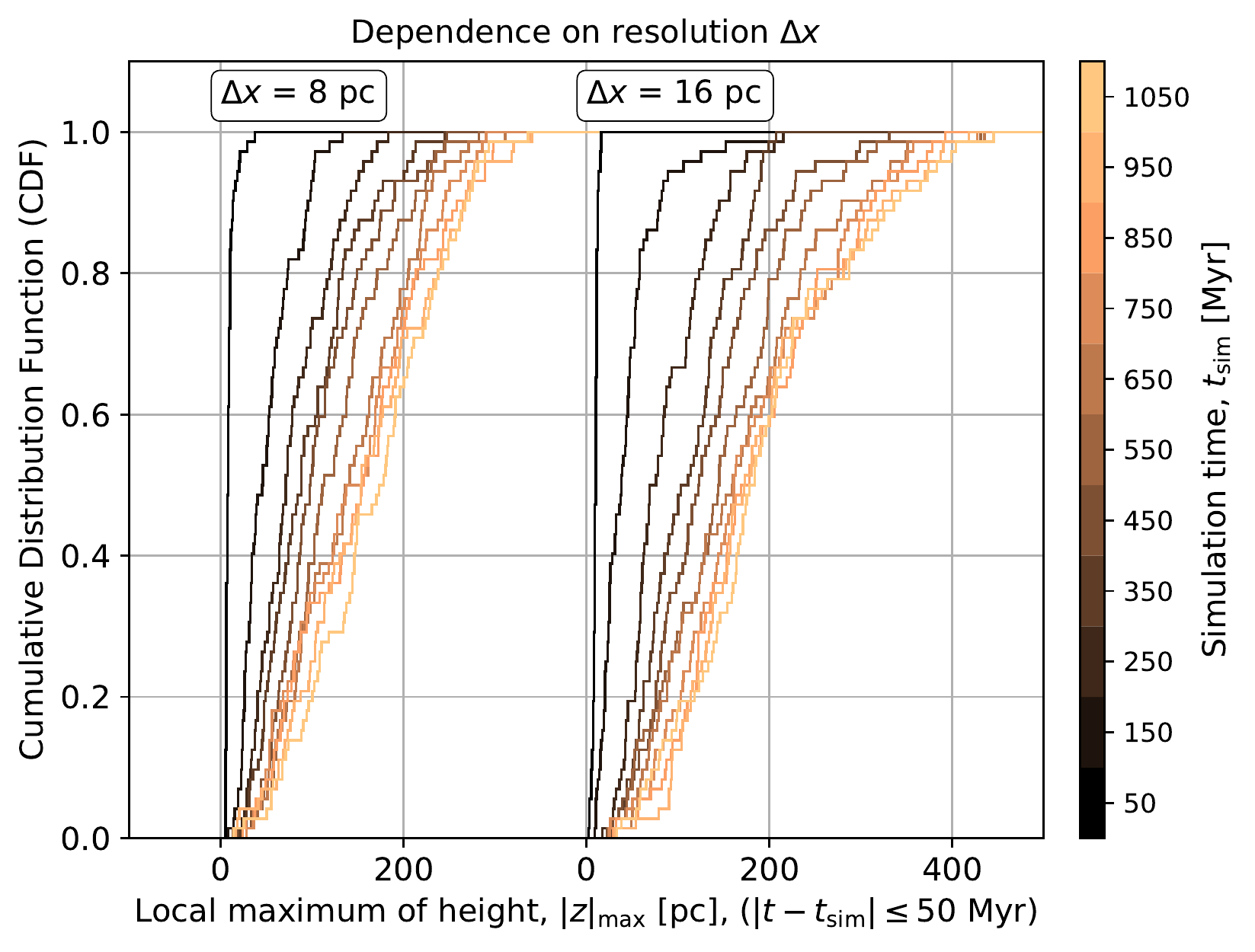}
    \caption{Dependence on the resolution $\Delta x$: time evolution of normalised cumulative distribution functions of $R$, $L_z$, $v_{\rm pec}$, and $|z|_{\rm max}$ for two cases with the spatial resolutions of 8\pc\ and 16\pc, corresponding to the simulation models of \textit{8pc\_0msun} and \textit{16pc\_0msun}, respectively. The initial radius is $R_{\rm init} = 6$\kpc, and the initial height is $z_{\rm init} = 0$\pc.}
    \label{fig: resolution dependence}
\end{figure*}

Fig.~\ref{fig: resolution dependence} shows the cumulative distribution functions, comparing the case with 8\pc\ resolution to the one with 16\pc\ resolution. The resolution of $\Delta x = 8$\pc\ has been used in the fiducial model of \textit{8pc\_0msun}. In the lower resolution case ($\Delta x = 16$\pc), we see higher or lower end tails for all four physical quantities, which might be because more massive and more extensive GMCs form in the lower resolution model, as shown in Section~\ref{subsec: distribution of giant molecular clouds}. However, most parts of the distribution functions are the same; for example, the mean values and the standard deviations are almost the same. Therefore, we conclude that there is little dependence on the resolution $\Delta x$.

\begin{figure*}
    \centering
	\includegraphics[width=\columnwidth]{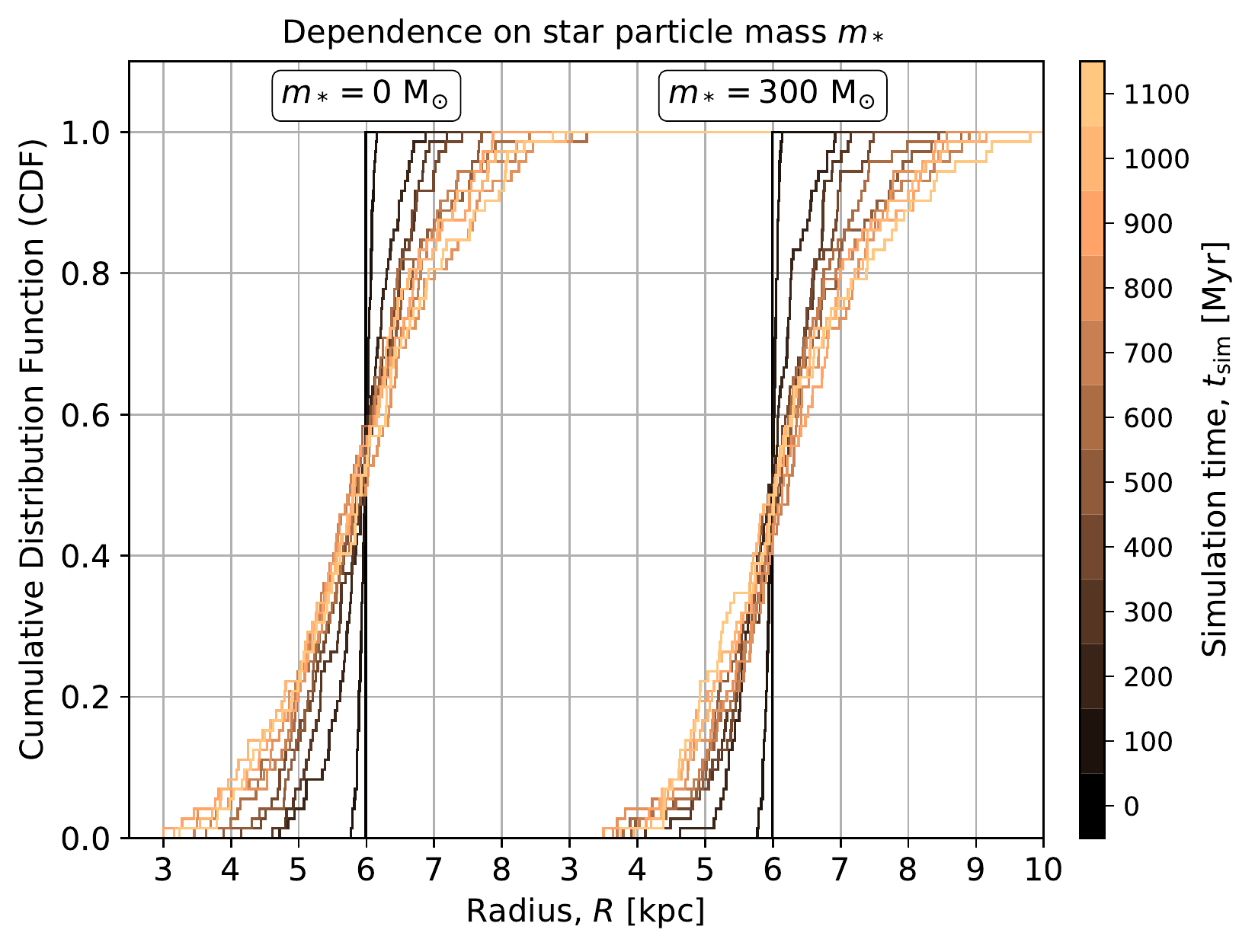}
	\includegraphics[width=\columnwidth]{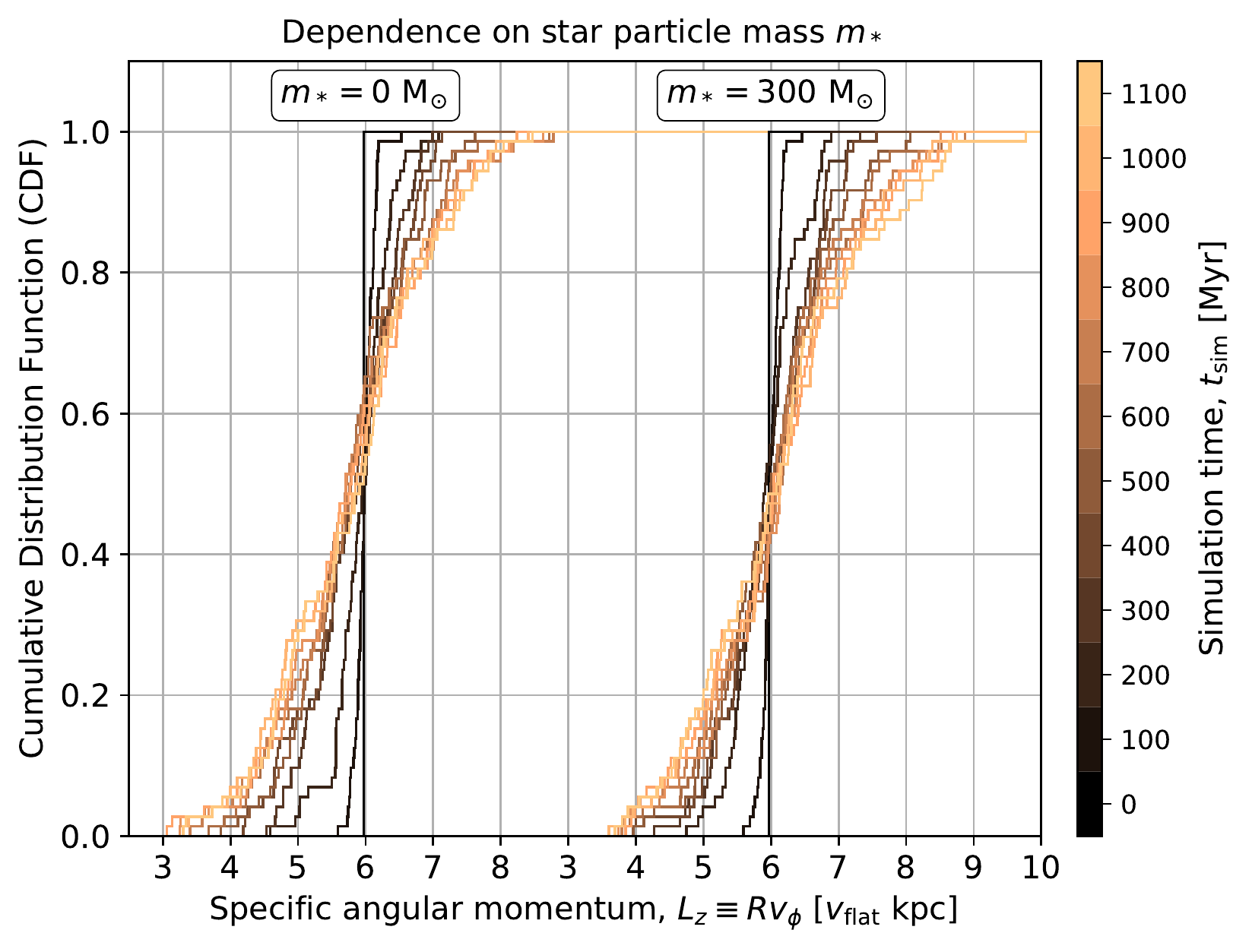}
	\includegraphics[width=\columnwidth]{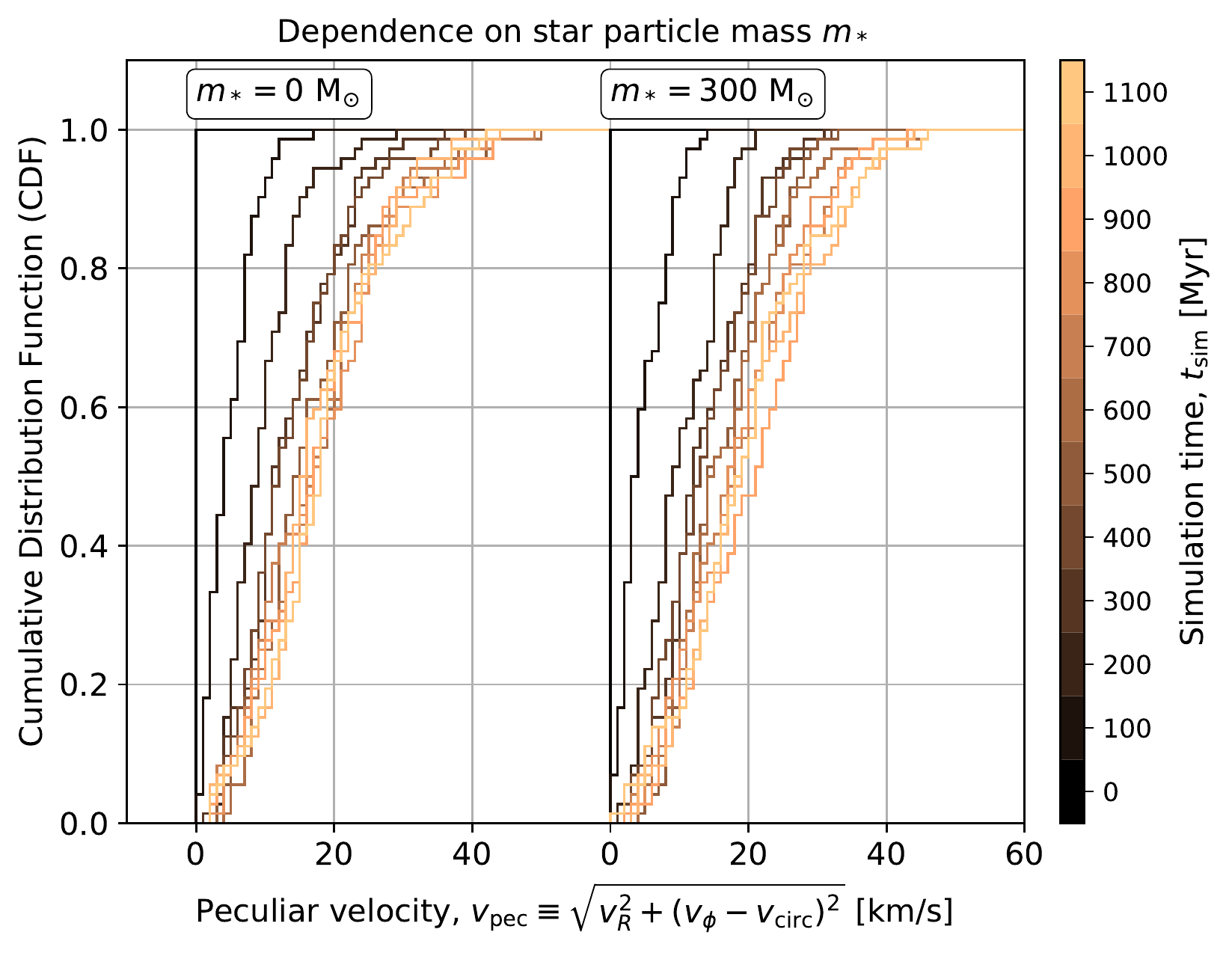}
	\includegraphics[width=\columnwidth]{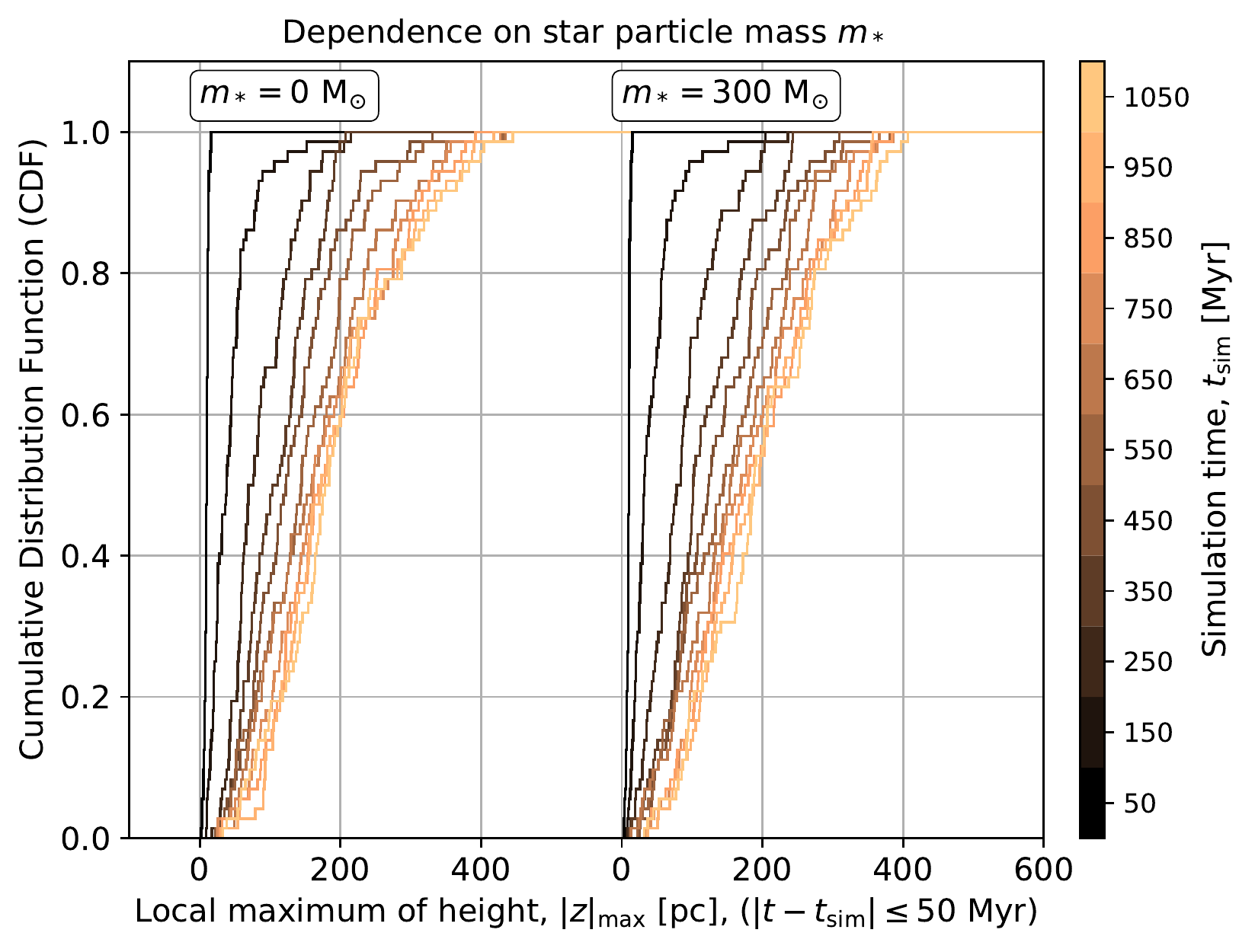}
    \caption{Dependence on the star particle's initial mass $m_*$: time evolution of normalised cumulative distribution functions of $R$, $L_z$, $v_{\rm pec}$, and $|z|_{\rm max}$ for two cases with the star particle's initial masses of 0\Msun\ and 300\Msun, corresponding to the simulation models of \textit{16pc\_0msun} and \textit{16pc\_300msun}, respectively. The initial radius is $R_{\rm init} = 6$\kpc, and the initial height is $z_{\rm init} = 0$\pc.}
    \label{fig: initial star particle mass dependence}
\end{figure*}

Fig.~\ref{fig: initial star particle mass dependence} shows the cumulative distribution functions, comparing the case with the star particle's initial mass of 0\Msun\ to the one with 300\Msun. The mass of $m_* = 0$\Msun\ has been used in the fiducial model of \textit{8pc\_0msun}. We do not see a significant difference between the two cases. It suggests that the effect of the unrealistic point mass of 300\Msun\ used in the star formation recipe is minor in terms of gravitational scattering of tracer particles by GMCs.

\section{Time evolution of significant outward migrator}
\label{appendix: time evolution of Sun-like motion particles}

\begin{figure*}
    \centering
	\includegraphics[width=\columnwidth]{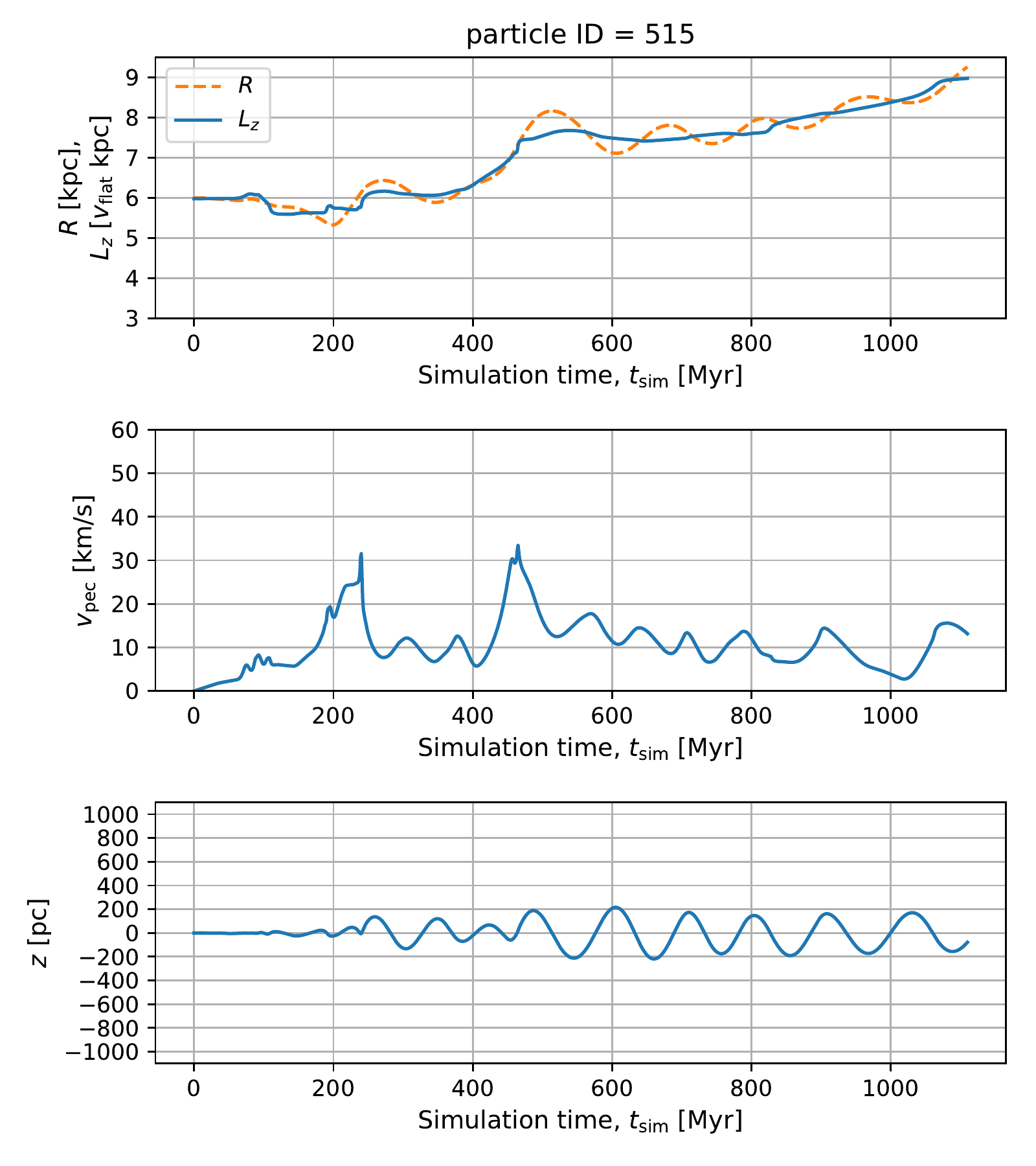}
	\includegraphics[width=\columnwidth]{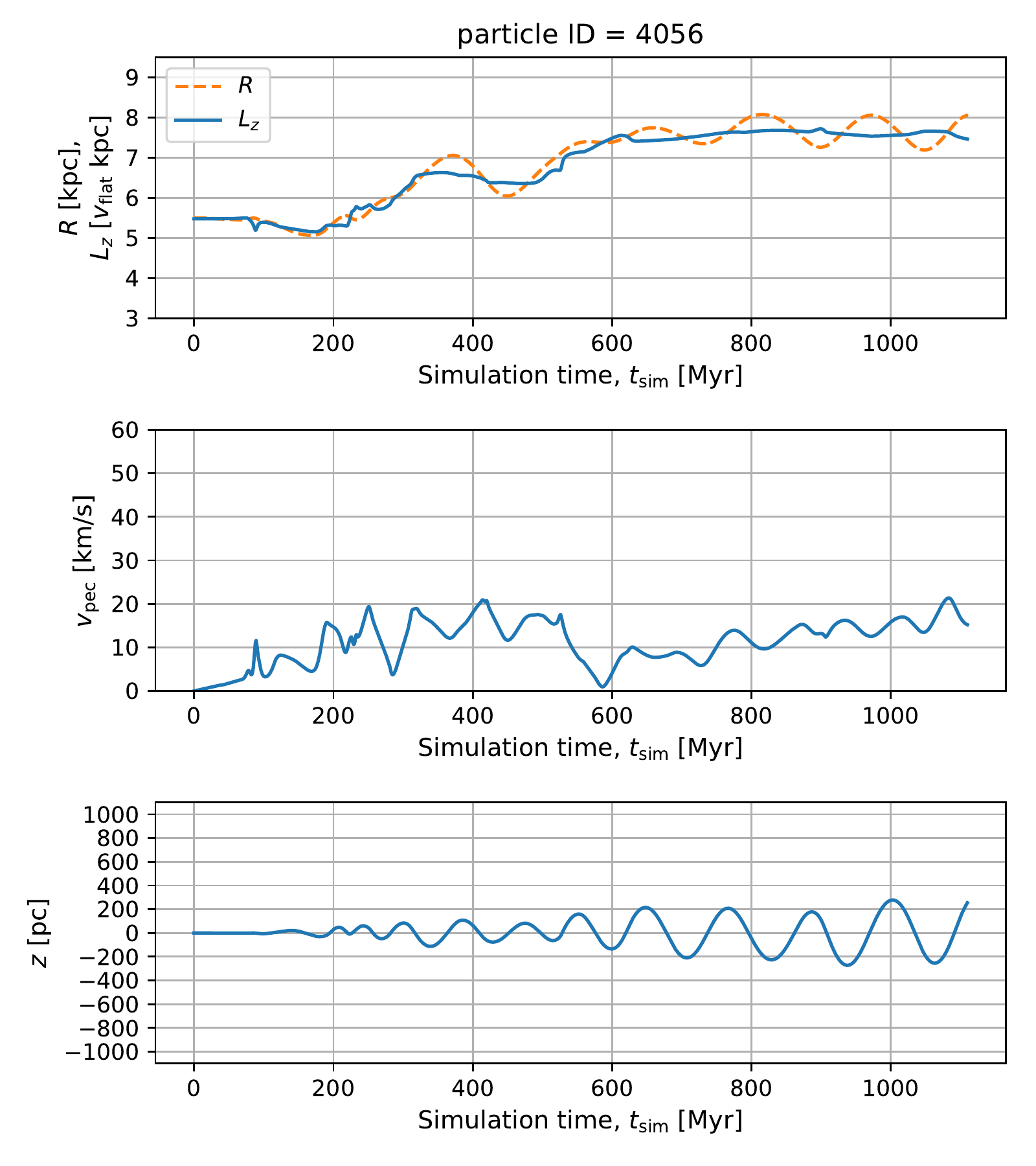}
    \caption{Time evolution of the radius, specific angular momentum (top row), peculiar velocity (middle row), and height (bottom row) for two particles (left and right columns for each) that show significant outward radial migration by the time of $t_{\rm sim} \sim 1$\Gyr\ without obtaining a large peculiar velocity, like the current motion of the Solar system. They are assigned the particle IDs of 515 and 4056, respectively, and Fig.~\ref{fig: scatter plot} shows their locations in the scatter plots.}
    \label{fig: time evolution each particle}
\end{figure*}

In Fig.~\ref{fig: scatter plot}, we see that two particles show significant outward radial migration without obtaining a large peculiar velocity, like the current motion of the Solar system. Fig.~\ref{fig: time evolution each particle} shows the time evolution of the radius, specific angular momentum, peculiar velocity, and height for those particles. These particles show an almost monotonic increase in their angular momentum and then reach large values in the end. On the other hand, they do not show a monotonic increase in their peculiar velocity, indicating that interactions with GMCs can sometimes decrease the peculiar motion.

%%%%%%%%%%%%%%%%%%%%%%%%%%%%%%%%%%%%%%%%%%%%%%%%%%

% Don't change these lines
\bsp	% typesetting comment
\label{lastpage}
\end{document}